\documentclass[reprint,prb,superscriptaddress,twocolumn,notitlepage,longbibliography,floatfix]{revtex4-2}
\usepackage[utf8]{inputenc}
\pdfoutput=1
\usepackage{times}
\usepackage[dvipsnames]{xcolor}
\usepackage{array}
\usepackage{verbatim}
\usepackage{multirow}
\usepackage{amsmath}
\usepackage{amssymb}
\usepackage{dsfont}
\usepackage{graphicx}
\usepackage{esint}
\usepackage[unicode=true,
 bookmarks=true,bookmarksnumbered=true,bookmarksopen=false,
 breaklinks=true,pdfborder={0 0 0},backref=false,colorlinks=true]
 {hyperref}
\usepackage{enumitem}
\usepackage[normalem]{ulem}
\usepackage{pifont}
\usepackage{circledsteps}

\hypersetup{
    linkcolor=purple,          
    citecolor=blue,        
    filecolor=magenta,      
    urlcolor=magenta           
}

\usepackage{zref-clever}
\newcommand{\cref}[1]{\zcref{#1}}

\zcRefTypeSetup{equation}{
    Name-sg=\textcolor{black}{Eq.},
    name-sg=\textcolor{black}{Eq.},
    Name-pl=\textcolor{black}{Eqs.},
    name-pl=\textcolor{black}{Eqs.}
}
\zcRefTypeSetup{section}{
    Name-sg=\textcolor{black}{Sec.},
    name-sg=\textcolor{black}{Sec.},
    Name-pl=\textcolor{black}{Secs.},
    name-pl=\textcolor{black}{Secs.}
}
\zcRefTypeSetup{appendix}{
    Name-sg=\textcolor{black}{App.},
    name-sg=\textcolor{black}{App.},
    Name-pl=\textcolor{black}{Apps.},
    name-pl=\textcolor{black}{Apps.}
}
\zcRefTypeSetup{figure}{
    Name-sg=\textcolor{black}{Fig.},
    name-sg=\textcolor{black}{Fig.},
    Name-pl=\textcolor{black}{Figs.},
    name-pl=\textcolor{black}{Figs.}
}
\zcRefTypeSetup{table}{
    Name-sg=\textcolor{black}{Table},
    name-sg=\textcolor{black}{Table},
    Name-pl=\textcolor{black}{Tables},
    name-pl=\textcolor{black}{Tables}
}

\renewcommand{\paragraph}[1]{\vspace{0.2cm}{\textit{#1}---\!}} 

\usepackage{simpler-wick}
\allowdisplaybreaks[1] 
\newcommand{\mbf}{\mathbf}
\newcommand{\mbb}{\mathbb}
\newcommand{\mcl}{\mathcal}

\newcommand{\mrm}{\mathrm}

\newcommand{\td}{\widetilde}
\newcommand{\ovl}{\overline}
\newcommand{\wht}{\widehat}

\def\pare#1{\left( #1 \right)}
\def\brak#1{\left[#1\right]}

\def\bra#1{\langle #1 |}
\def\ket#1{| #1 \rangle}

\def\inn#1{\langle #1 \rangle}
\def\Inn#1{\left\langle #1 \right\rangle}
\def\abs#1{\left| #1 \right|}
\def\Im{\mathrm{Im}}
\def\Re{\mathrm{Re}}
\def\sgn{\mathrm{sgn}}
\def\ii{\mathrm{i}}
\def\Tr{\mathrm{Tr}}

\def\SUt{{\mathrm{SU}(2)}}
\def\SUf{{\mathrm{SU}(4)}}
\def\Uo{{\mathrm{U}(1)}}

\def\Uf{{\mathrm{U}(4)}}
\def\Zt{{\mathrm{Z}_2}}

\def\mN{\mathcal{N}}

\def\DL{\frac{2\pi}{L}}
\def\Pbc{P_{\rm bc}}
\def\dd{\mathrm{d}}

\def\vxi{\vec{\xi}}
\def\vvphi{\vec{\varphi}}

\def\emp{|\mathrm{emp}\rangle}

\def\spin{\varsigma}

\def\Si{\mathrm{Si}}
\def\Ci{\mathrm{Ci}}
\def\PP{\mathbb{P}}

\def\OO{\mathcal{O}}

\def\TK{T_{\rm K}}

\def\vN{\vec{N}}

\begin{document}
\title{Bosonization Solution to Spin-Valley Kondo Problem: Finite-Size Spectrum and Renormalization Group Analysis}

\author{Yi-Jie Wang}
\thanks{These authors contributed equally to this work.}
\affiliation{International Center for Quantum Materials, School of Physics, Peking University, Beijing 100871, China}

\author{Geng-Dong Zhou}
\thanks{These authors contributed equally to this work.}
\affiliation{International Center for Quantum Materials, School of Physics, Peking University, Beijing 100871, China}

\author{Hyunsung Jung}
\affiliation{Department of Physics and Astronomy, Seoul National University, Seoul 08826, Korea}

\author{Seongyeon Youn}
\affiliation{Department of Physics and Astronomy, Seoul National University, Seoul 08826, Korea}
\affiliation{Center for Theoretical Physics, Seoul National University, Seoul 08826, Korea}

\author{Seung-Sup B.~Lee}
\email{sslee@snu.ac.kr}
\affiliation{Department of Physics and Astronomy, Seoul National University, Seoul 08826, Korea}
\affiliation{Center for Theoretical Physics, Seoul National University, Seoul 08826, Korea}
\affiliation{Institute for Data Innovation in Science, Seoul National University, Seoul 08826, Korea}

\author{Zhi-Da Song}
\email{songzd@pku.edu.cn}
\affiliation{International Center for Quantum Materials, School of Physics, Peking University, Beijing 100871, China}
\affiliation{Hefei National Laboratory, Hefei 230088, China}
\affiliation{Collaborative Innovation Center of Quantum Matter, Beijing 100871, China}

\date{\today}

\begin{abstract}
Spin-valley Anderson impurities (SVAIM) with (anti-)Hund's splitting provide a natural explanation to the origin of pairing potential and pseudogap in the magic-angle graphene \cite{wang_2025_solution}. 
In this work, we derive and analytically solve the low-energy Kondo theories for SVAIM at half-filling, with especial focus on the two anti-Hund's regimes: the impurity is either dominated by a valley doublet, or a trivial singlet. 
In the doublet regime, we reveal that a novel pair Kondo scattering $\lambda_x$ is required to flip the valley doublet, which involves a quartic operator of bath electrons. 
Our renormalization group (RG) calculation based on the Coulomb gas analog shows $\lambda_x$ drives a phase transition of the Berezinskii--Kosterlitz--Thouless type. 
One side of the transition is an anisotropic doublet phase, characterized by non-universal phase shifts of bath electrons and non-analytic impurity susceptibilities, while the other is a Fermi liquid formed by pair-Kondo resonance. 
The finite-size many-body spectrum, thermodynamic quantities, and correlation functions for both phases are analytically solved. 
Remarkably, the solution in the pair-Kondo Fermi liquid is achieved via the constructive approach of bosonization--refermionization along a solvable fixed line, where the many-body interaction $\lambda_x$ is mapped into a pseudo-fermion bilinear in a rigorous manner. 
Finally, we also apply the RG analysis to the singlet regime, and identify a second-order phase transition between the Kondo Fermi liquid and a local singlet phase. 
\end{abstract}

\maketitle

\section{Introduction}

Moir\'e systems provide a versatile platform to investigate novel correlation physics, including unconventional superconductors \cite{Cao_2018_SC, Lu_2019_superconductors, cao_pauli-limit_2021, xia_superconductivity_2025, guo_superconductivity_2025}, integer \cite{serlin_intrinsic_2020, Nuckolls_2020_strongly, chen_tunable_2020, li_quantum_2021} and fractional \cite{cai_signatures_2023, xu_observation_2023, zeng_thermodynamic_2023, han_signatures_2025, xie_tunable_2025} Chern insulators, etc.
From a theoretical perspective, these fruitful quantum phases arise from the interplay between tunable interactions and the rich topological and geometrical properties of single-particle electron wavefunctions, shaped by the moir\'e modulation. 
For fractional Chern insulators in moir\'e systems \cite{reddy_fractional_2023, yu_fractional_2024, wang_fractional_2024}, the low-energy single-particle wavefunctions can serve as analogs to the lowest Landau level \cite{tarnopolsky_origin_2019, ledwith_fractional_2020, wang_origin_2023, wu_topological_2019, yu_giant_2020}.
Furthermore, moir\'e electron orbitals can also emulate the Kondo lattices \cite{kumar_gate_2022} and multi-orbital Hubbard model \cite{fischer2024theory}.  
Specifically, for magic-angle twisted bilayer and trilayer graphene (MATBG/MATTG), it has been advocated that the correlation physics in topological flat bands is captured by the topological heavy fermion (THF) model \cite{song_magic-angle_2022, shi_heavy-fermion_2022, Yu_2023_THF_TSTG}.
This formalism features strongly correlated local $f$ orbitals \cite{haule_mott-semiconducting_2019, calderon_interactions_2020, Liu_2019_pseudoLL} hybridizing with itinerant Dirac $c$ bands, allowing applications of theoretical approaches such as dynamical mean-field theory (DMFT)~\cite{Georges1996} and slave-particle to MATBG/MATTG \cite{Chou_2023_Kondo, zhou_kondo_2024, Rai_2023_DMFT, Hu_2023_Kondo, Hu_2023_Symmetric, Chou_2023_scaling, Datta_2023_heavy, Lau_2023_topological, calugaru_thermoelectric_2024, youn_hundness_2024, herzog_2025_kekule, crippa_2025_dynamicalcorrelation, calugaru_2025_obtainingspectral}. 
Recent experimental progress has provided growing support for the THF picture \cite{xiao_2025_interacting, merino_interplay_2025, kim_2025_resolvingintervalleygapsmanybody, zhang_2025_heavyfermions}. 

The quantum impurity problem stands the central importance in the THF model, as well as in other strongly correlated problems with predominant local correlation and suitable for DMFT framework. 
Importantly, the artificial moir\'e orbitals in general host new effective internal degrees of freedom (flavors), which originate from the atomic-scale oscillations of electron wave-packets, such as the two $K$-valleys in graphene. 
In the companion work \cite{wang_2025_solution}, we have considered a spin-valley Anderson impurity model (SVAIM), which bears immediate relevance to MATBG/MATTG in the presence of hetero-strain \cite{herzog_2025_efficient}. 
We show that the SVAIM with (anti-)Hund's splitting provides a unified origin to the intriguing pairing potential \cite{wang_molecular_2024,zhao_2025_rvb} and pseudo-gap phenomenon \cite{youn_hundness_2024,youn_hundness_2025} in these systems \cite{Cao_2018_SC, Lu_2019_superconductors, Oh_2021_evidence, park_experimental_2025, kim_2025_resolvingintervalleygapsmanybody}. 

Complementing these phenomenological applications \cite{wang_2025_solution}, this work focuses on the analytical solutions to the phase transitions and fixed points in SVAIM at half-filling. 
Especially, we obtain the following results. 
(i) The doublet regime features a novel pair Kondo (PK) coupling $\lambda_x$, which is a highly non-trivial many-body interaction: it contains a quartic operator of bath electrons in product with a doublet flipping action. 
(ii) Via bosonization renormalization group (RG) calculations based the Coulomb gas analog, we show that through the interplay with the Ising coupling $\rho_z$ between the valley doublet and the bath valley densities, $\lambda_x$ can drive a Berezinskii--Kosterlitz--Thouless (BKT) transition. 
(iii) On one side of the transition, $\lambda_x \to 0$, and $\rho_z$ can flow to a continuous fixed line terminating at $\rho_z < \rho_z^c=\frac{1}{2} - \frac{1}{2\sqrt{2}}$. 
We dub this phase as the anisotropic doublet (AD) phase, and show it exhibits non-analytic impurity susceptibilities. 
(iv) On the other side, $\lambda_x$ grows into the strong-coupling regime. 
Crucially, using the constructive approach of bosonization--refermionization \cite{von_delft_bosonization_1998, vonDelft_1998_finitesize, zarand_analytical_2000}, we show that a solvable line exists at $\rho_z^\star=\frac{1}{4}$ (and arbitrary $\lambda_x$), where the highly non-trivial PK Hamiltonian can be mapped into a bilinear Hamiltonian of pseudo-fermions. 
We construct one-to-one correspondence between the physical Hilbert spaces in the original electron representation and the final pseudo-fermion representation, and use it to solve the finite-size spectrum. 
Solution shows that this strong-coupling phase is a Kondo Fermi liquid formed by PK resonance. 
We also compute the thermodynamic quantities and impurity susceptibilities in the thermodynamic limit, which all verify this result. 
(v) Finally, we apply the bosonization RG to the singlet regime, and identify the second-order transition between Kondo Fermi liquid and the local-singlet phase.

This work is organized as follows. 
In \cref{sec:Himp}, we introduce the SVAIM and derive the effective Kondo Hamiltonian at half-filling by integrating out the charge fluctuations. 
We also prepare the basics for bosonization. 
In \cref{sec:PK,sec:AD,sec:BKT,sec:refermFL}, we focus on the regime where a valley-doublet dominates the impurity. 
\cref{sec:PK} derives the effective model --- the pair-Kondo Hamiltonian $H_{\rm PK}$ with $\lambda_x$ --- in this regime, by further downfolding the Kondo model and integrating out the multiplet fluctuation. 
In \cref{sec:AD}, we show that $\lambda_x=0$ is a solvable line that exhibits power-law susceptibilities. 
This line will be the starting point for the bosonization RG calculations in \cref{sec:BKT}, which treats $\lambda_x$ in a perturbative manner. 
More importantly, $\rho_z < \rho_z^c = \frac{1}{2} - \frac{1}{2\sqrt{2}}$ will also be the stable continuous fixed line describing the AD phase. 
In \cref{sec:BKT}, we perform RG analysis based on the Coulomb gas analog and identify a BKT phase transition separating the AD phase from a strong-coupling FL phase with PK resonance. 
In \cref{sec:refermFL}, we obtain the exact solutions for the PK Fermi liquid at the fixed line $\rho_z^\star = \frac{1}{4}$ via refermionization, including the finite-size many-body spectrum, thermodynamic quantities, and correlation functions. 
In \cref{sec:singlet}, we extend the RG analysis to the singlet regime, where a trivial singlet dominates the impurity, and identify a second-order phase transition between the Kondo FL and a local singlet (LS) phase. 
We also relate our calculation to previous studies on the two-impurity Kondo problem. 
Finally, we briefly summarize the results in \cref{sec:summary}.

\section{The spin-valley quantum impurity}
\label{sec:Himp}

\subsection{The Anderson Hamiltonian}   \label{sec:Himp-Anderson}

We describe the spin-valley symmetries first. 
We label valleys by $l=\pm$ and spins by $s = \uparrow, \downarrow$, and dub the corresponding Pauli matrices as $[\sigma^{\mu}]_{l,l'}$ and $[\spin^{\nu}]_{s,s'}$, respectively ($\mu,\nu=0,x,y,z$). 
As in realistic systems, besides conservation of electric charge $\Uo_c$ generated by $\sigma^0 \spin^0$, we also assume a spin $\SUt_s$ symmetry generated by $\sigma^0 \spin^{x,y,z}$, and a valley symmetry of $D_{\infty} = \Uo_v \rtimes \Zt$. 
The $\Uo_v$ valley charge is generated by $\sigma^z \spin^0$, and $\Zt$ is generated by valley flipping $\sigma^x \spin^0$. 
Irreducible representations (irreps) of spin $\SUt_s$ will be labeled by the spin angular momentum $j = 0,\frac{1}{2},1,\cdots$, with degeneracy $2j+1$. 
Irreps of valley $D_{\infty}$ group will be labeled by the following notation $L$. 
$L=1,2,3,\cdots$ are two-fold degenerate with an absolute value $L$ of the $\Uo_v$ charges, as the $\pm L$ states are linked by the $\Zt$ action. 
$L=A_1, A_2$ labels two non-degenerate irreps of zero $\Uo_v$ charges, with even and odd $\Zt$ eigenvalues, respectively. 
We emphasize that, the ``valley'' degree of freedom can not only be realized by momentum valleys such as MATBG, but can also refer to degenerate orbital angular momenta (if continuous rotation is approximately preserved), layers (if inter-layer hopping is negligible), \textit{etc}, as long as the corresponding symmetries are intact. 

The Hamiltonian for the SVAIM reads, 
\begin{align} \label{eq:H_SVAIM}
    H &= H_0 + H_{\rm imp} + H_{\rm c} \ , \\\nonumber
    H_0 &= \int_{-L/2}^{L/2} \mrm{d}x : \psi_{l s}^\dagger(x) (\ii\partial_x) \psi_{ls} (x):  \ ,\\\nonumber
    H_{\rm imp} &= \epsilon_f \hat{N} +  U \frac{\hat{N}(\hat{N}-1)}2 + H_{\rm AH} \ , \\\nonumber
    H_{\rm c} &= \sqrt{2\Delta_0} \sum_{l s}  
    \pare{ \psi_{l s}^\dagger(0) f_{l s} + h.c.} \ ,
\end{align}
where $\psi_{ls}$ and $f_{ls}$ are the bath and impurity electron operators, respectively. 
$H_0$, $H_{\rm imp}$, and $H_{\rm c}$ denote the Hamiltonian for bath, impurity, and kinetic hybridization, respectively, where $H_{\rm AH}$ describes a general symmetry-allowed (anti-)Hund's interaction on the impurity. 
We describe each term in more detail below. 

To realize a constant hybridization function $\Delta_0$ for the impurity, the \textit{auxiliary} bath can be mapped to a one-dimensional chiral fermion $\psi_{ls}(x)$. 
By spin-valley symmetries, the Fermi velocities of all $ls$ flavors are degenerate, which we take as unity to rescale $x$ with the dimension of time $\tau$. 
By choosing $\hbar = 1$, $x$ and $\tau$ also have dimensions of inverse energy, and we denote it as $[x] = [\tau] = -1$. 
$L$ denotes the bath size. 
We are interested in properties in the thermodynamic limit $L \to \infty$, but we will also solve the finite-size many-body spectrum to the $\OO(\frac{1}{L})$ order to determine the nature of each quantum phase. 

We define the Fourier components of $\psi_{ls}(x)$ as
\begin{align}   \label{eq:psi_d}
    d_{ls}(k) &= \sqrt{\frac{1}{L}} \int_{-L/2}^{L/2} \mrm{d}x ~ \psi_{ls}(x) ~ e^{\ii kx} \\\nonumber
    \psi_{ls}(x) &= \sqrt{\frac{1}{L}} \sum_{k} d_{ls}(k) ~ e^{-\ii kx}
\end{align}
with a normalization convention $\{ \psi^\dagger_{ls}(x), \psi_{l's'}(x') \} = \delta(x-x') \delta_{ls,l's'}$, and $\{ d^\dagger_{ls}(k), d_{l's'}(k') \} = \delta_{k,k'} \delta_{ls,l's'}$. 
A general boundary condition of $\psi_{ls}(x)$ reads $\psi(-\frac{L}{2}) = \psi(\frac{L}{2}) e^{-\ii \pi \Pbc}$, where $\Pbc\in[0, 2)$. 
By definition \cref{eq:psi_d}, the momentum $k\in \frac{2\pi}{L} (\mathbb{Z} - \frac{\Pbc}{2})$. 
Normal-ordering $:\cdots :$ of bath electrons is defined with respect to such a reference Fermi sea $|0\rangle_0$ that obeys
\begin{align} \label{eq:vacuum-main}
    d_{\alpha}(k) |0\rangle_0 &= 0   \quad (\textrm{if}~ k>0) \ , \\\nonumber 
    d^\dagger_{\alpha}(k) |0\rangle_0 &= 0 \quad (\textrm{if}~ k \le 0) \ .
\end{align}
In this representation, 
\begin{align}   \label{eq:H0_momentum}
    H_0 &= \sum_k \sum_{ls} k :d^\dagger_{ls}(k) d_{ls}(k):  \ , \\  \label{eq:Hc_momentum}
    H_{\rm c} &= \sqrt{\frac{2\Delta_0}{L}} \sum_k \sum_{ls} d^\dagger_{ls}(k) f_{ls} + h.c. \ . 
\end{align}

For the impurity Hamiltonian, $\hat{N} = \sum_{ls} f^\dagger_{ls} f_{ls}$ counts the total impurity electron number, and $\epsilon_f$ and $U$ denote the impurity on-site potential and Hubbard repulsion, respectively. 
The symmetry-allowed (anti-)Hund's splittings $H_{\rm AH}$ can be parametrized with two parameters as follows
\begin{align} \label{eq:HAH}
    H_{\rm AH} &= - \frac{J_S}{2} \sum_{ll'} f_{l \uparrow}^\dagger f_{\ovl{l} \downarrow}^\dagger f_{\ovl{l'} \downarrow} f_{l' \uparrow}  - J_D \sum_{l} f_{l\uparrow}^\dagger f_{l \downarrow}^\dagger f_{l \downarrow} f_{l \uparrow} \ . 
\end{align}
In particular, the six two-electron states can be classified into a spin-triplet ($T$, forming irreps $[L,j] = [A_2,1]$), a valley doublet ($D$, $[2,0]$), and a trivial singlet ($S$, $[A_1, 0]$); see \cref{tab:2e}. 
In \cref{eq:HAH}, the first term lowers the energy of $S$ by $J_S$, while the second term lowers the energy of $D$ by $J_D$. 
If $J_S$ or $J_D$ is positive, the ground state(s) will be spin-singlet(s), thus the splitting is of anti-Hund's nature. 
If $J_S=J_D$, the valley symmetry group gets promoted to an $\SUt_v$ generated by $\sigma^{x,y,z} \spin^0$, and $S \oplus D$ span a valley-triplet. 
Therefore, $J_S \not= J_D$ can also be understood as a valley anisotropy. 
In reality, $J_{S,D}$ can arise from processes such as electron-phonon couplings \cite{wang_molecular_2024,wang_2025_epc}, which are much weaker compared to the Coulomb $U$, hence we focus on $|J_{S,D}| \ll U$. 

Finally, we remark that the SVAIM \cref{eq:H_SVAIM} also possesses an anti-unitary $C_2T$ symmetry and a particle-hole symmetry (PHS). 
These symmetries are not essential to the physics to be discussed, but are convenient for our analytical treatment. 
$C_2 T$ acts as $(C_2T) f_{ls} (C_2T)^{-1} = f_{l s}$ and $(C_2T) \psi_{l s}(x) (C_2T)^{-1} = \psi_{l s}(-x)$. 
It originates from the physical (Kramer's spinful) time-reversal symmetry, in product with $\SUt_s$ rotations and crystalline symmetries, and is hence made ``spinless'' and ``valley-less'' \cite{wang_2025_solution}. 
Therefore, we also refer to it as a time-reversal symmetry (TRS). 
If $\Pbc=0$ or $1$, we can define a (unitary) charge conjugation $f_{ls} \leftrightarrow f_{ls}^\dagger$, $d_{ls}(k) \leftrightarrow - d_{ls}^\dagger(-k)$, which leaves $H_0 + H_{\rm c}$ invariant [see \cref{eq:H0_momentum,eq:Hc_momentum}]. 
It transforms the operators contained in $H_{\rm imp}$ as $\hat{N} \leftrightarrow 4- \hat{N}$ and $H_{\rm AH} \leftrightarrow H_{\rm AH} - \left(\frac{J_S}{2} + J_D \right) (2-\hat{N})$. 
Due to these relations, $H_{\rm imp}$ will be invariant under charge conjugation if $\epsilon_f$ is tuned to the PHS point, $\epsilon_f = - \frac{3}{2} U + \frac{1}{4} J_S  + \frac{1}{2} J_D$. 
Readers may refer to \cref{app:imp} for more details. 

\begin{table}[tb]
    \centering
    \begin{tabular}{l|l|l}
    \hline
        $[L,j]$ & wave-function & energy of $H_{\rm imp}$ \\
    \hline
        $[A_1,0]$ & $|S\rangle = \frac{f_{+\uparrow}^\dagger f_{-\downarrow}^\dagger - f_{+\downarrow}^\dagger f_{-\uparrow}^\dagger }{\sqrt2} \emp$ & $E_S = 2\epsilon_f + U - J_S$ \\
    \hline
        $[2,0]$ & $|D,2\rangle = f_{+\uparrow}^\dagger f_{+\downarrow}^\dagger \emp$ & $E_D = 2\epsilon_f + U - J_D$ \\
        & $|D,\ovl{2}\rangle = f_{- \uparrow}^\dagger f_{- \downarrow}^\dagger \emp$ &  \\
    \hline
        $[A_2,1]$ & $|T,1 \rangle = f_{+ \uparrow}^\dagger f_{- \uparrow}^\dagger \emp$ & $E_T = 2\epsilon_f + U $ \\
        & $|T,0 \rangle = \frac{f^\dagger_{+\uparrow} f^\dagger_{-\downarrow} + f^\dagger_{+\downarrow} f^\dagger_{-\uparrow} }{\sqrt{2}}  \emp$ &  \\
        & $|T,\ovl{1}\rangle = f_{+ \downarrow}^\dagger f_{- \downarrow}^\dagger \emp$ &  \\
    \hline
    \end{tabular}
    \caption{Multiplet splitting among two-electron impurity states, induced by $H_{\rm AH}$. 
    By spin-valley symmetries, no splitting occurs among one-electron ($E_1 = \epsilon_f$) or three-electron states ($E_3 = 3\epsilon_f + 3 U - \frac{1}{2}J_S - J_D$), because both levels form the $[1,\frac{1}{2}]$ irrep of the valley and spin symmetries. 
    The empty impurity state $\emp$ and the fully occupied state $(\prod_{ls} f^\dagger_{ls}) \emp$ have energies $E_0=0$ and $E_4 = 4\epsilon_f + 6U - J_S - \frac{J_D}{2}$, respectively. }
    \label{tab:2e}
\end{table}

\subsection{The Kondo Hamiltonian}   \label{sec:Himp-Kondo}

In this paper, we will focus on the regime where the low-energy physics is dominated by two-electron impurity states ($\epsilon_f \approx - \frac{3}{2}U$), so that multiplet splitting plays a significant role. 
For convenience, we assume PHS. 

At an energy scale of $\omega \ll \OO(U)$, charge fluctuations get frozen, hence we integrate out charge fluctuations through a Schrieffer-Wolff (SW) transformation. 
The result is an effective Kondo model, 
\begin{align}   \label{eq:HK}
    H_{\rm K} = H_0 + H_{\rm imp}^{\rm (K)} + H_{\rm c}^{\rm (K)} \ .
\end{align}
The downfolded impurity Hamiltonian $H_{\rm imp}^{\rm (K)} = \sum_{\Gamma} E_\Gamma  \PP_\Gamma$, where $\PP_\Gamma$ is the projector to the $\Gamma=S,D,T$ manifolds listed in \cref{tab:2e}. 
In general, the SW transformation may generate corrections to $E_\Gamma$, which can be absorbed as a re-definition to $J_S$ and $J_D$, hence we neglect. 
By the spin-valley and $C_2T$ symmetries, we find the Kondo coupling $H^{\rm (K)}_{\rm c}$ must take the following form (\cref{app:imp}), 
\begin{align}   \label{eq:HK_c}
    H^{\rm (K)}_{\rm c} &= 2\pi \zeta_{0z} \sum_{\nu = x,y,z} \Theta^{0 \nu} \cdot \psi^\dagger \sigma^0 \spin^\nu \psi \Big|_{x=0} \\\nonumber
    &+ 2\pi \zeta_{xz} \sum_{\mu=x,y} \sum_{\nu=x,y,z} \Theta^{\mu \nu} \cdot \psi^\dagger \sigma^\mu \spin^\nu \psi \Big|_{x=0} \\\nonumber
    &+ 2\pi \zeta_{zz} \sum_{\nu = x,y,z} \Theta^{z\nu} \cdot \psi^\dagger \sigma^z \spin^\nu \psi \Big|_{x=0} \\\nonumber
    &+ 2\pi \lambda_z \cdot \Theta^{z0} \cdot \psi^\dagger \sigma^z \spin^0 \psi \Big|_{x=0} \\\nonumber
    &+ 2\pi \zeta_{x} \sum_{\mu=x,y} \Theta^{\mu 0} \cdot \psi^\dagger \sigma^\mu \spin^0 \psi \Big|_{x=0}  \ .
\end{align}
Here, $\Theta^{\mu \nu}  = \PP_2   \frac{f^\dagger \sigma^{\mu} \spin^{\nu} f }{2} \PP_2$ for $\mu\nu \not= 00$ are the fifteen $\SUf$ generators of spin-valley flavors, represented on the two-electron state manifold $\PP_2 = \sum_{\Gamma=S,D,T} \PP_\Gamma$. 
Therefore, the realistic Kondo coupling can be regarded as an anisotropic $\SUf$ moment-moment interaction between the bath and impurity. 
We also note that, without PHS, density-density interactions of the form $\PP_\Gamma :\psi^\dagger \sigma^0 \spin^0 \psi:$ would arise, but they are not relevant. 
Hereafter, we abbreviate $\psi^\dagger \sigma^\mu \spin^\nu \psi \big|_{x} = \sum_{ls,l's'} \psi^\dagger_{ls}(x) [\sigma^\mu \spin^\nu]_{ls,l's'} \psi_{l's'}(x)$. 

If $J_S = J_D = 0$, the model obeys the full $\SUf$ rotation among spin-valley flavors. 
In this limit, \cref{eq:HK_c} becomes isotropic, and we define $\zeta = \zeta_{0z} = \zeta_{xz} = \zeta_{zz} = \lambda_z = \zeta_x$. 
The sign of $\zeta$ is always anti-ferromagnetic ($> 0$), as it arises from the virtual super-exchange between the impurity and bath, and it is of order $\OO(\frac{\Delta_0}{U})$ \cite{zhou_kondo_2024}. 
For such $\SUf$ Kondo impurity, an exact Kondo screening develops at an energy scale below $\omega \lesssim \OO(\TK)$, the Kondo temperature. 
This phase is termed as the Kondo Fermi liquid (FL). 

\begin{figure}[t]
    \centering
    \includegraphics[width=0.7\linewidth]{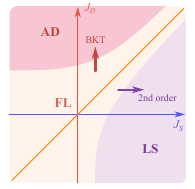}
    \caption{A schematic phase diagram of the spin-valley Anderson impurity model (SVAIM) at half-filling. AD, FL, LS stands for anisotropic doublet, (Kondo) Fermi liquid, and local singlet, respectively. }
    \label{fig:intro}
\end{figure}

Realistic splittings $|J_S| , |J_D| \ll U$ barely affect the super-exchange process, hence the Kondo coupling amplitudes $\zeta \sim \OO(\frac{\Delta_0}{U})$ are only perturbed. 
Therefore, although the five parameters in \cref{eq:HK_c} are no longer related by symmetries, their values will remain of the order $\OO(\frac{\Delta_0}{U})$, and the signs will remain anti-ferromagnetic. 
The major influence on the physics is due to the splittings in $H^{\rm (K)}_{\rm imp}$ itself. 
We have described a full phase diagram regarding $J_{S}$ and $J_D$ in Ref.~\cite{wang_2025_solution}, as summarized in \cref{fig:intro}. 
Concretely, if $T$ remains the impurity ground state ($J_S,J_D < 0$), then after further downfolding the Hilbert space from $\PP_2$ to $\PP_T$, there remains a spin-1 impurity with no valley degree of freedom, which can still be exactly screened by the two degenerate $l=\pm$ channels of bath electrons. 
The system still flows to a Kondo FL, which can smoothly cross over to the $\SUf$ FL described above. 
If $J_S=J_D>0$, the ground state is also a Kondo FL with the roles of spin and valley exchanged \cite{wang_2025_solution}. 

The most interesting physics appears when $D$ ($J_D > 0, J_D>J_S$) or $S$ ($J_S > 0, J_S>J_D$) becomes the impurity ground state. 
In these scenarios, a transition out of FL can eventually occur, which will be the main topic of this paper.

\subsection{Bosonization of bath}   \label{sec:Himp-bath}

To analytically study this problem, we will bosonize the bath electrons following Refs.~\cite{von_delft_bosonization_1998, vonDelft_1998_finitesize, zarand_analytical_2000}, 
\begin{align}   \label{eq:bosonization}
    \psi_{ls}(x) = \frac{F_{ls}}{\sqrt{2\pi x_c}} e^{-\ii \phi_{ls}(x)} e^{-\ii (N_{ls} - \frac{\Pbc}{2}) \frac{2\pi x}{L}}\ ,
\end{align}
where $x_c \to 0^+$ is an ultraviolet cutoff. 
One may refer to \cref{app:boson} for a brief review of the constructive approach of bosonization. 
Here $N_{ls}$ is the number of bath electrons in the $(ls)$ flavor  with respect to the vacuum defined in \cref{eq:vacuum-main}, $\phi_{ls}(x)$ is a bosonic field subject to periodic boundary condition, and $\Pbc$ specifies the boundary condition of the fermion operator. 
For a quick understanding of bosonization, note that for one-dimensional chiral fermions, any $\vN$-electron excited state (where $\vN$ collects all $N_{ls}$) can be constructed in two steps: first fill the lowest particle (or hole) excitations above $|0\rangle_0$ and obtain the lowest $\vN$-electron state $|\vN\rangle_0$ to match the electron charges, then generate particle-hole excitations (density fluctuations) above $|\vN\rangle_0$ to reach the excited state. 
Correspondingly, one can define two types of operators: the Klein factor $F_{ls}$ that links different $|\vN\rangle_0$ states and the bosonic fields $\phi_{ls}(x)$ that generate density fluctuations. 

$F_{ls}$ decreases $N_{ls}$ by $1$, $[N_{ls}, F_{ls}] = - F_{ls}$, and there is $F_{ls}^\dagger F_{ls} = F_{ls} F_{ls}^\dagger = 1$. 
$F_{ls}$ also encodes the anti-commutation between different flavors, as $\{ F_{ls}, F^\dagger_{l's'} \} = 2 \delta_{ls,l's'}$, $\{ F_{ls}, F_{l's'} \} = 2 F_{ls}^2 \delta_{ls,l's'}$. 

The bosonic fields are defined as
\begin{align} \nonumber
    \phi_{ls}(x) &= \sum_{q>0} - \sqrt{\frac{2\pi}{q L}} \left( e^{\ii q x} b^\dagger_{ls}(q) + e^{-\ii qx} b_{ls}(q) \right) e^{-\frac{x_cq}{2}}  \ ,\\
    \textrm{where} ~~ & b^\dagger_{ls}(q) = \ii \sqrt{\frac{2\pi}{qL}} \sum_{k} d^\dagger_{ls}(k+q) d_{ls}(k) \ .
\end{align}
The bosonic momentum $q \in \DL \mbb{Z}_{>0}$, and $b^\dagger$ operators create the desired particle-hole excitations, satisfying $[b_{ls}(q), b^\dagger_{l's'}(q')] = \delta_{q,q'} \delta_{ls,l's'}$. 
In particular, one can show that the electron density is given by 
\begin{align}  \label{eq:boson_dens}
    :\psi^\dagger_{ls}(x) \psi_{ls}(x): ~ = \frac{1}{2\pi}\partial_x \phi_{ls}(x) + \frac{N_{ls}}{L}
    + \mathcal{O}(L^{-2})\ ,
\end{align}
hence $\partial_x\phi_{ls}(x)$ represent density fluctuations in the $(ls)$ flavor.
Hereafter, we will always omit $\mathcal{O}(L^{-2})$ terms and keep $\mathcal{O}(L^{-1})$ unless specified otherwise. 
$e^{\ii \phi_{ls}(x)}$, termed as a vertex operator, serves as a Jordan-Wigner string that implements the anti-commutation relation within the same flavor. 
The following canonical commutation relation in real space will be used, 
\begin{align}   \label{eq:canonical_comm}
    [\phi_{ls}(x) , \partial_x\phi_{l's'}(x')] = 2\pi\ii \Big( \delta_{x_c}(x-x') - \frac{1}{L} \Big) \delta_{ls,l's'} \ ,
\end{align}
where $\delta_{x_c}(x) = \frac{x_c}{\pi} \frac{1}{x^2 + x_c^2}$ is a smoothened $\delta$ function. 
All Klein factors $F$ commute with all bosonic fields $b$ (hence $\phi$). 

By this construction, it is direct to bosonize $H_0$: 
\begin{align}   \label{eq:H0}
    H_0 = \sum_{ls} \Big[\DL \frac{N_{ls}(N_{ls}+1-\Pbc)}{2} + \sum_{q>0} q \!:\! b^\dagger_{ls}(q) b_{ls}(q) \!:\!  \Big],
\end{align}
where the first term measures the energy of $|\vN\rangle_0$ with respect to $|0\rangle_0$, and the second term indicates that acting $b^\dagger_{ls}(q)$ on $|\vN\rangle_0$ creates (a superposition of) particle-hole excitations that cost energy $q$. 
The normal-ordering $: \cdots :$ for boson fields brings $b$ to the right side of $b^\dagger$. 
Equivalently, we can also express the energy cost of density fluctuation in real space as
\begin{align}   \label{eq:H0_b_phi}
    \sum_{q>0} q : b^\dagger_{ls}(q) b_{ls}(q) : = \int_{-L/2}^{L/2} \frac{\dd x}{4\pi} :(\partial_x\phi_{ls}(x))^2: \ . 
\end{align}

To conveniently deal with the charge-flavor separation, we carry out a following canonical transformation to the boson fields, 
\begin{align}   \label{eq:phi-flavor}
    \begin{bmatrix}
        \phi_{c} \\
        \phi_{v} \\
        \phi_{s} \\
        \phi_{vs} \\
    \end{bmatrix} = \frac{1}{2}\brak{ \begin{array}{rrrr}
        1 & 1 & 1 & 1 \\
        1 & 1 & -1 & -1 \\
        1 & -1 & 1 & -1 \\
        1 & -1 & -1 & 1 \\
    \end{array}} \begin{bmatrix}
        \phi_{+\uparrow} \\
        \phi_{+\downarrow} \\
        \phi_{-\uparrow} \\
        \phi_{-\downarrow} \\
    \end{bmatrix} \ . 
\end{align}
$\chi = c,v,s,vs$ denotes the charge, valley, spin, and valley-times-spin densities, respectively. 
The transformation \cref{eq:phi-flavor} preserves the form of \cref{eq:canonical_comm,eq:H0}. 
Correspondingly, we can also examine the following $\Uo$ charges, 
\begin{align}   \label{eq:N-flavor-transformation}
    \begin{bmatrix}
        N_{c} \\
        N_{v} \\
        N_{s} \\
        N_{vs} \\
    \end{bmatrix} = \frac{1}{2} \brak{\begin{array}{rrrr}
        1 & 1 & 1 & 1 \\
        1 & 1 & -1 & -1 \\
        1 & -1 & 1 & -1 \\
        1 & -1 & -1 & 1 \\
    \end{array}} \begin{bmatrix}
        N_{+\uparrow} \\
        N_{+\downarrow} \\
        N_{-\uparrow} \\
        N_{-\downarrow} \\
    \end{bmatrix} \ ,
\end{align}
so that $N_{\chi} \in \frac{\mbb{Z}}{2}$. 
Notice that not all combinations of $N_{\chi} \in \frac{\mbb{Z}}{2}$ are physical, unless inverting \cref{eq:N-flavor-transformation} produces all $N_{ls} \in \mbb{Z}$. 
The sufficient and necessary condition is termed as the free gluing condition \cite{zarand_analytical_2000}. 
However, for our construction to be introduced, we do not need to explicitly write them down. 

\section{Doublet regime: pair-Kondo model}   \label{sec:PK}

We now specify to the parameter regime where $D$ is the ground states ($J_{D} > 0, J_D > J_S$). 
We will denote $J$ as the minimal multiplet excitation energy, hence $J = \min(J_D-J_S, J_D)$ in this doublet regime. 
Assuming that the Kondo resonance has \textit{not} formed at the energy scale of $\omega \lesssim J$, we can further downfold the impurity Hilbert space from $\PP_2$ to $\PP_D$. 
We will show that $\PP_D$ itself can support a Kondo FL, which smoothly crosses over to the other FL regimes described in \cref{sec:Himp-Kondo} \cite{wang_2025_solution}, hence we do not need to include more multiplets to describe the phase transition out of FL in this doublet regime. 

From now on, we will abbreviate $|D,L^z\rangle = |L^z\rangle$ with $L^z = \pm2$ without causing confusion. 
Also, we define Pauli matrices, 
\begin{align}  \nonumber
    &\Lambda_z = \ket{2}\bra{2} - \ket{\bar2} \bra{\bar2} , \qquad \Lambda_+ =  \Lambda_-^\dagger = \ket{2}\bra{\bar2} ,\\   \label{eq:Lambda-def-PK}
    &\Lambda_x = \Lambda_+ + \Lambda_- , \qquad  
    \Lambda_y = -\ii \Lambda_+ + \ii \Lambda_- \ . 
\end{align}
Note that $\Lambda_z$ is just a synonym to $\Theta^{z0}$, while $\Lambda_{\pm}$ are orthogonal to all $\Theta^{\mu \nu}$ operators defined below \cref{eq:HK_c}. 
Interestingly, as $\Lambda_\pm$ changes the $\Uo_v$ charges on impurity by $\pm 4$, they cannot be coupled to any bilinear operators of bath electrons, which at most change the bath $\Uo_v$ charges by $\pm 2$. 
Therefore, $\Lambda_\pm$ cannot appear in the Kondo Hamiltonian. 
However, their role in the doublet regime will be indispensable, due to the pair-Kondo couplings to be described below. 

\subsection{Pair-Kondo model}

For the Kondo Hamiltonian $H_{\rm K}$ obtained above [\cref{eq:HK}], the low-energy block only contains $\PP_D H_{\rm K} \PP_D = H_0 + H_z$. 
In particular, $\PP_D H_{\rm imp}^{\rm (K)} \PP_D$ only involves an energy constant that can be neglected, and 
\begin{align}   \label{eq:Hz}
    H_z = \PP_D H_{\rm c}^{\rm (K)} \PP_D = (2\pi \lambda_z) \Lambda_z :\psi^\dagger \sigma^z \spin^0 \psi:\big|_{x=0}
\end{align}
only involves the $\lambda_z$ term [\cref{eq:HK_c}]. 
Recall that $\lambda_z > 0$ is anti-ferromagnetic. 
We remark that, as a $\delta$-potential at $x=0$, $H_z$ simply serves to generate a phase shift to the bath electrons across $x=0$. 
Concretely, for the Hilbert subspaces that diagonalizes $\Lambda_z = \pm1$, and for the $l=\pm$ bath electron channels, single-electron levels of $H_0 + H_z$ contain an altered ``boundary condition'' at $x=0$, 
\begin{align}   \label{eq:phaseshift_0}
    \psi_{ls}(0^-) e^{\ii \pi \cdot l\Lambda_z \rho_z} = \psi_{ls}(0^+) e^{-\ii \pi \cdot l \Lambda_z  \rho_z}\ ,
\end{align}
where $\rho_z$ is a function of $\lambda_z$ that depends on the specific scheme to regularize the $\delta$ potential \cite{Andrei_1983_Solution, vonDelft_1998_finitesize, Krishnan_2024_kondo}; see \cref{app:boson-phase}. 
For our scheme, $\rho_z = \frac{\arctan \pi \lambda_z}{\pi} \in \pare{0,\frac{1}{2}}$. 
However, as the phase shift $\rho_z$ is the physical observable, instead of the ``bare'' parameter $\lambda_z$, one should adjust $\lambda_z$ across different regularization schemes to agree on $\rho_z$. 
Therefore, we regard $\rho_z$ as the more essential quantity. 

By matching this new ``boundary condition'' [\cref{eq:phaseshift_0}], the single-electron spectrum of $H_0 + H_z$ for the $ls$ flavor in the $\Lambda_z=\pm1$ sector can be directly obtained as
\begin{align}  \label{eq:k_value_AD}
    k \in \DL(\mbb{Z} - \frac{\Pbc}{2} + l \Lambda_z \rho_z) \ . 
\end{align}

Importantly, the $\zeta_{x}$ and $\zeta_{xz}$ terms in \cref{eq:HK_c} induce multiplet fluctuations away $\PP_D$ to higher-energy subspaces, $\PP_S$ and $\PP_T$, respectively. 
$\zeta_{0z}$ and $\zeta_{zz}$ act within the higher-energy subspace $\PP_S + \PP_T$. 
Therefore, a further SW transformation that integrates out these multiplet fluctuations should be implemented, which will bring about couplings of a quartic operator of bath electrons to the manifold of $D$. 

Of particular importance is the new quartic couplings to $\Lambda_\pm$. 
By matching the valley $\Uo_v$ and $\SUt_s$ symmetries, we find they must take the form
\begin{equation}   \label{eq:Hx}
    H_x = (2\pi)^2 \lambda_x x_c \cdot \Lambda_+ \cdot \psi_{-\downarrow}^\dagger \psi_{-\uparrow}^\dagger \psi_{+\uparrow} \psi_{+\downarrow} \Big|_{x=0} + h.c. \ .
\end{equation}
As the $\Zt$ action flips $l \leftrightarrow \ovl{l}$, the first term is swapped to its Hermitian conjugate ($h.c.$), hence $\lambda_x$ must be real-valued. 
The sign of $\lambda_x$ is not important, because it can be flipped by applying a gauge transformation $\ii \Lambda_z$ to the impurity subspace, while the physics should not depend on such a gauge transformation. 
In \cref{eq:Hx}, we have also defined the coupling strength $\lambda_x$ to be dimensionless. 
The second SW transformation will produce $\lambda_x \sim \mcl{O}(\frac{\Delta_0^2}{U^2} \frac{1}{J x_c})$, because $\zeta_{x},\zeta_{xz} \sim \OO(\frac{\Delta_0}{U})$. 
Since it takes an electron pair (that belongs to the $[L,j] = [2,0]$ irrep of the spin-valley symmetry group) to flip the impurity, we term $H_x$ [\cref{eq:Hx}] as a pair-Kondo (PK) coupling. 

At the classical level of dimension counting, $\lambda_x$ appears irrelevant under RG.
As the quartic bath operator at $x=0$ has a physical dimension of energy squared and hence a scaling dimension 2, {\it i.e.},  $[\psi^\dagger\psi^\dagger \psi \psi] = 2$, and the Hamiltonian has the dimension of energy, {\it i.e.}, $[H]=1$, there must be $[\lambda_x]=-1$, suggesting $\frac{\dd\lambda_x}{\dd\ell}=-\lambda_x$, where $\ell$ is the RG time. 
Nevertheless, we will show in \cref{sec:BKT} that with sufficient quantum corrections from an anti-ferromagnetic $\lambda_z$, $\lambda_x$ can grow into a strong-coupling regime, and eventually drive the system into a Kondo FL. 
Therefore, we keep this PK coupling. 
The effective low-energy theory in the doublet regime is given by 
\begin{equation}
H_{\rm PK} = H_0 + H_z + H_x\ . 
\end{equation}

On the other hand, the second SW transformation can also bring about quartic couplings to $\Lambda_z$ and $\PP_D$. 
However, these quartic couplings are irrelevant at the classical level, and are not expected to have significant interplay with $\lambda_z$ and $\lambda_x$ for the physics we are interested in. 
Therefore, we neglect them. 

\subsection{Bosonization}\label{sec:HPK-bosonization}

By bosonization, $H_z + H_x$ is mapped to 
\begin{align}  \label{eq:Hz_boson}
H_{z} &= \rho_z \Lambda_z 
    \pare{ 2 \partial_x \phi_{v}(x) \Big|_{x=0} + \frac{4\pi}{L} N_{v} } \ ,\\ \label{eq:Hx_boson}
H_{x} &= \frac{\lambda_x}{x_c} \pare{ \Lambda_+ \cdot F_v \cdot  e^{-\ii 2 \phi_v(0)} 
    + h.c. }\ ,
\end{align}
where we have used \cref{eq:boson_dens} to bosonize the density term in $H_z$ [\cref{eq:Hz}], substituted the ``bare'' $\lambda_z$ with the physical phase shift $\rho_z$, and exploited the flavor basis [\cref{eq:phi-flavor,eq:N-flavor-transformation}]. 
Finally, we have defined a composite Klein factor $F_v = F_{-\downarrow}^\dagger F_{-\uparrow}^\dagger F_{+\uparrow} F_{+\downarrow}$, which can be verified to decrease $N_v$ by $2$, namely, $[N_v, F_v] = - 2F_v$. 
Also, $F_v F_v^\dagger = F_v^\dagger F_v = 1$. 
Crucially, the impurity only couples to the valley fluctuation $\phi_v$, whereas $\chi=c,s,vs$ channels decouple. 
The following $\Uo$ charges are conserved by $H_{\rm PK}$: $N_c, N_s, N_{vs}$ and $N^{(\rm tot)}_v = N_v + \Lambda_z$. 

$H_z$ [\cref{eq:Hz_boson}] will generate a phase jump of $\phi_v(0^+) + 2\pi \Lambda_z\rho_z = \phi_v(0^-) - 2\pi \Lambda_z \rho_z$ to the bosonic fields $\phi_v$, which can be seen by completing the squares in $H_0 + H_z = \int\frac{\dd x}{4\pi} : (\partial_x \phi_v(x) + 4\pi \rho_z\Lambda_z \delta_{x_c}(0))^2 : + \cdots$, where energy constants in the doublet space proportional to $\PP_D = \Lambda_z^2$ and $\mcl{O}(L^{-1})$ terms are neglected in $\cdots$. 
This phase jump is also consistent with the phase shift in the original electron representation [\cref{eq:phaseshift_0}] by noting that $\psi_{ls}(x) \sim e^{-\ii \frac{l}{2} \phi_v(x) + \cdots}$. 

The same phase jump in $\phi_v$ can be manually generated by a gauge transformation $U = e^{\ii 2 \rho_z \Lambda_z \phi_v(0)}$. By \cref{eq:canonical_comm}, 
\begin{align}
    U \big(\partial_x\phi_v(x)\big) U^\dagger = \partial_x\phi_v(x) - 4\pi \rho_z \Lambda_z \delta_{x_c}(x) + \OO(L^{-1}) . 
\end{align}
Therefore, in the new gauge, $\ovl{H}_0 = U(H_0 + H_z)U^\dagger = \int\frac{\dd x}{4\pi} : (\partial_x \phi_v(x))^2 : + \cdots$ describes a free boson field with no additional $\delta$ potential. 
In other words, the longitudinal coupling to $\Lambda_z$ of $\phi_v$ has been removed. 
We thus work in this new gauge. 
Applying bosonization dictionary that also rigorously treats the $\OO(L^{-1})$ terms and the regularization of $\delta$ function [\cref{eq:Hbar-single-flavor}], we obtain the PK Hamiltonian in the new gauge $\ovl{H}_{\rm PK} = U H_{\rm PK} U^\dagger = \ovl{H}_{0} + \ovl{H}_x$, where
\begin{align}  \label{eq:ovlH0}
    \ovl{H}_0 &= U(H_0 + H_z)U^\dagger \\\nonumber
    &= \sum_{\chi=c,v,s,vs} \left( \sum_{q>0} q : b^\dagger_\chi(q) b_\chi(q) : + \DL \frac{N_{\chi}^2}{2} \right) \\\nonumber
    &+ \DL N_c(1-\Pbc) + \frac{4\pi}{L} \rho_z\Lambda_z N_v - \frac{4\rho_z^2}{x_c} \PP_D \left( 1 - \frac{\pi}{L} x_c \right) 
\end{align}
is a free Hamiltonian for boson fields. 
The $\PP_D = \Lambda_z^2$ term can be neglected in the doublet regime as an energy constant. On the other hand, 
{
\begin{align}
    \label{eq:ovlHx}
    \ovl{H}_x &= U H_x U^\dagger = \frac{\lambda_x}{x_c} \pare{  \Lambda_+ \cdot F_v \cdot e^{-\ii (2 - 4\rho_z) \phi_v(0)} + h.c. } , 
\end{align}}
which can be directly derived by noting that $[\Lambda_z, \Lambda_+] = 2\Lambda_+$ and hence 
\begin{align}   \label{eq:U_Lamb}
    U \Lambda_+ U^\dagger = \Lambda_+ e^{\ii 4 \rho_z \phi_v(0)} \ .
\end{align}
We will denote $\gamma = 2 - 4\rho_z$ for later convenience. Since $\rho_z \in (0,\frac{1}{2})$, $\gamma \in (0, 2)$ decreases monotonically with increasing $\rho_z$. 

At arbitrary $\rho_z$ and $\lambda_x$, $\ovl{H}_{\rm PK}$ cannot be directly diagonalized. 
However, exact solutions are available at two limits. 
When $\lambda_x=0$, $\ovl{H}_0$ alone can be directly diagonalized by the good quantum numbers $\Lambda_z$, $N_\chi$ for all $\chi$ (equivalently and more conveniently, $N_{ls}$ for all $ls$), and $n_\chi(q) = b_{\chi}^\dagger (q) b_\chi(q) \in \mbb{Z}_{\ge 0}$ for all $\chi$ and $q$ (equivalently, $n_{ls} (q) = b_{ls}^\dagger (q) b_{ls}(q) \in \mbb{Z}_{\ge 0}$). 
Alternatively, we can simply work in the original electron representation, where $H_0 + H_z$ generates a phase-shifted single-electron spectrum for a given $\Lambda_z$ eigenvalue, solved in \cref{eq:k_value_AD}. 
The many-body spectrum at $\lambda_x = 0$ with a fixed $\Lambda_z$ is simply given by filling electrons to this single-electron spectrum. 
We briefly show the two methods agree on this solution in \cref{sec:AD}. 

A more non-trivial solvable limit is at $\rho_z^{\star} = \frac{1}{4}$ and arbitrary $\lambda_x$. 
There, the vertex operator $e^{-\ii(2-4\rho_z^\star)\phi_v(0)} = e^{-\ii \phi_v(0)}$ appearing in $\ovl{H}_x$ [\cref{eq:ovlHx}] can be refermionized to a pseudo-fermion $\psi_v(x)$ by reversing \cref{eq:bosonization}. 
Then, $\ovl{H}_{\rm PK}$ will be a bilinear form involving $\psi_v(x)$, and can be directly diagonalized as well, which is presented in \cref{sec:refermFL}. 
Note that this solution is non-trivial, as it is a strongly interacting problem in the original electron representation. 
This limit is analogous to the Toulouse line in the classic anisotropic Kondo model \cite{toulouse_1969, giamarchi2003quantum}, and to the Emery-Kivelson line in the two-channel anisotropic Kondo model \cite{Mapping_Emery_1992, vonDelft_1998_finitesize, zarand_analytical_2000}. 

Hereafter, we will always denote $\ovl{X} = U X U^\dagger$ for any operator $X$ to stress that it is gauge transformed by $U$. 

\section{Exact solution to pair-Kondo model at \texorpdfstring{$\lambda_x=0$}{lx=0}}  \label{sec:AD}

In this section, we obtain exact solutions for $H_{\rm PK}$ at $\lambda_x=0$. 
The results will serve as the starting point for bosonization RG calculations in \cref{sec:BKT}, which treats $\lambda_x$ as perturbations away from this solvable line. 
Solutions obtained in this section for $\rho_z < \rho_z^c = \frac{1}{2} - \frac{1}{2\sqrt{2}}$ also describes the AD phase, which will be shown as a stable continuous fixed line under the RG flow. 

\subsection{Finite-size spectrum}   \label{sec:AD-spectrum}

We first express the many-body spectrum of $\ovl{H}_{\rm PK} = \ovl H_{0}$ using the commuting good quantum numbers $\Lambda_z = \pm 1$, $N_{ls} \in \mbb{Z}$, and $n_{ls}(q) = b_{ls}^\dagger (q) b_{ls}(q) \in \mbb{Z}_{\ge 0}$ as
\begin{align}  \label{eq:E_AD}
    E[\Lambda_z,\vN,\vec{n}] &= \sum_{ls} \Bigg[ \sum_{q\in \DL \mbb{Z}_{>0}} q \cdot n_{ls}(q) \\\nonumber
    &+ \DL \frac{N_{ls}(N_{ls}+1-\Pbc + 2l\Lambda_z\rho_z)}{2} \Big] , 
\end{align}
where $\vec{n}$ collects all $n_{ls}(q)$, and the energy constant term has been ignored. 
To see the spectrum with a fixed $\Lambda_z$ is equivalent to filling the phase-shifted single-electron spectrum in \cref{eq:k_value_AD}, notice that $U^\dagger(|2\Lambda_z\rangle \otimes |0\rangle_0)$ is an eigenstate of $H_0 + H_z$, as $|2\Lambda_z\rangle \otimes  |0\rangle_0$ diagonalizes the free and decoupled Hamiltonian $\ovl{H}_0 = U(H_0+H_z)U^\dagger$. 
Also, notice that the $p$-th lowest particle excitation above $|0\rangle_0$ costs energy $\DL(p-\frac{\Pbc}{2})$ for arbitrary $\Pbc$, therefore, the $p$-th lowest particle excitation above $U^\dagger (|2\Lambda_z\rangle \otimes |0\rangle_0)$ in the flavor $ls$ costs energy $\DL(p - \frac{\Pbc}{2} + l \Lambda_z \rho_z)$, hence filling $N_{ls}$ electrons at least costs $\sum_{p=0}^{N_{ls}} \DL(p - \frac{\Pbc}{2} + l \Lambda_z \rho_z) = \DL \frac{N_{ls}(N_{ls}+1-\Pbc+2l\Lambda_z\rho_z)}{2}$, which is the second term in \cref{eq:E_AD}. 
This filled state is $U^\dagger (|2\Lambda_z\rangle \otimes |\vN\rangle_0$. 
The first term in \cref{eq:E_AD} is simply the cost of particle-hole pairs with momentum $q$ above $U^\dagger (|2\Lambda_z\rangle \otimes |\vN\rangle_0$. 

If $\Pbc=0$, when $\rho_z=0$, all the $N_{l s}=0$ or $-1$ states (in both $\Lambda_z=\pm1$ sectors) are degenerate, leading to a $2^4 \times 2$-fold degeneracy. 
With an infinitesimal $\rho_z>0$, in \textit{each} $\Lambda_z=\pm1$ subspace, the $N_{l=\Lambda_z, s}=-1$ and $N_{l=-\Lambda_z, s}=0$ state becomes the only ground state, hence the total ground state degeneracy is two. 
If $\Pbc=1$, the ground state degeneracy is always two. Since $\rho_z < \frac{1}{2}$, no level crossing of single-electron levels occurs when increasing $\rho_z$.

\subsection{Impurity susceptibility}  
\label{sec:imp-susceptibility-main}

Although the many-body spectrum in each $\Lambda_z=\pm1$ sector in the $\lambda_x=0$ limit can be constructed by filling a non-interacting single-electron spectrum (but with phase shifts), the physical properties are not the same as a simple Fermi liquid, as a general phase shift $\rho_z$ will generate non-analytic correlation functions, and can lead to kinks in the spectral functions \cite{wang_2025_solution}. 
Here, we compute the impurity valley susceptibility, and show the susceptibility exhibits non-universal power laws. 

To evaluate the correlation functions in the thermodynamic limit, we can omit the $\mathcal{O}(L^{-1})$ terms in this subsection. In $\ovl{H}_0$ [\cref{eq:ovlH0}], all boson fields become free after the new gauge transformation. 

The partition function at $\lambda_x=0$ is given by $Z_0 = \Tr \big[ e^{-\beta (H_{0}+H_z)} \big] = \Tr \big[ e^{-\beta \ovl{H}_{0}} \big]$, where $\beta=1/T$ is the inverse temperature. 
We denote the thermal average over operator $X$ as 
\begin{align}   \label{eq:exp_X}
&\inn{X}_0 = \frac{1}{Z_0} \Tr\left[ X \cdot e^{-\beta (H_0+H_z)} \right] 
 \\\nonumber
=& \inn{\ovl X}_{\ovl 0}= \frac{1}{Z_0} \Tr\left[ \ovl{X} \cdot e^{-\beta \ovl{H}_0} \right] \ ,
\end{align}
where $\ovl{X} = UXU^\dagger$, $U = e^{\ii 2 \rho_z \Lambda_z \phi_v(0)}$, and the subscripts $0$ and $\ovl 0$ distinguish which gauge we are working in. 

Since $\Lambda_z$ is conserved, $\Lambda_z(\tau) = e^{\tau H} \Lambda_z e^{-\tau H} = \Lambda_z$, the longitudinal correlation function $\chi_z(\tau) = - \Inn{ T_\tau ~ \Lambda_z(\tau) ~ \Lambda_z(0) }_{0} = -1$ does not decay at any temperature, leading to a Curie's law of the static susceptibility,  $\chi_z \sim \frac{1}{T}$. 
For the transverse correlation function, $\chi_x(\tau) = - \langle T_\tau ~ \Lambda_-(\tau) ~ \Lambda_+(0) \rangle_0 = -\langle T_\tau ~ \ovl{\Lambda}_-(\tau) ~ \ovl{\Lambda}_+(0) \rangle_{\ovl{0}}$, there is
\begin{align}
    \ovl{\Lambda}_\pm(\tau) &=  U e^{\tau H_0} \Lambda_\pm e^{-\tau H_0} U^\dagger \\\nonumber
    &= e^{\tau \ovl{H}_0} (\Lambda_\pm e^{ \pm 4 \ii\rho_z \phi_{v}(0)}) e^{-\tau \ovl{H}_0} = \Lambda_\pm e^{\pm 4\ii \rho_z \phi_{v}(\tau,0)} \ . 
\end{align}
Here, $\phi_{v}(\tau,x) = e^{\tau \ovl{H}_0} \phi_{v}(x) e^{-\tau \ovl{H}_0}$ denotes the free evolution of boson fields, and $\ovl{H}_0$ commutes with $\Lambda_\pm$ in the thermodynamic limit. 
Therefore, 
\begin{align}  \label{eq:chix_AD}
    \chi_x(\tau) =& - \Big\langle \Lambda_-  \Lambda_+ \Big\rangle_{\ovl{0}}  \Big\langle T_\tau ~ e^{-4\ii  \rho_z \phi_{v}(\tau,0)} ~ e^{4\ii  \rho_z \phi_{v}(0,0)} \Big\rangle_{\ovl{0}} \nonumber\\
    \overset{T\to 0^+}{=} & - \frac{1}{2} \left( \frac{x_c}{|\tau| + x_c} \right)^{16 \rho_z^2} \ . 
\end{align}
The first step exploited the fact that the eigenstates are direct products of impurity and bath fields. 
$\langle \Lambda_- \Lambda_+ \rangle_{\ovl{0}} = \frac{1}{2}$ by definition, and the correlation functions of the bath vertex operators can be looked up in \cref{eq:vertex_corr_2}. 
One thus obtains that the static transverse susceptibility diverges as $\chi_x \sim T^{16\rho_z^2-1}$. 
Also, \cref{eq:chix_AD} implies that the spectrum of dynamic transverse susceptibility scales as $\Im[\chi_x(\omega+\ii0^+)] \sim - \sgn(\omega) |\omega|^{16\rho_z^2-1}$. 
We compute this Fourier transformation explicitly in \cref{app:exactAD-sus}. 
In particular, for $\rho_z<\rho_z^\star=\frac{1}{4}$, the power $\alpha = 16 \rho_z^2<1$, hence the transverse susceptibility always diverges.

\section{RG analysis of pair-Kondo model}    \label{sec:BKT}

We carry out RG analysis on the PK model $\ovl{H}_{\rm PK}$ [\cref{eq:ovlH0,eq:ovlHx}], to show that there is a Berezinskii-Kosterlitz-Thouless (BKT) phase transition at a critical $\rho_z^c = \frac{1}{2} - \frac{1}{2\sqrt{2}}$, $\lambda_x=0$. 
Since the $\phi_{c,s,vs}$ fields are decoupled and free, their partition functions can be factored out, hence we will only focus on the $\phi_v$ fields in $\ovl{H}_{\rm PK}$. 
Also, it suffices to work in the thermodynamic limit and we will ignore all $\OO(L^{-1})$ terms in this section. 
The relevant Hamiltonian thus reads $\ovl{H}_v = \ovl{H}_{0,v} + \ovl{H}_x$, where $\ovl{H}_{0,v} = \int_{-L/2}^{L/2} \dd x\ \frac{:(\partial_x \phi_{v}(x))^2:}{4\pi}$ is the free boson Hamiltonian, and $\ovl{H}_x$ is given by \cref{eq:ovlHx}. 

The correction to the partition function due to $\ovl H_x$ is  
\begin{align} \label{eq:delta-Z-def}
    \delta Z  = e^{-\beta \cdot \delta F} = \left\langle T_\tau ~ \exp\pare{-\int_{-\frac{\beta}{2}}^{\frac{\beta}{2}}\mrm{d}\tau~ \ovl{H}_x(\tau) } \right\rangle_{\ovl{0}} \ . 
\end{align}
$\delta F$ is the (additive) correction to the free energy, the subscript $\ovl 0$ represents average with respect to the free boson ensemble $e^{-\beta \ovl H_{0,v}}$, $T_\tau$ is the time-ordering operator, and 
{\small
\begin{align}  \label{eq:ovlHx_interacting}
& \ovl H_x (\tau) = e^{\tau \ovl H_{v,0}} \cdot \ovl H_{x} \cdot e^{-\tau \ovl H_{v,0}} \\\nonumber 
&=  \frac{\lambda_x}{ x_c} 
\pare{ \Lambda_+(\tau) \cdot F_v(\tau) \cdot  e^{-\ii \gamma \phi_v(\tau,0)} 
  + \Lambda_-(\tau) F_v^\dagger(\tau) \cdot   e^{ \ii \gamma \phi_v(\tau,0)}  } 
\end{align}}
is the PK coupling written in the interaction picture. $\gamma = 2-4\rho_z$. 
Notice that $\Lambda_\pm (\tau) = \Lambda_{\pm}$ and $F_v(\tau) = F_v$, because $\ovl H_{0,v}$ commutes with $\Lambda_\pm$ and $F_v$ in the thermodynamic limit. 
However, we still explicitly keep the $\tau$ index for the convenience of time ordering.

\subsection{Coulomb gas analog}
\label{sec:BKT-coulombgas}

The partition function in \cref{eq:delta-Z-def} can be expanded into a perturbative series summation in terms of $\lambda_x$, $\delta Z =  1 + \delta Z_{2} + \delta Z_{4} + \cdots$, where the $2n$-th order correction is 
\begin{align}  \nonumber  
    \delta Z_{2n} &= \frac{1}{(2n)!} \int_{-\frac{\beta}{2}}^{\frac{\beta}{2}}
        \prod_{i=1}^{2n} \dd \tau_i\ 
    \left\langle T_\tau  ~ \ovl{H}_x(\tau_{2n}) \cdots \ovl{H}_x(\tau_{1}) \right\rangle_{\ovl{0}} \\ \label{eq:deltaZ2n_0}
    &= \int_{(-\frac{\beta}{2}, \frac{\beta}{2})}^{>0} \mrm{d}^{2n}\tau ~ \left\langle \ovl{H}_x(\tau_{2n}) \cdots \ovl{H}_x(\tau_{1}) \right\rangle_{\ovl{0}} \ . 
\end{align}
For the average over $\ovl{0}$ to be non-vanishing, the operator string $\ovl{H}_x(\tau_{2n}) \cdots \ovl{H}_x(\tau_{1})$ cannot accumulate net $\Lambda_z$ or $N_v$ charges, which are preserved by $\ovl H_{0,v}$. 
Thus, odd-order terms $\delta Z_{2n+1}$ must vanish. 
In the last step, we explicitly chose one of the $(2n)!$ time-ordered integral domains, and introduced an abbreviation for the corresponding integral domain and integral measure, 
{
\begin{align}  \label{eq:integral-ordered-tau}
    \int_{(-\frac{\beta}{2}, \frac{\beta}{2})}^{>a} \mrm{d}^{2n}\tau = \int_{-\frac{\beta}{2}}^{\frac{\beta}{2}}\mrm{d}\tau_{2n} \int_{-\frac{\beta}{2}}^{\tau_{2n}-a}\mrm{d}\tau_{2n-1} \cdots \int_{-\frac{\beta}{2}}^{\tau_2-a}\mrm{d}\tau_1 ,
\end{align}}
where $a$ is chosen as $0$ in $\delta Z_{2n}$. 
The $\frac{1}{(2n)!}$ factor is canceled by adding up all such domains. 

To evaluate the expectation value over an operator $\ovl{X}$ that commutes with $C_{2x}$ (such as $\delta Z_{2n}$), it suffices to look at the $\Lambda_z=+$ sector of the eigenstates of $\ovl H_0$, because the other sector produces the same expectation value, {\it i.e.,}
$ \langle \ovl{X} \rangle_{\ovl{0}} 
 = \langle \ovl{X} \rangle_{\ovl{0},+}$. 
We will exploit this property henceforth, but will omit the  subscript  ${\ovl{0},+}$for brevity, unless otherwise mentioned. 
Also, as in this section we will only encounter $\phi$ fields located at $x=0$, we will omit their spatial argument. 

At each $\tau_j$ in $\delta Z_{2n}$, we should pick either the $\Lambda_+$ term or the $\Lambda_-$ term from $\ovl{H}_x(\tau_j)$ [\cref{eq:ovlHx_interacting}]. 
We now analyze the general structure of the non-vanishing terms in $\delta Z_{2n}$:
\begin{enumerate}[label=\arabic*.]
\item There is a common factor $\left(\frac{\lambda_x}{x_c}\right)^{2n}$. 
\item Starting with the $\Lambda_z=+$ state before $\tau_1$, we can only flip $\Lambda_z$ down and up alternately. 
Consequently,  we must choose the $\Lambda_-$ term at all $\tau_{2k+1}$ and the $\Lambda_+$ term at all $\tau_{2k}$. 
The staggered string of impurity operators can be factored out, and produces $\langle \Lambda_+ \Lambda_- \cdots \Lambda_+ \Lambda_- \rangle_{\ovl{0},+} = 1$. 
\item Due to (2), the operator string of vertex operators ($e^{\pm \ii \gamma \phi_{v}}$) is also fixed. It gives the factor 
{\small
\begin{align}
\Inn{ e^{-\ii\gamma \phi_{v}(\tau_{2n})} \cdots e^{\ii\gamma\phi_{v}(\tau_{1})} } 
    = \exp\bigg[ -\sum_{j>i} (-1)^{j-i} \gamma^2 &
\\\nonumber
  \times \ln \left( \frac{\pi T x_c}{ \sin\pare{ \pi T (\tau_{j} - \tau_{i}) + \pi T x_c}  }\right) \bigg] & ,
\end{align}}
where \cref{eq:vertex_corr_2n} is exploited. 
Remarkably, it is equivalent to the partition function of $2n$ classical particles on a line, carrying $\pm \gamma$ charges in a staggered pattern, and interacting through the logarithmic (two-dimensional) Coulomb force. 
This is known as the Coulomb gas analog \cite{Yuval_1970_exact_1, Anderson_1970_exact_2}. 
$x_c$ plays the role as a short-distance cutoff in this analog.
\item Due to (2), the Klein factors must compose a string of the form $F_v F_v^\dagger \cdots F_v F_v^\dagger=1$. 
\item Since $x_c$ simply plays the role of a short-distance (high-energy) cutoff, various ways to implement it should agree on the physical output. For convenience, we will change the integral from $\int^{>0}_{(-\frac{\beta}2,\frac{\beta}2)} \dd^{2n}\tau$ to $\int^{>x_c}_{(-\frac{\beta}2,\frac{\beta}2)} \dd^{2n}\tau$ and simultaneously replace all $\frac{\pi T x_c}{\sin\pare{ \pi T (\tau_{j} - \tau_{i}) + \pi T x_c} }$ with $\frac{\pi T x_c}{ \sin(\pi T (\tau_{j} - \tau_{i})) }$. 
\end{enumerate}
In summary, $\delta Z_{2n}$ can be written as 
{\small
\begin{align}  \label{eq:Z2n-def}
\delta Z_{2n} 
=&  \left(\frac{\lambda_x}{x_c} \right)^{2n} \int^{>x_c}_{(-\frac{\beta}{2}, \frac{\beta}{2})} \mrm{d}^{2n}\tau 
    \prod_{j>i} \pare{ \frac{ \pi T x_c}{ \sin(\pi T(\tau_{j} - \tau_i)) }}^{ -(-1)^{j-i} \gamma^2 } .
\end{align}}
It can be seen that $\delta Z$ in the $T\to 0^+$ limit is an analog to the 2D Coulomb gas problem \cite{Yuval_1970_exact_1, Anderson_1970_exact_2, Kotliar_toulouse_1996}, where particles (with charges $(-1)^i$) are restricted to the $\tau$-axis and interact with each other via the Coulomb interaction $\frac1{|\tau_j - \tau_i|^{\gamma^2}}$.

To gain some insights into the perturbation theory, let us calculate the lowest order correction $\delta Z_2$ in the zero-temperature limit ($T\to 0^+$):
\begin{align} \label{eq:Z2}
\delta Z_2 
&= \frac1{x_c T} \cdot \frac{\lambda_x^2}{\gamma^2 - 1} 
\pare{ 1 - (2x_c T)^{\gamma^2-1} }\ .
\end{align}
Correspondingly, the lowest order correction to the energy $\delta E_2 = - T \cdot \ln \delta Z $ is 
\begin{equation}
\delta E_2 
= - \frac1{x_c} \cdot \frac{\lambda_x^2}{\gamma^2 - 1} \pare{ 1 - (2x_c T)^{\gamma^2-1} }\ . 
\end{equation}
For $\rho_z < \rho_z^\star=\frac14$, $\gamma>1$ and $(2x_c T)^{\gamma^2 - 1}$ vanishes in the $x_c\to 0^+$, $T\to 0^+$ limit.
However, if $\gamma\le 1$, the $(2x_c T)^{\gamma^2 - 1}$ term diverges, suggesting invalidity of the $\lambda_x$-expansion. 
Therefore, $\gamma > 1$ or $\rho_z< \frac14 $ is necessary to validate the perturbation theory. 
Further, to justify the perturbation theory, the second-order correction should be smaller than the typical energy scale ($x_c^{-1}$) of the unperturbed system. 
Thus, the perturbative regime should be  
\begin{equation} \label{eq:perturbation-condition}
\text{perturbative regime}: \qquad \lambda_z< \frac12 , \qquad 
    \lambda_x^2 \lesssim \gamma^2-1    \ .  
\end{equation}
Interestingly, the model at $\rho_z^\star = \frac14$ can be mapped to a solvable free-fermion system by refermionizing $\frac{F_v}{\sqrt{2\pi x_c}}e^{-\ii \phi_v}$ to a new fermion operator. 
This limit is similar to the ``Toulouse line'' of the single-channel Kondo problem \cite{toulouse_1969,giamarchi2003quantum} and represents the strong coupling Fermi liquid phase.

\subsection{Flow equations}

We will take the order of limits $\lim_{x_c\to 0^+}$ $\lim_{T\to 0^+}$ $\lim_{L\to \infty}$ in the RG analysis in this subsection.
We can replace all the $\frac{\pi T x_c}{ \sin( \pi T (\tau_{j}-\tau_i)  ) }$ factors in \cref{eq:Z2n-def} by $\frac{x_c}{ \tau_{j}-\tau_i  }$ in this limit. 

Rescale all the coordinates as $\tau_i = b \tau'_i$, where $b= e^{\dd\ell}>1$, and then relabel $\tau_i'$ as $\tau$.
The partition function in \cref{eq:Z2n-def} becomes   
{\small
\begin{equation} \label{eq:Z2n-rescale}
\delta Z_{2n} = \frac{\lambda_x^{2n} \cdot b^{2n - \gamma^2 n}}{{x_c}^{2n}}
    \int_{(-\frac{\beta}{2b}, \frac{\beta}{2b})}^{>x_c b^{-1}} \dd^{2n} \tau\ 
    \prod_{j>i} 
    \pare{ \frac{x_c}{\tau_j -\tau_i} }^{-(-1)^{j-i} \gamma^2}   .
\end{equation}}
We can absorb the factor $b^{1-\frac{\gamma^2}2}$ into $\lambda_x$ and define it as the renormalized parameter, {\it i.e.}, $\lambda_{x}(\ell + \dd\ell) = \lambda_x(\ell) \cdot e^{\dd\ell (1-\frac{\gamma^2}2)}$. 
The flow equation for $\lambda_x$  immediately follows
\begin{equation}
    \frac{\dd\lambda_x}{d\ell} = \pare{1 - \frac{\gamma^2}2} \lambda_x \ .
\end{equation}
One can alternatively obtain this result by a simple power counting.
According to the discussions around \cref{eq:vertex_corr_2}, the scaling dimension of the vertex operator $[e^{\pm \ii \gamma \phi_{v}}] = \frac{\gamma^2}{2}$. 
To ensure the Hamiltonian $\ovl H_x$ has the correct scaling dimension $[\ovl H_x]=1$, the coupling constant must have $[\lambda_x]= 1 - \frac{\gamma^2}2$. 
$\lambda_x$ is relevant, marginal, and irrelevant for $\gamma<\sqrt2$, $=\sqrt2$, and $\sqrt2$, respectively. 

To obtain the flow of $\gamma$, we integrate out ``high-energy'' configurations as virtual processes. 
The $2n$-th order partition function has the form $\delta Z_{2n} = \delta Z_{2n,0} + \delta Z_{2n,1} + \mathcal{O}({\rm d}\ell^2)$, where all adjacent particles in  $\delta Z_{2n,0}$ are separated by at least a distance $x_c$, and one adjacent particle-pair (say $\tau_{i+1}, \tau_i$) in $\delta Z_{2n,1}$ has a distance $x_c b^{-1} < \tau_{i+1}-\tau_i < x_c$. 
Since adjacent particles carry opposite charges, we term the pair $(\tau_{i+1}, \tau_i)$ as a dipole. 
Multiple dipole excitations contribute the $\mathcal{O}(\mrm{d}\ell^2)$ term and will be neglected. 
We integrate out the dipoles and re-organize the low-energy terms into a new partition function, $\delta Z' = \sum_{n=0}^{\infty} \delta Z'_{2n}$, where $\delta Z'_{2n}$ contains $2n$ low-energy particles. 
We examine the free-energy
\begin{widetext}
\begin{align}
\delta F &= -T\cdot \ln\big[ \delta Z \big] = -T\cdot \ln\left[ 1 + \sum_{n=1}^{\infty} \delta Z_{2n,0} + \sum_{n=1}^{\infty} \delta Z_{2n,1} \right] 
    \\\nonumber
&= -T \cdot \ln\big( 1 +\delta Z_{2,1} \big) - T \cdot \ln\left[ 1 + \sum_{n=1}^{\infty} \left(\delta Z_{2n,0} + \delta Z_{2n+2,1} - \delta Z_{2n,0} \delta Z_{2,1} \right) + \mathcal{O}(\mrm{d}\ell^2) \right] \ .
\end{align}
Here $-T\cdot\ln(1 +\delta Z_{2,1})$ is the ``high-energy'' free-energy contributed by the inner degrees of freedom of the dipole. 
The second term is the ``low-energy'' free-energy, where Coulomb interaction is screened by the dipole. 
We thus conclude
\begin{align}  \label{eq:dZ2n_prime}
    \delta Z'_{2n} = \delta Z_{2n,0} + \delta Z_{2n+2,1} - \delta Z_{2n,0} \delta Z_{2,1}
\end{align}
serves as the effective $2n$-particle partition function for the low-energy particles.  

Let us first compute $\delta Z_{2,1}$, 
\begin{align}   \label{eq:dZ21}
\delta Z_{2,1} &= \pare{ \frac{\lambda_x}{x_c} }^2 \cdot b^{2-\gamma^2} \cdot \int^{\frac{1}{2 T b}}_{-\frac{1}{2 T b}} \mrm{d} \tau_2
    \int_{\tau_2-x_c}^{\tau_2-x_c b^{-1}} \mrm{d}\tau_1 ~ \left( \frac{x_c}{\tau_2 - \tau_1} \right)^{\gamma^2} 
    = \frac{\lambda_x^2}{x_c T}  \cdot \mrm{d}\ell  + \mathcal{O}(\dd\ell^2)  \ . 
\end{align}
Since the integral range over $\tau_1$ is proportional to $\mrm{d} \ell$, we can safely omit all the $\mathcal{O}(\mrm{d}\ell)$ factors elsewhere.

Next, we compute $\delta Z_{4,1}$ and see how it renormalizes $\delta Z_{2,0}$. 
The calculation for $\delta Z_{2n+2,1}$ with generic $n$ parallels with that for $\delta Z_{4,1}$, and will agree on the final RG equation. 
$\delta Z_{4,1}$ consists of three terms, $\delta Z_{4, 1} = \sum_{i=1}^{3} \delta Z_{4, 1}^{(i+1,i)}$, where $\delta Z_{2n+2, 1}^{(i+1,i)}$ has a dipole formed by $\tau_{i+1}$ and $\tau_i$.
The first term is 
{
\begin{align}
& \delta Z_{4,1}^{(2,1)} =  \frac{\lambda_x^4}{x_c^4 } 
    \int_{-\frac{1}{2bT}}^{\frac{1}{2bT}} \dd\tau_4 
    \int_{-\frac{1}{2bT}}^{\tau_4-x_c} \dd\tau_3 
    \int_{-\frac{1}{2bT}}^{\tau_3 - x_c} \dd\tau_2 
    \int_{\tau_2 - x_c}^{\tau_2 - x_c/b} \dd\tau_1 \  
    \exp\pare{ \gamma^2 \sum_{j>i} (-1)^{j-i} \ln \pare{\frac{\tau_j -\tau_i}{x_c}} }  \nonumber\\
=&  \frac{\lambda_x^2}{x_c^{2}} \cdot  \frac{ \lambda_x^2 \dd\ell}{x_c} 
    \int_{-\frac{1}{2bT}}^{\frac{1}{2bT}} \dd\tau_4 
    \int_{-\frac{1}{2bT}}^{\tau_4-x_c} \dd\tau_3 
    \int_{-\frac{1}{2bT}}^{\tau_3 - x_c} \dd\tau_2 ~
    \exp\pare{-\gamma^2 \ln\pare{ \frac{\tau_4 - \tau_3}{x_c} }
    -\frac{\gamma^2 x_c}{\tau_4-\tau_2} 
    +\frac{\gamma^2 x_c}{\tau_3 - \tau_2} + \mathcal{O}(x_c^2) }  \ . 
\end{align}}
Since this term is proportional to $\dd \ell$, we can omit all the $b$ factors elsewhere. 
We relabel $\tau_{3,4}$ as $\tau_{1,2}$, respectively, and relabel the original $\tau_{2}$ as $\tau'+\frac12 x_c $:
\begin{equation} \label{eq:dZ41-21-tmp1}
\delta Z_{4,1}^{(2,1)} =  \frac{\lambda_x^2}{x_c^{2}} \cdot  { \lambda_x^2 \dd\ell}
    \int_{-\frac{1}{2T}}^{\frac{1}{2T}} \dd\tau_2
    \int_{-\frac{1}{2T}}^{\tau_2-x_c} \dd\tau_1
    \int_{-\frac{1}{2T}}^{\tau_1 - \frac32 x_c} \dd\tau' ~
    e^{-\gamma^2 \ln\pare{ \frac{\tau_2 - \tau_1}{x_c}   } }
    \pare{ \frac1{x_c}
    -\frac{\gamma^2}{\tau_2-\tau'} 
    +\frac{\gamma^2}{\tau_1 - \tau'} + \mathcal{O}(x_c) } \ .
\end{equation}
The second and third terms in the parentheses can be viewed as the interaction between charges at $\tau_{1,2}$ and the dipole  at $\tau'<\tau_1$. 
After integrating out the dipole, we obtain 
\begin{align}
\delta Z_{4,1}^{(2,1)} =  \frac{\lambda_x^2}{x_c^2} 
    \int_{-\frac1{2T}}^{\frac1{2T}}  \dd\tau_2
    \int_{-\frac1{2T}}^{\tau_2 - x_c} \dd\tau_1 ~
    e^{-\gamma^2 \ln \pare{ \frac{\tau_2 - \tau_1}{x_c}} }
    \cdot { \lambda_x^2 d\ell}
    \Bigg[ \frac1{x_c} \pare{ \frac{1}{2T} + \tau_1 -\frac32 x_c } 
    + \gamma^2 \ln \pare{\frac{\tau_2 - \tau_1}{ \frac32 x_c}  }  
    + \mathcal{O}(x_c)  \Bigg]\ . 
\end{align}
We have omitted the term $\gamma^2 \ln \pare{ \frac{\frac1{2T} + \tau_1}{\frac1{2T}+\tau_2} }$ because it vanishes in the $T\to 0^+$ limit. 
Similarly, $\delta Z_{4,1}^{(3,2)}$ and $\delta Z_{4,1}^{(4,3)}$ are given by 
\begin{align} \label{eq:dZ41-32-tmp1}
\delta Z_{4,1}^{(3,2)} 
= &  \frac{\lambda_x^2}{x_c^2} 
    \int_{-\frac1{2T}}^{\frac1{2T}}  \dd\tau_2
    \int_{-\frac1{2T}}^{\tau_2 - 3x_c} \dd\tau_1 
    ~ e^{-\gamma^2 \ln \pare{ \frac{\tau_2 - \tau_1}{x_c}} } \cdot 
    { \lambda_x^2 \dd\ell}
    \pare{ \frac1{x_c}\pare{ \tau_2 - \tau_1 + 3x_c}
    + 2\gamma^2 \ln \pare{\frac{\tau_2 - \tau_1}{\frac32 x_c}} + \mathcal{O}(x_c)
    } \ ,
\\ \label{eq:dZ41-43-tmp1}
\delta Z_{4,1}^{(4,3)} 
=&  \frac{\lambda_x^2}{x_c^2} 
    \int_{-\frac1{2T}}^{\frac1{2T}}  \dd\tau_2
    \int_{-\frac1{2T}}^{\tau_2 - x_c} \dd\tau_1 
    ~ e^{-\gamma^2 \ln \pare{ \frac{\tau_2 - \tau_1}{x_c}} } \cdot 
    { \lambda_x^2 \dd\ell}
    \pare{ \frac1{x_c}\pare{\frac1{2T} - \tau_2 - \frac32x_c}
    + \gamma^2 \ln \pare{\frac{\tau_2 - \tau_1}{\frac32 x_c}} + \mathcal{O}(x_c)
    } \ ,
\end{align}
respectively. 
Notice that in $\delta Z_{4,1}^{(3,2)}$ the least distance between $\tau_2$ and $\tau_1$ is $3x_c$. 
We manually change the least distance back to $x_c$, which will lead to an error of the same order as an $\mathcal{O}(1)$ term in the parentheses. 
Adding up the three terms, we obtain the total $\delta Z_{4,1}$ 
\begin{equation}
\delta Z_{4,1} = 
 \frac{\lambda_x^2}{x_c^2} 
    \int_{-\frac1{2T}}^{\frac1{2T}}  \dd\tau_2
    \int_{-\frac1{2T}}^{\tau_2 - x_c} \dd\tau_1 
    ~ e^{-\gamma^2 \ln \pare{ \frac{\tau_2 - \tau_1}{x_c}} } \cdot 
    { \lambda_x^2 \dd\ell} 
    \pare{ \frac1{x_c T}
    + 4 \gamma^2 \ln \pare{\frac{\tau_2 - \tau_1}{x_c}} + \mathcal{O}(1)
    } \ . 
\end{equation}

According to \cref{eq:dZ2n_prime}, the renormalized two-particle partition function is $\delta Z_2' = \delta Z_{2,0} + \delta Z_{4,1} - \delta Z_{2,1}\delta Z_{2,0}$, where $\delta Z_{2,0}$ is rescaled as explained after \cref{eq:Z2n-rescale} and $\delta Z_{2,1} = \frac{\lambda_x^2}{x_c T}$ is given in \cref{eq:dZ21}. 
The $\frac1{x_c T}$ term in $\delta Z_{4,1}$ is exactly canceled by $\delta Z_{2,1}\delta Z_{2,0}$.
Thus, 
\begin{align}
& \delta Z_2'  
=  \frac{\lambda_x^2 e^{ \dd\ell (2-\nu)}}{x_c^{2}}
\int_{-\frac1{2T}}^{\frac1{2T}} \dd\tau_2 \int_{-\frac1{2T}}^{\tau_2 - x_c} \dd\tau_1 ~  
    e^{-\gamma^2 \ln \pare{ \frac{\tau_1 - \tau_2}{x_c}}} \cdot 
    \pare{ 1 + 
        4\gamma^2  \lambda_x^2  \dd\ell 
        \cdot \ln \pare{ \frac{\tau_1 - \tau_2}{x_c} }
        + \mathcal{O}( \lambda_x^2 \dd\ell \cdot 1)
     }
    + \mathcal{O}( \dd\ell ^2) + \mathcal{O}(\lambda_x^6)\ . 
\end{align}
\end{widetext}

Comparing it to the original form of the partition function in \cref{eq:Z2n-def}, one can immediately read off the parameter renormalization, 
\begin{align}
\lambda_x^2 (\ell +  \dd\ell ) &= \lambda_x^2 (\ell) e^{ \dd\ell  (2 - \gamma^2)} \ , \\ \nonumber
\gamma^2 (\ell +  \dd\ell ) &= \gamma^2(\ell) - 4 \gamma^2(\ell) \cdot \lambda_x^2 \cdot \dd\ell  \ .
\end{align}
The omitted $\mathcal{O}(\lambda_x^2 \dd\ell \cdot 1)$ term will lead to an $\mathcal{O}(\lambda_x^4)$ correction to $\lambda_x^2$. 
Recall that $\gamma = 2 - 4\rho_z$, we derive the flow equations
\begin{align} \label{eq:flow-1}
    \frac{\dd\lambda_x}{ \dd\ell } 
    &= (-1 + 8\rho_z - 8\rho_z^2) \lambda_x + \mathcal{O}(\lambda_x^3)
\ ,\\ \label{eq:flow-2}
\frac{\dd \rho_z}{ \dd\ell } &= (1- 2\rho_z) \lambda_x^2 + \mathcal{O}(\lambda_x^3)
\ . 
\end{align}
We will omit the $\mathcal{O}(\lambda_x^3)$ terms. 

\subsection{Phase diagram and BKT transition}

\begin{figure}[t]
\includegraphics[width=1\linewidth]{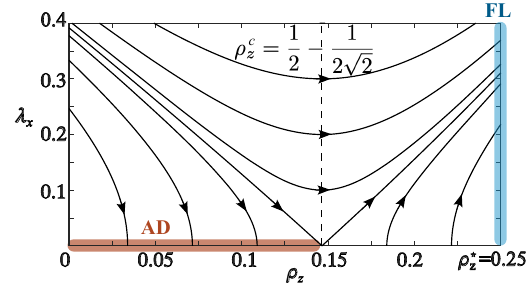}
\caption{RG flow of the pair-Kondo model in the doublet regime. Red and blue fixed lines represent the anisotropic doublet (AD) and Kondo Fermi liquid (FL) phases, respectively. }
\label{fig:flow}
\end{figure}

To simplify the calculation, we want to find an invariant that is unchanged under the flow. 
We observe 
\begin{align}
 \dd\ell &= \frac{ \dd\lambda_x}{(1-\gamma^2/2) \lambda_x}
    = - \frac{ \dd \gamma^2}{4\gamma^2  \lambda_x^2} \ .
\end{align}
Integrating both sides, we obtain the invariant 
\begin{align}\label{eq:flow-invariant}
    c
      &=  \lambda_x^2 + \ln (1-2\rho_z) - (1-2\rho_z)^2 + \frac{1+\ln 2}2 
\end{align}
up to a constant. 
A given $c$ value defines a curve in the $\rho_z$-$\lambda_x$ plane, and the RG flow follows these curves. 
We hence obtain the flow diagram in \cref{fig:flow}. 

There are two types of stable fixed lines: the blue one at $\rho_z^\star=\frac14$ and the red one at $\lambda_x=0$, $\rho_z< \frac12 - \frac1{2\sqrt2}$. 
We have chosen convention that $c=0$ at the critical point $\lambda_x=0$, $\rho_z^c = \frac12 -\frac1{2\sqrt2}$.
The red fixed line corresponds to the AD phase discussed in the last section. 
However, the flow in \cref{eq:flow-1,eq:flow-2} seems not stop at the blue fixed point. 
This is due to the invalidity of the perturbation theory (\cref{eq:perturbation-condition}) around the strong coupling line. 
As will be clear soon, the low-energy physics at $\rho_z^\star=\frac14$ is equivalent to a free-fermion system with a phase shift. 
Thus, $\rho_z^\star=\frac14$ represents a free-fermion fixed point. 
A straightforward analysis shows the phases diagram: 
\begin{equation}
    \text{Kondo Fermi liquid:} \qquad c> 0  \;\; \text{or} \;\;  
    \rho_z  > \frac12 -\frac1{2\sqrt2}\ ,
\end{equation}
\begin{equation}
    \text{anisotropic doublet:} \qquad c< 0 \;\; \text{and} \;\;  \rho_z<\frac12 - \frac1{2\sqrt2}\ . 
\end{equation}

We also obtain the Kondo temperature from the RG analysis. 
$c$ is the controlling parameter for the phase transition.
If $c=0$, the RG flow will take an infinite RG time ($\ell \to \infty$) to achieve the critical point at $\lambda_x = 0$, $\rho_z^c = \frac12 - \frac1{2\sqrt2}$. 
This is because the flow velocity approaches zero at the critical point. 
If $c$ is positive but small, the renormalized parameters will eventually hit the Fermi liquid fixed line, but the flow is extremely slow around the critical point at $\lambda_x = 0$, $\rho_z^c = \frac12 -\frac1{2\sqrt2}$. 
Thus, to estimate the RG time it takes to achieve the Fermi liquid fixed line, it suffices to examine the flow equations around the critical point:
\begin{equation}\label{eq:flow-around-BKT}
    \frac{\dd \lambda_x}{ \dd\ell } = t \lambda_x ,\qquad 
    \frac{\dd t}{ \dd\ell } = 4 \lambda_x^2\ ,
\end{equation}
where $t=1-\gamma^2/2 = -1 + 8\rho_z - 8 \rho_z^2$ and only quadratic and bilinear terms in $t$ and $\lambda_x$ are kept.
To the same order, $c= \lambda_x^2 - \frac14 t^2$. 
The flow equation for $t$ is $\frac{\dd t }{  \dd\ell } = 4c + t^2$. 
Since $c$ is invariant under the flow, we have the solution 
\begin{equation}
    \ell = \ell_0  + \frac1{\sqrt{4c}} \arctan \frac{t}{\sqrt{4c}} \ .
\end{equation}
Here $\ell_0$ is the initial RG time for the energy scale $D_{\rm PK}$ where the pair-Kondo model is justified. 
The Fermi liquid fixed line is characterized by $t=\frac12$ ($\gamma=1$, $\rho_z=\rho_z^\star=\frac14$). 
Given $c$ being small, $t/\sqrt{4c} \to \infty$, the RG time from the energy scale $D_{\rm PK}$ to the Kondo temperature is $\ell - \ell_0 \approx \frac{\pi}{4 \sqrt{c}}$.
Therefore, the Kondo temperature is determined as 
\begin{equation} \label{eq:Kondo-temperature}
    T_{\rm K} \sim D_{\rm PK} \cdot \exp\pare{ - \frac{\pi}{4 \sqrt{c}}} \ , \qquad (1\gg c >0)
  \ . 
\end{equation}

BKT transition also exists in the exemplary anisotropic Kondo problem (AK) \cite{Yuval_1970_exact_1, Anderson_1970_exact_2, giamarchi2003quantum}. 
Our AD line resembles the ferromagnetic line in AK, except that $\rho_z^c$ in AK is zero. 
The difference arises because the PK coupling is a quartic operator, hence $\lambda_x$ is irrelevant at the tree level, while the transverse Kondo coupling in AK is marginal. 
Therefore, an infinitesimal anti-ferromagnetic $\rho_z$ in AK suffices to drive the system into strong-coupling, while a threshold anti-ferromagnetic $\rho_z^c$ in PK is required. 
It is the finite $\rho_z^c$ that allows AD to appear in an Anderson model, where the effective $\rho_z$ is always anti-ferromagnetic. 

\section{Exact solution to pair-Kondo model at strong coupling fixed line (\texorpdfstring{$\rho_z^\star=\frac14$}{rhoz=1/4})}   \label{sec:refermFL}

In this section, we refermionize $\ovl{H}_{\rm PK}$ [\cref{eq:ovlH0,eq:ovlHx}] along the line $\rho_z^\star = \frac{1}{4}$ \cite{von_delft_bosonization_1998, zarand_analytical_2000, vonDelft_1998_finitesize} to calculate the finite-size many-body spectrum, thermodynamic quantities, and various correlation functions. 
The results will confirm the strong-coupling phase in the BKT phase diagram is a Kondo FL, where the doublet is exactly screened by the bath due to a resonance of PK coupling $\lambda_x$. 
Especially, in \cref{sec:refermFL-referm} we will construct one-to-one mapping between the original physical Hilbert space to the pseudofermion Hilbert space, which will allow us to obtain the correct finite-size spectrum. 
For the sake of calculating finite-size spectrum, $\mcl{O}(\frac{1}{L})$ terms will be kept in \cref{sec:refermFL-referm,sec:finite-size-spectrum}. 

\subsection{Refermionization} \label{sec:refermFL-referm}

Since $e^{-\ii \phi_v(x)}$ has the same scaling dimension as a fermion operator, we aim to find an exact construction with form of \cref{eq:bosonization} to map it into a pseudo-fermion $\psi_v(x) \sim  \frac{F_v}{\sqrt{2\pi x_c}} e^{-\ii \phi_v(x)}$.  
Simultaneously, for the Hamiltonian to respect pseudo-fermion parity, we will also map $\Lambda_\pm$ to a local fermion operator $f_{v}$. 

We construct $f_v$ first. 
To enforce the anti-commutation between $\psi_v(x)$ and $f_v$, we introduce a further gauge transformation $U_2 = e^{\ii \frac{\pi} 4 N_v \Lambda_z}$. 
$U_2$ serves to dress $\Lambda_+$ and $F_v$ with 
\begin{align}   \label{eq:U2_Lamb_Fv}
    U_2 \Lambda_+ U_2^\dagger = e^{\ii\frac{\pi}{2} N_v}  \Lambda_+ \ , \quad 
    U_2 F_v U_2^\dagger = e^{-\ii\frac{\pi}{2} \Lambda_z}  F_v  \ .
\end{align}
Here, we have exploited \cref{eq:BCH} and $[\Lambda_z, \Lambda_+] = 2 \Lambda_+$, $[N_v, F_v] = - 2F_v$. 
We define the local fermions as 
\begin{align}   \label{eq:fo}
    f^\dagger_v = e^{\ii \frac{\pi}{2} N_v}   \Lambda_+  e^{- \ii \frac{\pi}{2} \Lambda_z} , \quad 
    f_v = e^{-\ii \frac{\pi}{2} N_v}  e^{ \ii  \frac{\pi}{2} \Lambda_z}   \Lambda_- \ .
\end{align}
They satisfy  $\{f_v, f_v^\dagger\} = 1$ and 
\begin{align}
    f^\dagger_v f_v = \frac{\Lambda_z + \Lambda_0}{2} , \qquad f_v f^\dagger_v = \frac{\Lambda_0 - \Lambda_z}{2}\ . 
\end{align}
Due to the $e^{\pm \ii \frac{\pi}2 N_v}$ factor, which can be understood as a Jordan-Wigner string that counts the total valley charges in the bath, $f_v$ also anti-commutes with the composite Klein factor $F_v$ in the bath, 
\begin{equation}
    \{ f_v, F_v \} = \{ f_v^\dagger, F_v\} = 
    \{ f_v, F_v^\dagger \} = \{ f_v^\dagger, F_v^\dagger\} = 0\ . 
\end{equation}
which will eventually guarantee $f_v$ to anti-commute with $\psi_v(x)$. 

Hamiltonian in this new gauge is denoted as $\widehat{H}_{\rm PK} = U_2 \ovl{H}_{\rm PK} U_2^\dagger = \widehat{H}_{0} + \widehat{H}_x$, where $\widehat{H}_{0} = U_2 \ovl{H}_{0} U_2^\dagger = \ovl{H}_0$, and, by \cref{eq:U2_Lamb_Fv}, 
\begin{align}   \label{eq:wdhHx_0}
\widehat{H}_{x} = U_2 \widehat{H}_{x} U_2^\dagger =&  
\frac{\lambda_x}{x_c} \pare{  f_v^\dagger F_v \cdot e^{-\ii \phi_v(0)}  
    + h.c. } \ .
\end{align}

Next, we construct $\psi_v(x)$. One may attempt to write $\psi_v(x)$ as $\frac{F_v}{\sqrt{2\pi x_c}} e^{-\ii \phi_v(x)} e^{-\ii (N_v - \frac{\Pbc'}2) \frac{2\pi}{L} x}$ in analogy to \cref{eq:bosonization}, where $N_v$ plays the role of the total charge of pseudo-fermions, and $\Pbc'$ determines the boundary condition. 
But this construction is invalid because $F_v$ changes $N_v$ by $-2$ rather than $-1$, which is crucial for the anti-commutation relations such as $\{\psi_v(x),\psi_v(x')\}=0$. 
(See calculations around \cref{eq:anti-commutation-bosonization} for more details.)
In order to resolve this issue, we introduce a new basis for particle numbers 
\begin{align}  \label{eq:N_to_mN}
\begin{bmatrix}
    \mN_v \\ \mN_1 \\ \mN_2 \\ \mN_3
\end{bmatrix}
=   \brak{
    \begin{array}{rrrr}
    1 & 0 & 0 & 0 \\
   -1 & 1 & 0 & 0 \\
    1 & 0 & 1 & 0 \\
    1 & 0 & 0 & 1 
    \end{array}}
\begin{bmatrix}
    N_{+\uparrow} \\ N_{+\downarrow} \\ N_{-\uparrow} \\ N_{-\downarrow}
\end{bmatrix} \ ,
\end{align}
which satisfy 
\begin{equation}  \label{eq:mN_charge}
    [\mN_v, F_v] = - F_v,\qquad 
    [\mN_{1,2,3}, F_v] = 0 \ . 
\end{equation}
Note that the transformation from $N_{ls}$ to $\mN$ quantum numbers is unimodular, meaning that {\it any integer-valued} $\mN$ are physical. 
Also, by \cref{eq:mN_charge}, $\mN_v$ will play the same role as $N_v$ to keep track of the bath valley charges (although $N_v \in \frac{\mbb{Z}}{2}$ while $\mN_v \in \mbb{Z}$), as long as $\mN_{1,2,3}$ are conserved. 
We will explain this aspect in more detail around \cref{eq:mN_to_Nvtot}. 
We also tabulate the inverse of \cref{eq:N_to_mN}, 
\begin{equation} \label{eq:mN-to-N}
\begin{bmatrix}
    N_{+\uparrow} \\ N_{+\downarrow} \\ N_{-\uparrow} \\ N_{-\downarrow}
\end{bmatrix}
=  \brak{
    \begin{array}{rrrr}
    1 & 0 & 0 & 0 \\
    1 & 1 & 0 & 0 \\
   -1 & 0 & 1 & 0 \\
   -1 & 0 & 0 & 1
    \end{array}}
\begin{bmatrix}
        \mN_v \\ \mN_1 \\ \mN_2 \\ \mN_3
\end{bmatrix} \ . 
\end{equation}
Combining \cref{eq:N-flavor-transformation,eq:mN-to-N}, there is also
\begin{equation}  \label{eq:mN-to-Nflavors}
\begin{bmatrix}
    N_c \\ N_v \\ N_s \\ N_{vs}
\end{bmatrix}
= \frac12 \left[
\begin{array}{rrrr}
0 & 1 & 1 & 1 \\
4 & 1 & -1 & -1 \\
0 & -1 & 1 & -1 \\
0 & -1 & -1 & 1 \\
\end{array}
\right]
\begin{bmatrix}
    \mN_v \\ \mN_1 \\ \mN_2 \\ \mN_3
\end{bmatrix} \ . 
\end{equation}

Therefore, we define the pseudo-fermion operator as 
\begin{align}    \label{eq:psi_v}
\psi_{v}(x) &= \frac{F_v}{\sqrt{2\pi x_c}} e^{-\ii \left( \mN_v - \frac{1}{2} \right) \frac{2\pi x}{L}}  e^{-\ii \phi_v(x)} \ , 
\end{align}
where we have chosen $\Pbc' = 1$. 
Nonetheless, as $\Pbc'$ is independent of the physical boundary condition $\Pbc$, it will just be a gauge choice, and does not affect the final results. Similar to \cref{eq:psi_d}, we define Fourier components, 
\begin{align}  
\psi_{v}(x) &= \sqrt{\frac{1}{L}} \sum_{{k}} d_v(k) ~ e^{-\ii {k} x} 
\end{align}
with ${k} \in \frac{2\pi}{L} \left( \mathbb{Z} - \frac{1}{2} \right)$. 
Following calculations in \cref{app:boson-oprt}, one can explicitly verify that the canonical anti-commutation relations hold, 
\begin{align}
    \{\psi_{v}(x) ,  \psi_{v}(x') \} = 0, \quad  \{\psi_{v}(x) ,  \psi^\dagger_{v}(x') \} = \delta(x-x')\ .
\end{align}

The many-body Hilbert space will be completely indexed by the integers
\begin{align}   \label{eq:index_quantum_number}
    &\{ \mN_1, ~~ \mN_2, ~~ \mN_3, ~~ f_v^\dagger f_v,  \\\nonumber
    &~~ b_{c}^\dagger(q) b_c(q), 
    ~~ b_{s}^\dagger(q) b_s(q), 
    ~~ b_{vs}^\dagger(q) b_{vs}(q), 
    ~~ d_v^\dagger(k) d_v(k) \}\ .
\end{align}
In particular, $\mN_v = \sum_{k} :d_v^\dagger(k) d_v(k):$ according to the bosonization-refermionization dictionary (\cref{app:boson}), where the normal-ordering for pseudo-fermions will be defined with respect to such an auxiliary Fock state $\ket{\Omega_0'}$ that satisfies
\begin{equation} \label{eq:vacuum00}
    \inn{ \Omega_0'| f_v^\dagger f_v | \Omega_0'} =0,\quad  
    \inn{ \Omega_0'| d_v^\dagger(k) d_v(k) | \Omega_0'} = \theta(k<0). 
\end{equation}
Recall that $k \in \frac{2\pi}{L}(\mathbb{Z} - \frac12)$.

Next, we identify the good quantum numbers conserved by $\widehat{H}_{\rm PK}$ among \cref{eq:index_quantum_number}. 
In fact, only $f_v^\dagger f_v$ and $d^\dagger_{v}(k) d_v(k)$ will be broken by $\widehat{H}_{x}$ [\cref{eq:wdhHx_0}]. 
However, in analogy to $N_{v}^{\rm (tot)}$, we can construct the total pseudo-fermion charge
\begin{equation} \label{eq:Npf-def}
\mN_{\rm pf} =  f_v^\dagger f_v + \mN_v
   = f_v^\dagger f_v + \sum_k : d_v^\dagger(k) d_v(k) :  \ , 
\end{equation}
which is conserved by $\widehat{H}_{\rm PK}$. 
Using \cref{eq:N-flavor-transformation,eq:N_to_mN}, and noting that $\Lambda_z = 2f_v^\dagger f_v - 1$, we can express $N_v$ and $N_v^{\rm (tot)}$ in terms of $\mN_v$, $\mN_{\rm pf}$, and $\mN_{1,2,3}$, 
\begin{align}   \label{eq:mN_to_Nv}
N_v &= 2\mN_{v}  + \frac12 \pare{ \mN_1 - \mN_2 - \mN_3 } \ , \\  \label{eq:mN_to_Nvtot}
N_v^{\rm (tot)} &= 2\mN_{\rm pf} - 1 + \frac12 \pare{ \mN_1 - \mN_2 - \mN_3 } \ . 
\end{align}
It can thus be seen that, within each symmetry block of fixed $\mN_{1,2,3}$, $\mN_v$ and $\mN_{\rm pf}$ play the same roles as $N_{v}$ and $N_v^{\rm (tot)}$, respectively. 
The full set of good quantum numbers is thus given by
\begin{align}   \label{eq:good_quantum_number}
    &\{ \mN_1, ~~ \mN_2, ~~ \mN_3, ~~ \mN_{\rm pf}, \\\nonumber
    &~~ b_{c}^\dagger(q) b_c(q), 
    ~~ b_{s}^\dagger(q) b_s(q), 
    ~~ b_{vs}^\dagger(q) b_{vs}(q) \}\ .
\end{align}

Referring to \cref{eq:H0}, the kinetic energy of $\phi_v$ can be directly refermionized, 
\begin{align} 
& \sum_{q} q :b^\dagger_{v}(q) b_v(q): = \sum_{k} k ~ :d_v^\dagger(k) d_v(k): - \frac{2\pi}{L} \frac{\mN_v^2}2 \\\nonumber
&= \sum_{k} k ~ :d_v^\dagger(k) d_v(k): - \frac{2\pi}{L} \left( \frac{\mN_{\rm pf}^2 + f_v^\dagger f_v}2 - \mN_{\rm pf} f_v^\dagger f_v \right). 
\end{align}
In the second line, we have exploited \cref{eq:Npf-def} and $(f_v^\dagger f_v)^2 = f_v^\dagger f_v$. 
Now, re-writing other terms in $\widehat{H}_{0} = \ovl{H}_0$ [\cref{eq:ovlH0}, with $\rho_z^\star = \frac{1}{4}$, and dropping the energy constant proportional to $\PP_D$] that involve $N_{\chi}$ in terms of the good quantum numbers $\mN_{1,2,3}$ and $\mN_{\rm pf}$, we arrive at $\widehat{H}_{\rm PK} = \widehat{H}_{0} + \widehat{H}_x$, with
{\small
\begin{align} \label{eq:wdhH0}
& \widehat{H}_{0} = \frac{2\pi}{L} \Bigg( \sum_{i=1}^3 \brak{ \frac{(1- \Pbc)}{2} \mN_i + \frac{\mN_i^2}2}  - \mN_{\rm pf}  \\\nonumber
& - \frac14\brak{\mN_1 - \mN_2 - \mN_3} 
    + \mN_{\rm pf} \brak{ \frac32 \mN_{\rm pf} + \mN_1 - \mN_2 -\mN_3}
    \Bigg) \\ \nonumber
& + \sum_{\chi=c,s,vs} \sum_{q} q~ b_{\chi}^\dagger(q) b_\chi(q) 
+ \varepsilon_f ~ f_v^\dagger f_v + \sum_{k} k~ : d_v^\dagger(k) d_v(k) :\ ,
\end{align}
}
\begin{align}
\label{eq:wdhHx}
\widehat{H}_x &= \sqrt{\frac{2\pi \Gamma}{L}} \sum_{k} \pare{ f_v^\dagger d_v(k) + d_v^\dagger (k) f_v } ,
\end{align}
where we have defined 
\begin{equation}
    \varepsilon_f = \frac{2\pi}{L} \pare{ \frac12 - \mN_{\rm pf} - \frac12 \brak{\mN_1 - \mN_2 - \mN_3} },\quad 
    \Gamma = \frac{\lambda_x^2}{x_c} \ .
\end{equation}
Recall again that $k$ for $d_v(k)$ runs over $\DL(\mbb{Z}-\frac{1}{2})$, while bosonic $q \in \DL \mbb{Z}_{> 0}$. 

We will soon show that $\Gamma$ is to be identified as the Kondo energy scale along this solvable line. 
The single-fermion spectrum of the bilinear form involving pseudo-fermions $f_v$ and $d_v$ will be directly diagonalized in the next subsection. 
The many-body spectrum is given by filling $\mN_{\rm pf}$ pseudo-fermions to this single-fermion spectrum, so as to preserve the constraint \cref{eq:Npf-def}.

Now we are ready to diagonalize $\widehat{H}_{\rm PK}$. 
We first enumerate the conserved quantum numbers \cref{eq:good_quantum_number}, where $\mN_{\rm pf}$ and $\mN_{1,2,3}$ take values in integers, and $b^\dagger_\chi(q) b_\chi(q)$ for $\chi=c,s,vs$ take values in non-negative integers. 
For a given set of conserved quantum numbers [\cref{eq:good_quantum_number}], $\widehat{H}_{\rm PK}$ [consisting of \cref{eq:wdhH0,eq:wdhHx}] is a free-fermion Hamiltonian of pseudo-fermions $f_v$ and $d_v(k)$, hence its many-body eigenstates are just Fock states of the eigenmodes of the hopping Hamiltonian. 
However, these states live in an {\it extended} Hilbert space indexed by 
\begin{align}
    \{&  \mN_{\rm pf}, ~~ \mN_1, ~~ \mN_2, ~~ \mN_3, ~~ b_{c}^\dagger(q) b_c(q), 
    ~~ b_{s}^\dagger(q) b_s(q), \nonumber\\
    &  ~~ b_{vs}^\dagger(q) b_{vs}(q),
    ~~ f_v^\dagger f_v, 
    ~~ d_v^\dagger(k) d_v(k) \}\ .
\end{align}
Among these states, if and only if \cref{eq:Npf-def} is satisfied, they belong to the physical Hilbert space \cref{eq:index_quantum_number}. 
Therefore, we should discard states violating this constraint in the end. 

\subsection{Finite-size many-body spectrum} 
\label{sec:finite-size-spectrum}

For later convenience, we choose a new Fock state $|\Omega_0\rangle$ as the reference state in the extended Hilbert space. It differs from $|\Omega_0'\rangle$ only in the occupation of $f_v$ fermions, 
\begin{align} \label{eq:vaccum0}
    \inn{ \Omega_0| f_v^\dagger f_v | \Omega_0} &=\theta(\varepsilon_f\le 0)  \ , \\\nonumber
    \inn{ \Omega_0| d_v^\dagger(k) d_v(k) | \Omega_0} &= \theta(k<0)\ .
\end{align}
The advantage of $|\Omega_0\rangle$ over $|\Omega_0'\rangle$ is that, $|\Omega_0\rangle$ does not contain any ``hole'' below Fermi surface, while $|\Omega_0'\rangle$ does if $\varepsilon_f < 0$. 
Therefore, for given good quantum numbers in \cref{eq:good_quantum_number}, $|\Omega_0\rangle$ is the ground state in the extended Hilbert space at $\lambda_x = 0$, and it adiabatically changes to the ground state upon turning on $\lambda_x$. 

With the pseudo-fermion normal-ordering respecting $\ket{\Omega_0}$, 
\begin{align}   \label{eq:wdhH0_1}
    & \widehat{H}_{0} = \frac{2\pi}{L} \Bigg( \sum_{i=1}^3 \brak{ \frac{(1- \Pbc)}{2} \mN_i + \frac{\mN_i^2}2}  - \mN_{\rm pf}  \\\nonumber
    &\;  - \frac14\brak{\mN_1 - \mN_2 - \mN_3} 
    + \mN_{\rm pf} \brak{ \frac32 \mN_{\rm pf} + \mN_1 - \mN_2 -\mN_3}
    \Bigg)  \\ \nonumber
&\; + \sum_{\chi=c,s,vs} \sum_{q} q~ b_{\chi}^\dagger(q) b_\chi(q) + \theta(\varepsilon_f\le0) \varepsilon_f \\\nonumber
&\; + \varepsilon_f : f_v^\dagger f_v : + \sum_{k} k~ : d_v^\dagger(k) d_v(k) : \ ,
\end{align}
while $\widehat{H}_x$ remains unchanged. 
Formally written, the normal-ordering with respect to $|\Omega_0\rangle$ has subtracted a ``vacuum'' energy of 
\begin{equation} \label{eq:vaccum-energy0-HPK}
    E[\Omega_0] = \theta(\varepsilon_f\le 0)\cdot \varepsilon_f + \sum_{n=-\infty}^0 \frac{2\pi}{L} \pare{ n- \frac12 }\ . 
\end{equation}

Next, we diagonalize the bilinear hopping Hamiltonian in the last row of $\widehat{H}_{0}$ [\cref{eq:wdhH0_1}] plus $\widehat{H}_x$ [\cref{eq:wdhHx}]. 
Suppose the eigen mode is given by $d_n^\dagger = u_n f_v^\dagger + \sum_{k} v_{k,n} d_v^\dagger (k)$. The hopping Hamiltonian in the last row of \cref{eq:wdhH0_1} is diagonalized to  
\begin{equation}
    \sum_{n} \epsilon_n ~ \substack{\star \\ \star}~ d_n^\dagger d_n ~\substack{\star \\ \star}~ + \delta E[\Omega]  \ ,
\end{equation}
where $~\substack{\star \\ \star}~ \cdots ~\substack{\star \\ \star}~$ is the normal ordering with respect to the ground state $\ket{\Omega}$ in the presence of $\lambda_x$, which also contains no hole, and has the same pseudo-fermion number as $|\Omega_0\rangle$. 
$\delta E[\Omega]$ is the change of vacuum energy from $\ket{\Omega_0}$ to $\ket{\Omega}$. 

The equation of motion for each single-fermion eigen mode is given by 
\begin{align}
& f_v^\dagger \pare{ \varepsilon_f \cdot u_n 
    + \sqrt{\frac{2\pi \Gamma}{L}} \sum_k v_{k,n} } \\\nonumber
& + \sum_k d_v^\dagger(k) \pare{  k \cdot v_{k,n} +  \sqrt{\frac{2\pi \Gamma}{L}} \cdot u_n } \\\nonumber
= & \epsilon_n \Big[ f_v^\dagger \cdot u_n 
 + \sum_{k} d_v^\dagger(k) \cdot \epsilon_n v_{k,n} \Big]\ ,
\end{align}
where the left-hand side is the commutator of the hopping Hamiltonian with $d_n^\dagger$, and the right-hand side is the commutator $[\epsilon_n ~ \substack{\star \\ \star}~ d_n^\dagger d_n~ \substack{\star \\ \star}~ , d_n^\dagger]$. 
The eigenenergy $\epsilon_n$ is solved by
\begin{align}
& \epsilon_n -\varepsilon_f = 
    \frac{2\pi \Gamma}{L} \sum_{k} \frac{1}{\varepsilon_n - {k}} 
    = - \pi \Gamma \cdot \tan \frac{L\epsilon_n}{2} \ ,
\end{align}
where $k\in \frac{2\pi}{L}(\mathbb{Z}-\frac12)$, and we have used the Mittag-Leffler expansion, $\tan z= \sum_{j=-\infty}^{+\infty} \frac{-1}{z - \pi (j - \frac{1}{2})}$, as $\tan z$ has poles of residue $-1$ at $z = \pi (n-\frac{1}{2})$. 
For $|\epsilon_n - \varepsilon_f|\ll \Gamma$, there must be $\epsilon_n \in \frac{2\pi}{L} \mathbb{Z}$. 
For $|\epsilon_n - \varepsilon_f|\gg \Gamma$, there must be $\epsilon_n \in \frac{2\pi}{L} (\mathbb{Z} + \frac12)$.

\begin{figure}[t]
    \centering
    \includegraphics[width=0.95\linewidth]{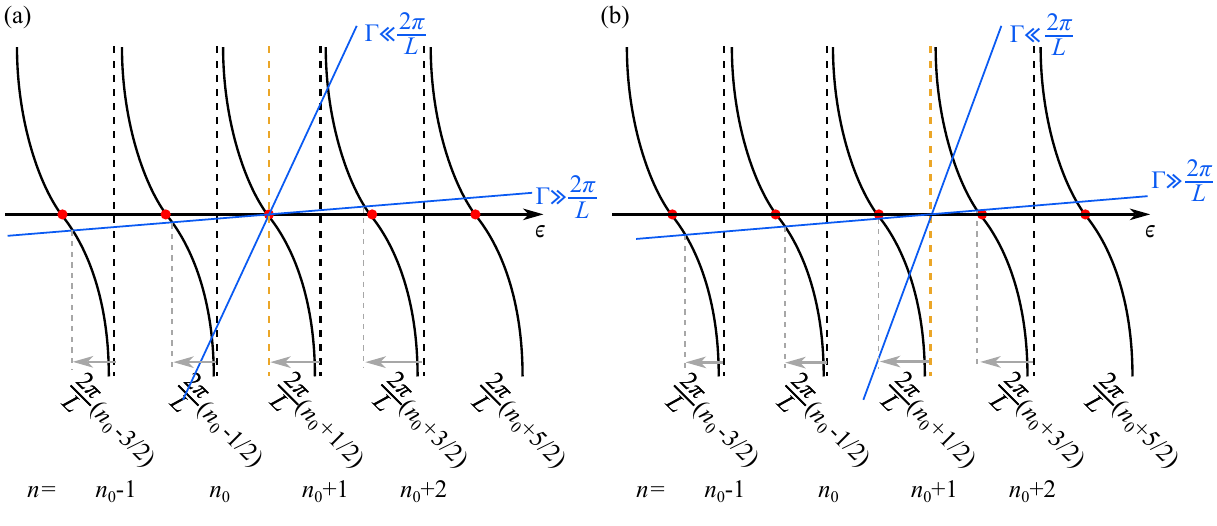}
    \caption{Phase shift of energy levels of the pair-Kondo model at the Fermi liquid fixed point $\rho^\star=\frac14$. 
    Black vertical dashed lines indicate $\epsilon=\frac{2\pi}{L} \pare{n-\frac12}$ ($n\in \mathbb{Z}$), and yellow vertical dashed lines represent $\varepsilon_f$, which is $\frac{2\pi}{L} n_0$ and $\frac{2\pi}{L}\pare{n_0 + \frac12}$ in (a) and (b), respectively. 
    Red dots indicate $\epsilon=\frac{2\pi}{L}n$. 
    The blue lines are $\frac{\epsilon-\varepsilon_f}{\Gamma}$. 
    The black curves are the function $-\tan\pare{\frac{L}{2} \epsilon }$, their crossings with the blue lines give the eigenvalues $\epsilon_n = \frac{2\pi}{L}\pare{ n - \frac12 + \delta_n }$.
    $\frac{2\pi}{L} \delta_n$ is shown by the gray arrows.
    }
    \label{fig:phase-shift}
\end{figure}

We denote the levels as 
\begin{equation} \label{eq:delta-n-def}
\epsilon_n = \frac{2\pi}{L} \left( n - \frac{1}{2} + \delta_n \right) 
\end{equation}
with 
\begin{align} \label{eq:delta-n-def-explicit}
    \delta_n = \begin{cases}
                \frac{1}{\pi} \arctan \frac{\pi \Gamma}{\epsilon_n - \varepsilon_f - 0^+}, \qquad \epsilon_n\le \varepsilon_f\\
                -1 + \frac{1}{\pi} \arctan \frac{\pi \Gamma}{\epsilon_n - \varepsilon_f},\qquad \epsilon_n> \varepsilon_f
    \end{cases}
\end{align}
being the phase shift, as shown in \cref{fig:phase-shift}.
In the $\Gamma \gg \frac{2\pi}{L}$ limit, as $n$ increases from $-\infty$ to $0$, $\delta_n$ changes from $0$ to $-\tfrac{1}{2}$. As $n$ further increases to $\infty$, $\delta_n$ decreases to $-1$.

As shown in \cref{fig:phase-shift}, no level crossing for the single-fermion levels happens as $\Gamma$ (or $\lambda_x$) is turned on. 
Thus, $\ket{\Omega}$ is adiabatically connected to $\ket{\Omega_0}$ [\cref{eq:vaccum0}]. 
For $\varepsilon_f \le 0$, the highest occupied level in $\ket{\Omega}$ is $\epsilon_1$;
for $\varepsilon_f > 0$, the highest occupied level in $\ket{\Omega}$ is $\epsilon_0$. 
Thus, the vacuum energy subtracted in normal-ordering with respect to $|\Omega\rangle$ reads $E[\Omega]=\sum_{n=-\infty}^1 \epsilon_n$ if $\varepsilon_f\le 0$, and reads $\sum_{n=-\infty}^0 \epsilon_n$ if $\varepsilon_f > 0$. 

We focus on the limit $\Gamma \gg \frac{2\pi}{L}$. 
Since $\epsilon_1=0$ in this limit, we can always formally express $E[\Omega]$ as  
\begin{equation} \label{eq:E-Omega-def}
    E[\Omega]  = \sum_{n=-\infty}^0 \frac{2\pi}{L} \pare{ n - \frac12 + \delta_n }  \ ,
\end{equation}
regardless of the sign of $\varepsilon_f$, where we have substituted $\epsilon_n$ using \cref{eq:delta-n-def}. 
Comparing it to \cref{eq:vaccum-energy0-HPK}, we find the variation of vacuum energy as 
\begin{align}
\delta E[\Omega] &= E[\Omega]  \!-\! E[\Omega_0]
= -\theta(\varepsilon_f\le 0) \cdot \varepsilon_f
+ \! \sum_{n=-N}^{0} \frac{2\pi}{L} \delta_n \ ,
\end{align}
where $N \approx \frac{L}{2\pi} D$ and $D$ is a high-energy cutoff representing the bandwidth of bath electrons. 
We evaluate the second term $\delta E'[\Omega] = \sum_{n=-N}^{0} \frac{2\pi}{L} \delta_n$ in \cref{app:dE}. 
The result is
\begin{equation}
    \delta E[\Omega] =
    -\theta(\varepsilon_f\le 0) \cdot \varepsilon_f + 
    \frac12 \varepsilon_f - \Gamma - \Gamma \ln \pare{ \frac{D}{\Gamma}} + \cdots\ ,
\end{equation}
where $-\Gamma-\Gamma \ln\frac{D}{\Gamma}$ represents the binding energy; these are energy constants that do not enter the many-body spectrum. 
The omitted $\cdots$ include $\mathcal{O}(L^{-1})$ terms that are independent of $\varepsilon_f$, $\mathcal{O}(D^{-1})$ terms, and $\mathcal{O}(L^{-2})$ terms. 

Up to constant energy terms that do not affect excitation energies, 
\begin{align}  \label{eq:HPK-refermionized-simpe}
& \widehat{H}_{\rm PK} = \frac{2\pi}{L} \Bigg( \sum_{i=1}^3 \brak{ \frac{(1- \Pbc)}{2} \mN_i + \frac{\mN_i^2}2}   - \frac32 \mN_{\rm pf}  \\\nonumber
& \;  - \frac12\brak{\mN_1 - \mN_2 - \mN_3} 
    + \mN_{\rm pf} \brak{ \frac32 \mN_{\rm pf} + \mN_1 - \mN_2 -\mN_3}
    \bigg) \\\nonumber
& \; + \sum_{\chi=c,s,vs} \sum_{q>0} q~ b_{\chi}^\dagger(q) b_\chi(q) \\\nonumber
& \; + \sum_{n \in \mathbb{Z}} \frac{2\pi}{L}\pare{ n - \frac12 + \delta_n } 
    ~\substack{\star \\ \star}~ d_n^\dagger d_n ~\substack{\star \\ \star}~ \ . 
\end{align}
Note that the first term in $\delta E [\Omega]$ cancels the $\theta(\varepsilon_f\le 0)\varepsilon_f$ term in the third row of \cref{eq:wdhH0_1}, and the second term in $\delta E[\Omega]$ also changes the $\mN_{\rm 1,2,3,pf}$-dependent terms in the first two rows of \cref{eq:wdhH0_1}. 
As we have explained in the end of \cref{sec:refermFL-referm}, $\widehat{H}_{\rm PK}$ is defined in an extended Hilbert space, and we must impose the constraint $\mN_{\rm pf} = f_v^\dagger f_v + \sum_{k} :d_v^\dagger d_v:$ to select out the physical states. 
In normal-ordering with respect to $\ket{\Omega_0}$ or $\ket{\Omega}$, this constraint reads 
\begin{align}    \label{eq:constraint}
    \mN_{\rm pf} &= \theta(\varepsilon_f \le 0) + :f_v^\dagger f_v: + \sum_{k} :d_v^\dagger d_v: \\\nonumber
    &= \theta(\varepsilon_f \le 0) + \sum_n ~\substack{\star \\ \star}~ d_n^\dagger d_n ~\substack{\star \\ \star}~ \ ,
\end{align}
respectively, because $|\Omega_0\rangle$ is adiabatically connected to $|\Omega\rangle$. 
It is also worth emphasizing that the reference state $\ket{\Omega}$ used in the normal ordering occupies levels with $n \leq 1$ when $\varepsilon_f \leq 0$, and levels with $n \leq 0$ when $\varepsilon_f > 0$, as explained above \cref{eq:E-Omega-def}. 

We now derive the lowest many-body states for given $\mN_{\rm pf,1,2,3}$ in the $\Gamma \gg \frac{2\pi}{L}$ limit, for the PK Hamiltonian \cref{eq:HPK-refermionized-simpe}.
First, these states have $b_{\chi}^\dagger (q) b_\chi(q)=0$ for $\chi=c,s,vs$. 
Besides, to be physical, they must further obey \cref{eq:constraint}. 
Therefore, they should occupy the lowest $\mN_{\rm pf} -\theta (\varepsilon_f\le 0)$ pseudo-fermion levels, in addition to those already occupied in $\ket{\Omega}$. 
No particle-hole excitations in the pseudo-fermions should be generated in order to save energy. 
To be more specific, if $\varepsilon_f\le 0$, then $\epsilon_{n\le 1}$ are already occupied in $\ket{\Omega}$, and the low-energy physical state must further occupy $\epsilon_2, \epsilon_3 \cdots \epsilon_{\mN_{\rm pf}}$, which costs an energy $\tfrac{2\pi}{L} \frac{\mN_{\rm pf}(\mN_{\rm pf}-1)}{2}$. 
If $\varepsilon_f>0 $, then $\epsilon_{n\le 0}$ are occupied in $\ket{\Omega}$, and the low-energy physical state must further occupy $\epsilon_1, \epsilon_2 \cdots \epsilon_{\mN_{\rm pf}}$, which costs the same amount of energy $\tfrac{2\pi}{L} \frac{\mN_{\rm pf}(\mN_{\rm pf}-1)}{2}$, as $\epsilon_1=0$. 
Substituting the last line of \cref{eq:HPK-refermionized-simpe} with this expression, we conclude the lowest many-body energy for given $\mN_{\rm pf,1,2,3}$ is 
\begin{align}  \label{eq:E_Npf123}
E[\mN_{\rm pf,1,2,3}] &=  \frac{2\pi}{L} \Bigg( \sum_{i=1}^3 \brak{ \frac{(1- \Pbc)}{2} \mN_i + \frac{\mN_i^2}2}  \\\nonumber
    &- 2 \mN_{\rm pf} - \frac12\brak{\mN_1 - \mN_2 - \mN_3} \\\nonumber
    &+ \mN_{\rm pf} \brak{ 2 \mN_{\rm pf} + \mN_1 - \mN_2 -\mN_3}
    \Bigg) \ . 
\end{align}
The lowest state is non-degenerate in each sector of given $\mN_{\rm pf,1,2,3}$, because the boson excitations ($\chi=c,s,vs$) and particle-hole excitations of the pseudo-fermions cost at least an energy of $\frac{2\pi}{L}$. 

According to \cref{eq:mN-to-Nflavors}, $\mN_{\rm pf}$ ($=\mN_v + f_v^\dagger f_v$) and $\mN_{1,2,3}$ fully  determine the quantum numbers $N_{v}^{\rm (tot)}$, $N_{c,s,vs,}$:
\begin{equation}
\begin{bmatrix}
    N_c \\ N_v^{\rm (tot)} + 1 \\ N_s \\ N_{vs}
\end{bmatrix}
= \frac12 \left[
\begin{array}{rrrr}
0 & 1 & 1 & 1 \\
4 & 1 & -1 & -1 \\
0 & -1 & 1 & -1 \\
0 & -1 & -1 & 1 \\
\end{array}
\right]
\begin{bmatrix}
    \mN_{\rm pf} \\ \mN_1 \\ \mN_2 \\ \mN_3
\end{bmatrix} \ ,
\end{equation}
where the second row has exploited \cref{eq:mN_to_Nvtot}. 
Expressing $\mN_{\rm pf,1,2,3}$ in terms of $N_{c,s,vs}$ and $N_v^{\rm (tot)}$, \cref{eq:E_Npf123} simplifies into
\begin{align}
    E[N_{c,s,vs}, N_{v}^{\rm (tot)}]
&= \frac{2\pi}{L} \bigg( (1-\Pbc) N_c - \frac12 \\\nonumber
& \quad + \frac12 N_{v}^{\rm (tot)2}
     + \frac12 \sum_{\chi=c,s,vs} N_\chi^2 \bigg)\ . 
\end{align}
For $-1< \Pbc < 1$, the ground state is given by $N_c = -1$, $N_{s}=N_{vs}=N_v^{\rm (tot)}=0$ ($\mN_{\rm pf}=\mN_1=1$, $\mN_2=\mN_3=0$), and the ground state energy is $E =  \frac{2\pi}{L}(\Pbc-1)$. 
We introduce $\Delta N_c = N_c + 1$, $\Delta N_{s} = N_s$, $\Delta N_{vs} = N_{vs}$, and $\Delta N_v = N_v^{\rm (tot)}$ as the deviation of quantum numbers from the ground state. 
Then the minimal excitation energy in each quantum number sector (with respect to the ground state) is given by 
\begin{equation}
    \Delta E = \frac{2\pi}{L}
    \pare{ (-\Pbc) \Delta N_c +  \frac12  \sum_{\chi=c,v,s,vs} \Delta N_\chi^2  }\ . 
\end{equation}
Importantly, $\Delta N_{c,v,s,vs}$ satisfy the free-gluing condition as $N_{c,v,s,vs}$, since they also correspond to integer $\mN_{\rm pf,1,2,3}$ in the same way as $N_{c,v,s,vs}$ [\cref{eq:mN-to-Nflavors}]: 
\begin{equation}
\begin{bmatrix}
    \Delta N_c \\ \Delta N_v \\ \Delta N_s \\ \Delta N_{vs}
\end{bmatrix}
= \frac12 \left[
\begin{array}{rrrr}
0 & 1 & 1 & 1 \\
4 & 1 & -1 & -1 \\
0 & -1 & 1 & -1 \\
0 & -1 & -1 & 1 \\
\end{array}
\right]
\begin{bmatrix}
    \mN_{\rm pf} \\ \mN_1 \\ \mN_2+1 \\ \mN_3+1
\end{bmatrix} \ .
\end{equation}
Comparing $\Delta E$ to $H_{\rm PK}$ in the initial bosonized form [$H_0+H_z$, given by \cref{eq:H0,eq:Hz_boson}], we find it equivalent to the quantum number part of $H_{\rm PK}$ with a new boundary condition $\td{P}_{\rm bc} = \Pbc +1$, suggesting that the Kondo screening only introduces a $\pi$ phase shift.

We now prove that {\it the many-body spectrum below the energy scale of $\Gamma$ is equivalent to a free-fermion system with a $\pi$ phase shift}, when all the particle-hole excitations are also included. 
For $\chi=c,s,vs$, there are $\sum_{\chi=c,s,vs} \sum_{q>0} q~ b_\chi^\dagger(q) b_\chi(q)$. 
Meanwhile, particle-hole excitations in the pseudo-fermion part $\sum_{n} \frac{2\pi}{L}(n-1) ~\substack{\star\\ \star}~ d_n^\dagger d_n ~\substack{\star\\ \star}~ $ with a fixed $\mN_{\rm pf}$ can be equivalently written as $\sum_{q>0} q \td{b}_{v}^\dagger(q) \td{b}_{v}(q): $ according to the bosonization dictionary [\cref{eq:boson_H0}], where $\td{b}_v(q)$ should be distinguished from $b_v(q)$ that stands for the particle-hole excitations of $d_v(k)$ pseudo-fermions. 
Thus, the effective Hamiltonian 
\begin{align}   \label{eq:Heff_belowGm_boson}
    H_{\rm eff} &= \frac{2\pi}{L}
    \pare{ (1 - \td{P}_{\rm bc}) \Delta N_c + \frac12  \sum_{\chi=c,v,s,vs} \Delta N_\chi^2  } \\\nonumber
    &+ \sum_{\chi=c,s,vs} \sum_{q>0} q ~ b_\chi^\dagger(q) b_\chi(q) + \sum_{q>0} q \td{b}_v^\dagger(q) \td{b}_v(q)
\end{align}
generates all many-body levels below the energy scale of $\Gamma$. 
It has the same form as $H_{\rm PK}$ [$H_0+H_z$, given by \cref{eq:H0,eq:Hz_boson}] with $\lambda_x=\rho_z=0$, but with a phase-shifted boundary condition $\td{P}_{\rm bc} = \Pbc +1$. 
This result indicates that the doublet has been completely absorbed into the bath via PK resonance. 
Reversing \cref{sec:Himp-bath}, one can write \cref{eq:Heff_belowGm_boson} as 
\begin{equation}
    H_{\rm eff} = \sum_{l s} \sum_{k \in \frac{2\pi}{L} \pare{ \mathbb{Z} - \frac{\td{P}_{\rm bc}}2 }} k ~: c_{ls}^\dagger (k) c_{ls}(k) :\ ,
\end{equation}
where $c_{ls}(k)$ are fermion operators constructed from $\Delta N_{c,s,vs,v}$, and $b_{c,s,vs}(q)$ and $\td{b}_{c,s,vs}(q)$, which have absorbed the doublet degree of freedom into the bath.

\subsection{Thermodynamic quantities}   \label{app:exactFL-thermo}

$\widehat{H}_{\rm PK}$ describes a fermionic level $f_v$ subject to a hybridization bath of $\psi_v$. 
In this section, we integrate out the bath fields $\psi_v$, and obtain an effective theory for the fermionic level $f_v$ only (which represents the impurity doublet), from which we can derive the impurity entropy, $S_{\rm imp}$, and the static susceptibility of $\Lambda_z$ operator, $\chi_z(0)$. 

For this purpose, let us derive the impurity free-energy $F_{\rm imp}$. 
In the thermodynamic limit, we can neglect all the $\mathcal{O}(L^{-1})$ terms in $\widehat{H}_{\rm PK}$ [\cref{eq:wdhH0,eq:wdhHx}]. 
We write the partition function of $\widehat{H}_{\rm PK}$ in terms of a path integral over Grassmann variables $f^\dagger_v(\tau), f_v(\tau), d^\dagger_v(k,\tau), d_v(k,\tau)$, with Fourier components over fermionic Matsubara frequencies $\ii\omega$, 
\begin{align}
    f_v(\tau) &= \frac{1}{\sqrt{\beta}} \sum_{\ii\omega} f_v(\ii\omega) e^{-\ii \omega\tau} \ ,
    \\\nonumber
    d_v(k,\tau) &= \frac{1}{\sqrt{\beta}} \sum_{\ii\omega} d_v(k,\ii\omega) e^{-\ii \omega\tau} \ . 
\end{align}
We will not distinguish a Grassmann variable from a fermionic operator in the notation. 
Since the $c,s,vs$ bath fields are decoupled, their partition function can be simply factored out. 

The remaining path integral over the $v$ fields reads
\begin{align}
    &Z = \Tr[e^{-\beta \widehat{H}_{\rm PK}}] = \int_{\mcl{D}[f^\dagger,f,d^\dagger,d]}  e^{-\pare{S_f + S_d + S_{fd}}} \ ,
\end{align}
where
\begin{align}
    S_f &= \sum_{\ii \omega} f^\dagger_v(\ii\omega) (-\ii\omega + h) f_v(\ii \omega) , \\\nonumber
    S_d &= \sum_{\ii\omega, k} d^\dagger_{v}(k,\ii\omega) (-\ii\omega + k) d_{v}(k,\ii\omega)  , \\\nonumber
    S_{fd} &= \sqrt{\frac{2\pi \Gamma}{L}} \sum_{\ii\omega, k} \Big( d^\dagger_v(k,\ii\omega) f_{v}(\ii\omega) + f^\dagger_v(\ii\omega) d_v(k,\ii\omega) \Big) \ .
\end{align}
Here, we also introduce a Zeeman field $\frac12 h \Lambda_z$, which in the pseudo-fermion language corresponds to an on-site energy of $f_v$. 

We carry out the Gaussian integrals over $\int_{\mcl{D}[d^\dagger,d]}$ and obtain an effective action for $f_v$:
\begin{equation}  \label{eq:Simp}
    S_{\rm imp} = \sum_{\ii \omega} f^\dagger_v(\ii\omega) (-\ii\omega + h + \Delta(\ii \omega)) f_v(\ii \omega)\ ,
\end{equation}
where 
\begin{align}
    \Delta(\ii\omega) &= \DL \Gamma \sum_k \frac{1}{\ii\omega - k} = \Gamma \int\mrm{d}k ~ \frac{1}{\ii\omega - k} \\\nonumber
    &= - \ii \cdot  \sgn(\omega) \cdot \pi \Gamma
\end{align}
is the hybridization function. 
The partition function for the impurity is hence 
\begin{equation}
    Z_{\rm imp} = \int_{\mcl{D}[f^\dagger,f]} e^{-S_{\rm imp}} = \prod_{\ii\omega} (-\ii \omega + h - \ii \pi\Gamma \cdot \sgn(\omega))\ . 
\end{equation}

The free-energy of impurity ($F_{\rm imp} = -\frac{1}{\beta} \ln Z_{\rm imp} $) is hence given by
\begin{align}   \label{eq:Fimp_0}
    F_{\rm imp} &= -\frac{1}{\beta} \sum_{\ii\omega} \ln\left[-\ii\omega + h - \ii \pi \Gamma \cdot \sgn(\omega) \right] e^{\ii \omega 0^+} \\\nonumber
    &= - \int_{-\infty}^{\infty} \frac{\mrm{d}\omega}{2\pi \ii} f(\omega) \ln\left[-\omega + h + \ii \pi\Gamma \right] e^{\omega 0^+} \\\nonumber
    &\qquad + \int_{-\infty}^{\infty} \frac{\dd \omega}{2\pi \ii} f(\omega) \ln\left[-\omega + h - \ii \pi\Gamma \right]e^{\omega 0^+}  \ ,
\end{align}
where we have added the factor $e^{\ii \omega 0^+}$ for convergence. 
In the second line, the 1st (2nd) term is deformed from the contour in the lower (upper) half of the complex plane that generates the Matsubara summation over negative (positive) $\ii\omega$, respectively. $f(\omega) = 1/(e^{\frac{\omega}{T}} + 1)$ is the Fermi-Dirac function. 
Also, we need to specify the branches of the $\ln$ and $\mrm{arccot}$ functions. 
To do this, we consider the limit $\Gamma \to 0^+$, where
\begin{align}   \label{eq:Fimp_Gm_0}
    F_{\rm imp} &=  \int_{-\infty}^{\infty} \frac{\dd \omega}{2\pi \ii} f(\omega) \ln\left[ \frac{\omega - h + \ii 0^+}{\omega - h - \ii 0^+} \right] \\\nonumber
    &= -\int_{h}^{\infty} \dd\omega ~ f(\omega) = -T \ln [1 + e^{-h/T}] \ . 
\end{align}
In the second equality, we have specified $\ln (\omega-h+\ii0^+) = -\ii 2\pi \cdot \theta(\omega-h) - \ii \pi \theta(h-\omega)$, $\ln (\omega-h-\ii0^+) = - \ii \pi \theta(h - \omega)$. 
In the third equality, we have exploited 
\begin{align}   \label{eq:int_FD}
    f(\omega) = - T \partial_\omega [\ln(1+e^{- \omega/T})]\ . 
\end{align}
With the current choice of branch cut, \cref{eq:Fimp_Gm_0} recovers the partition function of a standard two-level system with energies $0$ and $h$, where $Z = e^{-F_{\rm imp}/T} = 1 + e^{-h/T}$.  

\subsubsection{Impurity entropy \texorpdfstring{$S_{\rm imp}$}{Simp}}

At $\Gamma=0$, with \cref{eq:Fimp_Gm_0}, it can then be computed that
\begin{align}
    S_{\rm imp} &= -\frac{\partial F_{\rm imp}}{\partial T} = \ln[1 + e^{-h/T}] + \frac{h}{T} \frac{1}{e^{h/T} + 1} \\\nonumber
    &= \ln[e^{h/2T} + e^{-h/2T}] - \frac{h}{2T} \tanh\frac{h}{2T} \ . 
\end{align}
If $\frac{h}{T} \to \infty$, $S_{\rm imp}=0$; if $\frac{h}{T} \to 0$, $S_{\rm imp} = \ln 2$. 
At $\frac{h}{T} \gtrsim 1$, $S_{\rm imp}$ vanishes exponentially. 

At finite $\Gamma$, based on \cref{eq:int_FD}, integral-by-part for \cref{eq:Fimp_0} gives 
\begin{align}   \label{eq:Fimp_Gm_finite}
    &F_{\rm imp} = T \int_{-\infty}^\infty \frac{\dd\omega}{2\pi \ii } \ln\pare{ 1 + e^{- \frac{\omega}{T}} } \\\nonumber
    &\qquad \qquad \times \pare{ \frac{-1}{\omega - h - \ii \pi\Gamma} + \frac1{\omega - h + \ii \pi\Gamma} } e^{\omega 0^+}  \nonumber\\
=& - T \int_{-\infty}^{\infty} \dd \omega \ 
    \ln\pare{ 1 + e^{- \frac{\omega}{T}} }  \ \frac{1}{\pi} \frac{\pi\Gamma}{(\omega - h)^2 + (\pi\Gamma)^2} e^{\omega 0^+}  .  \nonumber
\end{align}
By comparing \cref{eq:Fimp_Gm_finite} and \cref{eq:Fimp_Gm_0}, one finds that, the free-energy at finite $\Gamma$ and fixed $h$ is equivalent to an ``average'' over an ensemble of two-level systems, where  $h$ follows a Lorentz distribution with a spread $\pi \Gamma$. 
Accordingly, $S_{\rm imp}$ will follow the same ``ensemble average''. 
Therefore, at $h=0$, if $\pi\Gamma \gg T$, only a small fraction ($\sim\frac{T}{\pi \Gamma}$) of systems in the ensemble has $S_{\rm imp}=\ln 2$, while the remaining systems have $S_{\rm imp}=0$, hence the ensemble average will asymptote to $0$. 
On the other hand, if $\pi\Gamma \ll T$, almost all systems in the ensemble will have $S_{\rm imp}=\ln 2$, hence the ensemble average will asymptote to $\ln 2$. 

\subsubsection{Static longitudinal susceptibility \texorpdfstring{$\chi_z(0)$}{chiz}}

We also start with the $\Gamma=0$ case. 
The magnetization can be computed as $M = \frac12\inn{\Lambda_z} = \inn{f_v^\dagger f_v} -\frac12 = \frac{\partial F_{\rm imp}}{\partial h} - \frac12 = \frac{1}{e^{h/T}+1} - \frac12$. 
At $h=0$, we get a background value $M=0$. 
The longitudinal susceptibility is given by 
\begin{align}
    \chi_z(0) = -\frac{\partial M}{\partial h} = \frac{1}{T} \frac{1}{e^{h/T} + e^{-h/T} + 2} \ . 
\end{align}
Without $h$ (and since without $\Gamma$), it scales as $1/T$, the Curie's law. 
When $h$ is larger than $T$, $\chi_z(0)$ gets frozen to 0 exponentially. 

At finite $\Gamma$, we have
\begin{align}
    &\chi_z(0) = -\frac{\partial^2 F}{\partial h^2}  \\\nonumber
    &= - T \int_{-\infty}^{\infty} \dd \omega \ 
    \ln\pare{ 1 + e^{- \beta \omega} }  (-\partial_h^2) \frac{1}{\pi} \frac{\pi\Gamma}{(\omega - h)^2 + (\pi\Gamma)^2} \\\nonumber
    &= \int_{-\infty}^{\infty} \dd \omega  \frac{1}{T} \frac{1}{e^{\omega/T}+e^{-\omega/T}+2}  \frac{1}{\pi} \frac{\pi\Gamma}{(\omega - h)^2 + (\pi\Gamma)^2} \ ,
\end{align}
which is given by an ``ensemble average'' of susceptibilities of two-level systems likewise.  
At $h=0$, if $\pi\Gamma \gg T$, then only a fraction ($\sim\frac{T}{\pi \Gamma}$) of systems in the ensemble exhibits unfrozen susceptibility $\sim \frac{1}{T}$, hence the total susceptibility averages to $\chi^{z}(0) \sim \frac{1}{\pi\Gamma}$, which is the typical behavior of FL at low temperature. 
If $T \gg \pi \Gamma$, on the other hand, the whole ensemble exhibits Curies' law, hence $\chi_z(0) \sim \frac{1}{T}$.

\subsection{Impurity correlation functions}

We also compute correlation functions (dynamic susceptibilities) in thermodynamic limit. 
Finite-size terms of $\mathcal{O}(L^{-1})$ will be omitted. 
It is useful to work out the correlation functions of the pseudo-fermions first, as physical correlation functions will eventually be expressed in terms of them. 
From \cref{eq:Simp} (see \cref{app:exactFL-thermo}), setting $h=0$, it is direct to read off the Green's function for $f_v$ fermions as
\begin{align}   \label{eq:Gimp}
    \mcl{G}_f(\ii \omega) &= \frac{1}{\ii \omega + \ii \cdot \sgn(\omega) \cdot \pi \Gamma} \ . 
\end{align}
We use $\mcl{G}_f$ for the Green's function of the pseudo-fermion $f_v$, in order to distinguish from the Green's function of the physical $f$ electron. 

At zero temperature, Fourier transforming \cref{eq:Gimp} to the imaginary time axis $\tau$, we obtain 
\begin{align} \nonumber
    \mcl{G}_f(\tau) &= \int_{-\infty}^{\infty} \frac{\mrm{d}\omega}{2\pi} ~ \mcl{G}_f(\ii\omega) ~ e^{- \ii \omega \tau} = - \int_{0}^{\infty} \frac{\mrm{d}\omega}{\pi} ~ \frac{\sin(\omega \tau)}{\omega + \pi \Gamma} \\ \nonumber
    &= - \left( \frac{1}{2} - \frac{\Si(\pi\Gamma\tau)}{\pi} \right) \cos(\pi\Gamma\tau) - \frac{\Ci(\pi\Gamma \tau)}{\pi} \sin(\pi \Gamma \tau) \\
    &= -\frac{1}{\Gamma \tau} + \mcl{O}\left(\frac{1}{(\Gamma\tau)^2} \right)  \mrm{as} ~~ \Gamma \tau \gg 1 \ . 
\end{align}
Here we have exploited
{\small
\begin{align}
    \Si(z) &= \int_0^{z} \mrm{d}t \frac{\sin(t)}{t} \overset{z\to\infty}{=} \frac{\pi}{2} - \frac{\cos(z)}{z} + \mcl{O} \left(\frac{1}{z^2}\right) \ ,\\\nonumber
    \Ci(z) &= -\int_{z}^{\infty}\mrm{d}t \frac{\cos(t)}{t}  \overset{z\to\infty}{=} \frac{\sin(z)}{z} + \mcl{O}\left(\frac{1}{z^2}\right) \ .
\end{align}}

Now we compute the longitudinal correlation function
\begin{align}
    &\chi_z(\tau) = - \left\langle T_\tau \   e^{\tau H} \cdot \Lambda_z \cdot e^{-\tau H} \cdot \Lambda_z \right\rangle_0 \\\nonumber
    &= -\left\langle T_\tau \  e^{\tau \widehat{H}_{\rm PK}} \!\cdot\! (2f^\dagger_v f_v-1) \cdot e^{-\tau \widehat{H}_{\rm PK}} \!\cdot\! (2f^\dagger_v f_v-1)  \right\rangle_{\widehat{0}} . 
\end{align}
Here, states $\ket{G}$ in the $\widehat{0}$ gauge are transformed to $0$ gauge according to $\ket{G} = U^\dagger U^\dagger_2 |\widehat{G}\rangle$. 
Recall that $U_2 = e^{\ii \frac{\pi}{4} N_v \Lambda_z}$, and $U = e^{\frac{\ii }{2} \Lambda_z \phi_v(0)}$ at $\rho_z^\star=\frac{1}{4}$. 
In terms of the pseudo-fermions, $2f^\dagger_v f_v-1$ measures the density fluctuation of $f_v$ fermions. 
As they are non-interacting, Wick's theorem applies:  
\begin{align}
    \chi_z(\tau) &= - \mcl{G}_f(\tau) \mcl{G}_f(-\tau) \overset{\Gamma\tau\gg1}{=} -\frac{1}{(\Gamma \tau)^2} + \mcl{O}\left( \frac{1}{(\Gamma\tau)^3} \right) . 
\end{align}
By comparing to calculations in \cref{app:exactAD-sus}, we conclude that the scaling in the imaginary time domain as $\frac{1}{\tau^2}$ will imply that the dynamic susceptibility to scale as  $\Im \chi^R_z(\omega) \sim -\frac{\omega}{\Gamma^2}$, which is the Fermi liquid behavior.

Let us also examine the transverse correlation function, 
\begin{align}
    & - \chi_x(\tau) = \left\langle T_\tau ~ e^{\tau H} \cdot \Lambda_+ \cdot e^{-\tau H} \cdot \Lambda_- \right\rangle_{0} \\\nonumber
    &= \left\langle T_\tau ~ e^{\tau \ovl{H}} \cdot e^{\ii \phi_v(0)} \Lambda_+ \cdot e^{-\tau \ovl{H}} \cdot \Lambda_- e^{-\ii \phi_v(0)} \right\rangle_{\ovl{0}} \\\nonumber
    &= \left\langle T_\tau ~ e^{\tau \ovl{H}} \cdot F_v^\dagger e^{\ii \phi_v(0)} \Lambda_+ \cdot e^{-\tau \ovl{H}} \cdot \Lambda_- F_v e^{-\ii\phi_v(0)} \right\rangle_{\ovl{0}} \\\nonumber
    &= \Big\langle T_\tau ~ e^{\tau \widehat{H}_{\rm PK}}  \Big(F_v^\dagger e^{\ii \frac{\pi}{2} \Lambda_z}  e^{\ii \phi_v(0)}  \Lambda_+ e^{\ii \frac{\pi}{2} N_v} \Big)  e^{-\tau \widehat{H}_{\rm PK}}  \Big( h.c. \Big)  \Big\rangle_{\wht{0}} \\\nonumber
    &= (2\pi x_c) \left\langle T_\tau ~~ e^{\tau \widehat{H}_{\rm PK}} \cdot \psi^\dagger_v(0)  f_v^\dagger  \cdot e^{-\tau \widehat{H}_{\rm PK}} \cdot f_v \psi_v(0)  \right\rangle_{\widehat{0}} . 
\end{align}
In the third line we inserted an identity $1 = F_v^\dagger F_v$, and used the fact that $F_v$ commutes with $\ovl{H}$. 
In the fourth line, we have carried out the gauge transformation $U_2$, and used that $U_2 F_v U_2^\dagger = F_v e^{- \ii \frac{\pi}{2} \Lambda_z}$, and $U_2 \Lambda_\pm U_2^\dagger = \Lambda_\pm e^{\ii \frac{\pi}{2} N_v}$. 
In the last line, we have exploited the definition \cref{eq:fo}, $f_v^\dagger = e^{\ii \frac{\pi}{2} N_v} \Lambda_+ e^{-\ii \frac{\pi}{2} \Lambda_z} = e^{\ii \frac{\pi}{2} N_v} e^{\ii \frac{\pi}{2} \Lambda_z} \Lambda_+ $, because $e^{\ii \frac{\pi}{2} \Lambda_z} = \ii \cdot \Lambda_z$, which anti-commutes with $\Lambda_\pm$. 
At this stage, we find that the transverse susceptibility is given by a pairing correlation function of the pseudo-fermions. 
We do not explicitly evaluate the expression, but only remark that, at $\omega \ll \mcl{O}(\Gamma)$, the system is nothing but a non-interacting Fermi liquid with $\pi$ phase shift. 
Consequently, the pairing correlation function will also scale as $\frac{1}{\tau^2}$, implying that $\Im \chi_x^R(\omega) \sim -\omega$.

\section{Singlet regime}
\label{sec:singlet}

We now specify to the parameter regime where the singlet $S$ is the impurity ground state, and the doublet is the second lowest state ($U \gg J_{S} \ge J_D > 0$). 
We assume the Kondo resonance has \textit{not} formed at the energy scale of $\omega \lesssim J_S$, so that we can further downfold the Hamiltonian to the subspace of $\PP_{S,D} = \PP_S + \PP_D$. 
We derive the effective Hamiltonian, and investigate the RG flow near the phase transition between the Kondo FL and the LS phase. 

\subsection{Effective Hamiltonian}

The Kondo Hamiltonian in \cref{sec:Himp-Kondo} restricted to the $\PP_{S,D}$ subspace reads
\begin{equation}   \label{eq:H_singlet_regime}
    H^{(S,D)} = H_0 + \PP_{S,D} (H_{\rm imp}^{\rm (K)} + H_{\rm c}^{\rm (K)}) \PP_{S,D}\ , 
\end{equation}
where 
\begin{equation}
\PP_{S,D} \ H_{\rm imp}^{\rm (K)} \ \PP_{S,D} = J\cdot\PP_D + \mrm{const} .,\quad (J=J_S-J_D), 
\end{equation}
\begin{align}
&\PP_{S,D} \ H_{\rm c}^{\rm (K)} \ \PP_{S,D} = (2\pi \lambda_z) \cdot \Theta^{z0} \cdot \psi^\dagger \sigma^z \spin^0 \psi \nonumber\\
&\;\; + (2\pi \frac{\zeta_{x}}{\sqrt{2}}) \cdot \Big( \Theta^{x0} \cdot \psi^\dagger \sigma^x \spin^0 \psi + \Theta^{y0} \cdot \psi^\dagger \sigma^y \spin^0 \psi \Big). 
\end{align}
We have assumed PHS to avoid terms such as $\PP_{S,D} \cdot\psi^\dagger \psi$ for simplicity. 
In principle, we still need to apply a second SW transformation to eliminate the off-diagonal elements $(\PP_S+\PP_D) H_{\rm c}^{\rm (K)} \PP_T$, which can induce quartic terms, including the pair-Kondo term $\lambda_x$ [\cref{eq:Hx}], and terms like $\PP_S \cdot \psi^\dagger\psi^\dagger \psi\psi$, etc. 
However, as the Kondo-LS transition will be governed by the interplay between $\lambda_z$, $\zeta_x$, $J$, while these quartic terms are in general more irrelevant, it suffices to discard them in the RG analysis. 
We also mention that, as the bilinear couplings and the impurity Hamiltonian are already at their most general form, any corrections arising from the second SW transformation can be simply absorbed as a re-definition to $\lambda_z, \zeta_x$, or $J$. 
Therefore, $H^{(S,D)}$ in \cref{eq:H_singlet_regime} is the total effective Hamiltonian that we will consider. 

We abbreviate $|D,L^z\rangle = |L^z\rangle$ for $L^z = 2,\ovl{2}$, and abbreviate $|S\rangle = |0\rangle$, to stress that they carry different $\Uo_v$ charges. 
We introduce the operators
\begin{align}
    \Lambda_z =& |2\rangle \langle 2| - |\ovl{2}\rangle \langle \ovl{2}| = \Theta^{z0}  , \\ \Theta_+ =& \Theta_-^\dagger = |2\rangle \langle 0| + |0\rangle \langle \ovl{2}| = \frac{\Theta^{x0} + \ii \Theta^{y0}}{\sqrt{2}} \ . 
\end{align}
Notice that $\PP_D = \Lambda_z^2$, $\Lambda_+ = \Theta_+^2$. 
With these notations, we re-write the Hamiltonian as 
\begin{align}\label{eq:HSD}
& H^{(S,D)} = \sum_{ls} \sum_k k : d^\dagger_{ls}(k) d_{ls}(k) : 
    + J\cdot \Lambda_z^2 \nonumber\\
&\; + (2\pi \lambda_z) \Lambda_z \sum_{ls} l \cdot \psi^\dagger_{ls}(0) \psi_{ls}(0) \nonumber\\
&\; + (2\pi \zeta_{x}) \left( \Theta_+ \sum_{s} \psi^\dagger_{-s}(0) \psi_{+s}(0) + h.c.  \right)\ .
\end{align}
Following \cref{app:boson}, it can be bosonized as  
\begin{align}
&\sum_{ls}  \int\frac{\mrm{d}x}{4\pi} :(\partial_x \phi_{ls}(x))^2:  
    + J \cdot \Lambda_z^2 + \rho_z \Lambda_z \sum_{ls} l : \partial_x \phi_{ls}(0) :  \nonumber\\
&+ \frac{\zeta_x}{x_c} \left(  \Theta_+  \sum_{s} F^\dagger_{-s} F_{+s} e^{\ii (\phi_{-s}(0) - \phi_{+s}(0))} + h.c.  \right)\ . 
\end{align}
We have ignored the finite-size terms of order $\mcl{O}(\DL)$, since they should not affect the RG flow. 

For later convenience, we introduce the following basis, 
\begin{align}
    \begin{bmatrix}
        \varphi_\uparrow \\
        \varphi_\downarrow \\
    \end{bmatrix} = \begin{bmatrix}
        \frac{1}{\sqrt{2}} & 0 & -\frac{1}{\sqrt{2}} & 0 \\
        0 & \frac{1}{\sqrt{2}} & 0 & -\frac{1}{\sqrt{2}}  \\
    \end{bmatrix} \begin{bmatrix}
        \phi_{+\uparrow} \\
        \phi_{+\downarrow} \\
        \phi_{-\uparrow} \\
        \phi_{-\downarrow} \\
    \end{bmatrix}\ . 
\end{align}
$\varphi_{s}$ for $s = \uparrow,\downarrow$ are simply superpositions of $\phi_{v}$ and $\phi_{vs}$, and represent the valley density fluctuations per spin sector. 
They should not be confused with the spin fluctuation $\phi_s$. 
We also define the corresponding composite Klein factors $F_s = F_{-s}^\dagger F_{+s}$ for $s=\uparrow,\downarrow$. 
Using the fact that $F_{ls}$ anti-commutes with $F_{l's'}^\dagger$ for $ls\not=l's'$, there is $F_{\uparrow} F_{\downarrow} = F_v$. 

Omitting boson fields decoupled from the impurity, we arrive at the effective Hamiltonian
\begin{align}
H =&  \sum_{s} \int\frac{\mrm{d}x}{4\pi}  :\!(\partial_x \varphi_{s})^2 \!: 
+ J \!\cdot\! \Lambda_z^2 + \rho_z \Lambda_z \sqrt{2} \sum_{s} : \!\partial_x \varphi_{s} \!:\Big|_{x=0}  \nonumber \\
& + \frac{\zeta_x}{x_c} \cdot \left(  \Theta_+ \sum_{s} F_s \cdot e^{-\ii \sqrt{2} \varphi_{s}(0)} + h.c.  \right)\ .
\end{align}
Similar to the discussions in \cref{sec:HPK-bosonization}, the $\rho_z$ coupling can be absorbed by a gauge transformation $U =  e^{\ii \sqrt2 \rho_z \Lambda_z \varphi_{\uparrow}(0) + \ii \sqrt2 \rho_z \Lambda_z \varphi_{\downarrow}(0)}$. 
Let us denote the transformed Hamiltonian as $\ovl{H} = U H U^\dagger$, and divide $\ovl{H}$ into a free-boson part $\ovl{H}_0$ and a transversal coupling $\ovl{H}_x$.
Following the same calculations in \cref{sec:HPK-bosonization} and \cref{app:boson-oprt}, we find
\begin{align}   \label{eq:singlet_H0}
\ovl H_0 
&= \sum_{s=\uparrow\downarrow} \int \frac{\dd x}{4\pi} 
    :(\partial_{x} \varphi_{s})^2 :
    + \frac{\varepsilon_D}{x_c} \cdot \Lambda_z^2 \ ,
\end{align}
where $\frac{\varepsilon_D}{x_c} = J - \frac{4\rho_z^2}{x_c}$ is the further lowered energy of the doublet state due to the longitudinal coupling $\rho_z$. 
We have chosen $\varepsilon_D$ dimensionless for later convenience. 
Since we are interested in the phase transition to the LS regime, we will assume $\varepsilon_D>0$.  

Using $U \Theta_\pm U^\dagger = e^{\pm \ii \sqrt2 \rho_z (\varphi_{\uparrow}(0) + \varphi_\downarrow(0))} \cdot \Theta_\pm$, the transversal coupling is found 
\begin{align}    \label{eq:singlet_Hx_1}
\ovl H_x =& \frac{\zeta_x}{x_c}  \Theta_+\Big[ 
    F_\uparrow \cdot e^{-\ii\sqrt2(1-\rho_z) \varphi_\uparrow(0) + \ii \sqrt2 \rho_z \varphi_\downarrow(0) } \nonumber\\
& + F_\downarrow \cdot e^{\ii \sqrt2\rho_z \varphi_\uparrow(0) - \ii\sqrt2(1-\rho_z) \varphi_\downarrow(0)} \Big] + h.c.  \nonumber\\
=& \frac{\zeta_x}{x_c} \sum_{\nu=\pm}  \Theta_{\nu} \cdot \sum_{s=\uparrow,\downarrow} 
    F_s^{(-{\nu})} \cdot e^{-\ii \nu \vxi_s \cdot \vvphi(0) }\ ,
\end{align}
where  $F^{(-)}_s = F_s $,  $F^{(+)}_s = F_s^\dagger$, $\vvphi = (\varphi_{\uparrow}, \varphi_{\downarrow})$ and 
\begin{align}
    \vxi_s = \sqrt{2}(\delta_{s,\uparrow},\delta_{s,\downarrow}) - \sqrt{2}(\rho_z, \rho_z) \ . 
\end{align}
where $\delta_{s,s'}$ is the Kronecker delta. 
Basically, $\nu$ keeps track of how the $\Uo_v$ charge is exchanged between the impurity and bath, which is conserved in total, while $s=\uparrow,\downarrow$ indicates which spin sector (channel) of the bath has participated in the exchange. 
We also define $\frac{\xi^2}{2} = \frac{\nu^2\vxi_s^2}{2} = 2\rho_z^2-2\rho_z+1$, which is the scaling dimension of the vertex operators.  
For $0<\rho_z<\frac{1}{2}$, $1>\frac{\xi^2}{2} > \frac{1}{2}$ (monotonically) respectively. 
Also, note that $\vxi_\uparrow$ and $\vxi_{\downarrow}$ are linearly independent for all $\rho_z$.

We analyze the RG flow of \cref{eq:singlet_H0,eq:singlet_Hx_1} below. 

\subsection{Coulomb gas analog}

The model is solvable if $\zeta_x=0$, and we denote the partition function at $\zeta_x =0$ as $Z_0$. 
The (multiplicative) correction to the partition function $Z$ at finite $\zeta_x$ is given by a perturbative expansion, 
\begin{equation}
\delta Z = 
    \Inn{T_\tau \exp\pare{- \int_{-\frac1{2T}}^{\frac1{2T}}  \ovl H_x(\tau) } }_{\ovl 0} = \sum_{n=0}^{\infty} \delta Z_{2n} \ ,  
\end{equation}
where
\begin{align}
\delta Z_{2n} = \int^{>0}_{(-\frac{1}{2T}, \frac{1}{2T})} \!\mrm{d}^{2n}\tau ~ \Big\langle \ovl{H}_x(\tau_{2n}) \cdots \ovl{H}_x(\tau_{1})  \Big\rangle_{\ovl{0}} \ . 
\end{align}
The subscript $\ovl 0$ represents average with respect to the equilibrium ensemble of $\ovl{H}_0$ [\cref{eq:singlet_H0}]. 
$T_\tau$ is the time-ordering operator. 
$\int_{(-\frac1{2T},\frac1{2T})}^{>0} \dd^{2n}\tau$ [\cref{eq:integral-ordered-tau}] represents the integral in the domain $\tau_{2n}> \tau_{2n-1} \cdots >\tau_2>\tau_1$. 
\begin{align}
    \ovl H_x(\tau) =& e^{\tau \ovl{H}_0} \ovl{H}_x e^{-\tau \ovl{H}_0} \nonumber\\
    =& \frac{\zeta_x}{x_c} \sum_{\nu=\pm}  \Theta_{\nu}(\tau) \sum_{s=\uparrow,\downarrow} 
    F_s^{(-\nu)}(\tau)  \cdot e^{-\ii \nu \vxi_s \cdot \vvphi(\tau)}
\end{align}
writes the operators in the interacting picture. 
Hereafter in this section, we omit the spatial argument of $\varphi_s$, since they are located at $x=0$. 
Crucially, while $F_s(\tau) = F_s$, the impurity operators are time-evolved as 
\begin{align}  \label{eq:Theta-tau-singlet}
\Theta_+(\tau) &= 
    e^{\frac{\varepsilon_D}{x_c} \tau }  \cdot \ket{2}\bra{0} 
    +  e^{-\frac{\varepsilon_D}{x_c} \tau } \cdot \ket{0}\bra{\bar2}  \ , \nonumber\\
\Theta_-(\tau) &= 
    e^{-\frac{\varepsilon_D}{x_c} \tau }  \cdot \ket{0}\bra{2} 
    +  e^{\frac{\varepsilon_D}{x_c} \tau } \cdot \ket{\bar2}\bra{0}\ .  
\end{align}

Following the discussions in \cref{sec:BKT-coulombgas}, as the impurity and bath are decoupled in $\ovl{H}_0$ [\cref{eq:singlet_H0}], the ensemble average $\ovl{0}$ can be factorized into an average over impurity operators and an average over the bath, as
\begin{align}
& \delta Z_{2n} = \frac{\zeta_x ^{2n}}{x_c^{2n}}
    \int_{(-\frac{1}{2T},\frac{1}{2T})}^{>0} \dd^{2n}\tau
    \sum_{\{\nu\}} \Big\langle \Theta_{\nu_{2n}}(\tau_{2n}) \cdots\Theta_{\nu_1}(\tau_1) \Big\rangle_{\bar 0}
    \nonumber \\
&\;  \times \sum_{\{s\}} \Inn{ F_{s_{2n}}^{(\ovl{\nu}_{2n})}
    \cdots F_{s_1}^{(\ovl{\nu}_1)} }_{\bar0} 
    \left\langle e^{-\ii \sum_{i=1}^{2n} \nu_{i} \vxi_{s_{i}} \cdot \vvphi(\tau_{i})}  \right\rangle_{\ovl{0}} \ .
\end{align}
Here, $\sum_{\{\nu\}}$ and $\sum_{\{s\}}$ indicates the summation over all $\nu_{i}=\pm$ and $s_{i} = \uparrow,\downarrow$, respectively. 
We now analyze the general structure of all non-vanishing terms in the summation $\sum_{\{\nu\}} \sum_{\{ s \}}$. 

For the impurity average, as $\ovl{0}$ only contains the $|0\rangle$ state in the zero-temperature limit ($T \ll \frac{\varepsilon_D}{x_c}$), $\Theta_{\nu_1}$ should only excite it to \textit{either} $|2\rangle$ \textit{or} $|\ovl{2}\rangle$, while $\Theta_{\nu_2}$ should then lower it back to $|0\rangle$. 
Therefore, there must be $\nu_2 = \ovl{\nu}_1$, where $\nu_1$ can be either $+$ or $-$. 
The same analysis can be recursively applied to any $\Theta_{\nu_{2k}} \Theta_{\nu_{2k-1}}$. 
Therefore, the non-vanishing terms in $\sum_{\{ \nu \}}$ are given by all configurations of $\nu$ that satisfies $\nu_{2k}=\ovl{\nu}_{2k-1} = +$ or $\nu_{2k}=\ovl{\nu}_{2k-1} = -$. 
In total, there are $2^n$ different such configurations in the summation over $\{\nu\}$. 
Between $\tau_{2k}-\tau_{2k-1}$, the impurity stays at the high-energy $D$ manifold, hence it accumulates a factor $e^{-\frac{\varepsilon_D}{x_c}(\tau_{2k} - \tau_{2k-1})}$ according to \cref{eq:Theta-tau-singlet}. The full impurity average value thus always equals to
\begin{align} \label{eq:nu-summation-singlet}
   \Big\langle \Theta_{\nu_{2n}}(\tau_{2n}) \cdots \Theta_{\nu_1}(\tau_1 ) \Big\rangle_{\bar 0}  =
    e^{ - \frac{\varepsilon_D}{x_c} \sum_{i}^{n} (\tau_{2i}-\tau_{2i-1}) }   \ . 
\end{align}

For each configuration of $\{\nu\}$, one can assign $s_i$ for $i=1,\cdots,2n$ independently, with the only constraint that the bath valley charge per spin flavor (accumulated by $F_\uparrow$ and $F_{\downarrow}$, respectively) are both zero. 
In other words, this is equivalent to requiring the total bath charge is zero, $\sum_{i} \nu_i=0$, which is already satisfied, and requiring that the ``spin-contrasting'' bath charge is also zero, $\sum_{i} \nu_i s_i =0$. 
It is convenient to regard the quantities $\nu_i s_i$ as independent variables, and $n$ out of $2n$ of them should be $+$, with the remaining ones being $-$. 
Thus, for a given configuration of $\{\nu\}$, there are $\binom{2n}{n}$ configurations in the summation over $\{s\}$. 

It can be shown that $F_\uparrow$ commutes with $F_{\downarrow}$ and $F_{\downarrow}^\dagger$. Therefore, the product of Klein factors is trivially $1$. 
The remaining average over the vertex operators is given by \cref{eq:vertex_corr_2n}:
{\small
\begin{align} 
&\left\langle e^{-\ii \sum_{i=1}^{2n} \nu_{i} \vxi_{s_{i}} \cdot \vvphi(\tau_{i})}  \right\rangle_{\ovl{0}}  = \\ \nonumber
&   \exp\pare{  - \sum_{j>i} \nu_j \nu_i \,(\vec\xi_{s_j}\cdot\vec\xi_{s_i}) 
        \ln \pare{ \frac{\pi T x_c}{ \sin\pare{ \pi T (\tau_{j} - \tau_{i}) + \pi T x_c}  } }}  \ . 
\end{align}}
As we did in \cref{sec:BKT-coulombgas}, we can replace the factor $\ln \pare{ \frac{\pi T x_c}{ \sin\pare{ \pi T (\tau_{j} - \tau_{i}) + \pi T x_c}  } }$ by $\ln \pare{ \frac{x_c}{ \tau_{j} - \tau_{i}}  }$ in the zero-temperature limit, and correspondingly change the integral range $\int_{(-\frac1{2T},\frac1{2T})}^{>0} \dd^{2n}\tau$ to $\int_{(-\frac1{2T},\frac1{2T})}^{>x_c} \dd^{2n}\tau$. 
The factor $e^{-\frac{\varepsilon_D}{x_c}(\tau_{2i}-\tau_{2i-1})}$ also changes to $e^{-\frac{\varepsilon_D}{x_c}(\tau_{2i}-\tau_{2i-1})} \cdot e^{\varepsilon_D}$. 

To sum up, we have 
\begin{widetext}
\begin{align} \label{eq:Z2n-singlet}
\delta Z_{2n} &= \frac{\zeta_x^{2n}}{x_c^{2n} } e^{n\cdot \varepsilon_D}
    \int_{(-\frac1{2T},\frac1{2T})}^{>x_c} \dd^{2n}\tau \ 
    e^{ -\frac{\varepsilon_D}{x_c} \sum_{i=1}^{n} (\tau_{2i}-\tau_{2i-1}) }
    \sum_{\{\nu\}}' \sum_{\{s\}}'
    \exp\pare{  - \sum_{j>i} \nu_j \nu_i \,(\vxi_{s_{j}}\cdot\vxi_{s_i}) \,
        \ln\pare{ \frac{x_c}{\tau_j-\tau_i} }
    }\ ,
\end{align}
\end{widetext}
where $\sum_{\{\nu\}}' \sum_{\{s\}}'$ only selects out the non-vanishing configurations. 

The partition function in \cref{eq:Z2n-singlet} describes $2n$ particles on a line, interacting through two types of Coulomb forces: a 1D Coulomb potential proportional to $\frac{\varepsilon_D}{x_c}|\tau_{j'}-\tau_{j}|$, and a 2D logarithmic Coulomb potential proportional to $\ln\frac{x_c}{|\tau_{j'}-\tau_{j}|}$. 
The 1D Coulomb charge of the particles is given by $(-1)^{j}$, while the 2D Coulomb charge of the particles is given by a vector $\nu_{j} \vxi_{s_j}$. 
Whether the 2D Coulomb force is repulsive or attractive depends on the inner product between two ``vector charges''. 
Below we explain this analog in detail.

First, $\frac{\varepsilon_D}{x_c} (\tau_{2i} - \tau_{2i-1}) = \frac{\varepsilon_D}{x_c} |\tau_{2i} - \tau_{2i-1}|$ can be interpreted as a 1D Coulomb potential between two particles with distance $|\tau_{2i} - \tau_{2i-1}|$, because its derivative with respect to the distance, the electric field strength, will be constant. 
The two particles possess opposite 1D Coulomb charge, as the potential energy grows with increasing distance $|\tau_{2i} - \tau_{2i-1}|$. 
More generically, we can assign the particle at $\tau_j$ with a 1D Coulomb charge $(-1)^j$, and the total energy will simplify to 
\begin{align}
    -\sum_{j'>j} (-1)^{j'-j} (\tau_{j'}-\tau_{j}) = \sum_{i=1}^{n} (\tau_{2i}-\tau_{2i-1}) \ . 
\end{align}
The proof is simple. For a particle at $\tau_{2i-1}$, all the particles $\tau_{j}$ to its right ($2i-1>j$) possess a vanishing total charge. 
As the 1D Coulomb force does not decay, the vanishing total charge also implies an exactly vanishing total Coulomb force. 
On the other hand, all the particles $\tau_{j'}$ to its left ($j'>2i-1$) possess a $+1$ total charge, hence attracting $\tau_{2i-1}$ to its left. 
Similarly, one can show that $\tau_{2i}$ is attracted to its right. 
The net effect will thus be equivalent to only counting the interaction between $\tau_{2i}$ and $\tau_{2i-1}$. 

We also remark that, since the two particles have a minimal distance $x_c$, the minimal energy cost of the 1D Coulomb interactions correspond to a factor of $e^{-n\varepsilon_D}$ in the partition function, which will be compensated by the pre-factor $e^{n\varepsilon_D}$. 
Therefore, $\zeta_x$ still represents the fugacity. 

The 2D Coulomb force takes a similar form with that in \cref{sec:BKT}, with the difference that the 2D Coulomb charge behaves as a vector $\nu_{j} \vxi_{s_j}$, and the Coulomb potential between a particle pair is proportional to the inner product of the ``vector charges''.

To gain some insights into the perturbation theory, let us calculate the lowest order correction $\delta Z_2$, 
\begin{align}
\delta Z_2 = 4\frac{\zeta_x ^{2}}{x_c^{2}} e^{ \varepsilon_D}
    \int_{(-\frac1{2T},\frac1{2T})}^{>x_c} \dd^2\tau \ 
    e^{ -\frac{\varepsilon_D}{x_c} (\tau_2 - \tau_1) } 
    \pare{ \frac{x_c}{\tau_2-\tau_1} }^{\xi^2}\ .
\end{align}
where $\xi^2=2-4\rho_z + 4\rho_z^2$. 
The factor $4=2\times 2$ originates from the summation $\sum'_{\{\nu\}} \sum'_{\{s\}}$, where $\nu_1=\pm$ and $s_1 = \uparrow,\downarrow$ in total contribute four equal terms. 
Since $e^{\varepsilon_D - \frac{\varepsilon_D}{x_c}(\tau_2-\tau_1)}\le 1$ for $\varepsilon_D\ge 0$, any finite $\varepsilon_D$ will suppress the partition function correction from $\zeta_x$, and will guarantee the integral over $\mrm{d}\tau_1$ to be convergent. 
Integrating over $\mrm{d}\tau_1$ yields 
$x_c \mathrm{E}_{2-4\rho_z + 4\rho_z^2}(\varepsilon_D)$, where $\mathrm{E}_n(x) = \int_{1}^\infty \dd t \ e^{-x t}t^{-n} $ is exponential integral function. 
Therefore, 
$\delta Z_2 = 4\frac{\zeta_x^2}{T x_c} \cdot e^{\varepsilon_D} \cdot \mathrm{E}_{2-4\rho_z + 4\rho_z^2} (\varepsilon_D) $, and the leading-order correction to the ground state energy is 
\begin{align}
  &  \delta E_2 = - T \cdot \delta Z_2 
= - 4\frac{\zeta_x^2}{x_c} e^{\varepsilon_D} \cdot \mathrm{E}_{2-4\rho_z + 4\rho_z^2} (\varepsilon_D) \nonumber\\
=& - 4\frac{\zeta_x^2}{x_c} e^{\varepsilon_D}
\bigg( \frac1{(1-2\rho_z)^2} + \varepsilon_D^{(1-2\rho_z)^2} \cdot \Gamma (-(1-2\rho_z)^2) \nonumber\\
& + \frac{\varepsilon_D}{4\rho_z(1-\rho_z)} + \mathcal{O}(\varepsilon_D^2)
\bigg)\ ,
\end{align}
where $\Gamma(x)$ is the $\Gamma$-function. 

\subsection{Flow equations}

To obtain the RG flow, we coarse-grain the temporal coordinate by rewriting $\tau = b \tau'$, where $b= e^{\dd \ell} > 1$, and then relabeling $\tau'$ as $\tau$.
Then the partition function in \cref{eq:Z2n-singlet} becomes 
\begin{align} \label{eq:Z2n-rescale-singlet}
& \delta Z_{2n} = \frac{\zeta_x^{2n}}{x_c^{2n}} b^{2n} 
    b^{-2n(1-2\rho_z+2\rho_z^2)}
    e^{n\varepsilon_D}
    \int_{(-\frac1{2T}, \frac1{2T})}^{>x_c b^{-1}} \dd^{2n}\tau \nonumber\\
&   e^{ -b \frac{\varepsilon_D}{x_c} \sum_{i=1}^{n} (\tau_{2i}-\tau_{2i-1}) }
    \sum_{\{\nu\} }' \sum_{\{s\}}'
    \prod_{j>i} \pare{ \frac{x_c}{\tau_j-\tau_i} }^{-\nu_j \nu_i \,\vec\xi_{s_j}\cdot\vec\xi_{s_i}}  \!\!. 
\end{align}
The factor $b^{2n}$ originates from rescaling the integral measure, and $b^{\sum_{j>i}\nu_j \nu_i \,\vec\xi_{s_j}\cdot\vec\xi_{s_i}} = b^{-2n(1-2\rho_z+2\rho_z^2)}$ originates from rescaling the 2D Coulomb factors. 
To be more concrete, since for all configuration $\{\nu\},\{s\}$, there is $\sum_{i=1}^{2n} \nu_i \vxi_{s_i} = 0$, we obtain 
$ \sum_{j>i}  \nu_j \nu_i \,\vxi_{s_j} \cdot\vxi_{s_i} = \frac12 \sum_{j,i} (\nu_j \vec\xi_{s_j}) \cdot (\nu_i \vxi_{s_i}) - \frac{1}{2} \sum_i \nu_i^2 \vec\xi_i^2 = - 2n (1 - 2\rho_z + 2\rho_z^2)$. 
For the 1D Coulomb interaction, the rescaled effective coupling reads $\varepsilon_D' = \varepsilon_D e^{\dd\ell}$, while the rescaled fugacity satisfies $\zeta_x' \cdot e^{\frac{\varepsilon_D'}{2}}  = \zeta_x \cdot b^{1 - (1-2\rho_z+2\rho_z^2)} e^{\frac{\varepsilon_D}{2}}$. 
Therefore, $\zeta_x' = \zeta_x \cdot e^{\dd\ell(2\rho_z-2\rho_z^2)} \cdot e^{\frac{1}{2}(\varepsilon_D-e^{\dd\ell}\varepsilon_D)} = \zeta_x \cdot e^{\dd\ell(2\rho_z-2\rho_z^2-\frac{\varepsilon_D}{2})}$. 
These relations imply the tree-level RG flow equations 
\begin{equation}
\frac{\dd \zeta_x}{\dd \ell} = \pare{  2\rho_z - 2\rho_z^2 -\frac{1}{2} \varepsilon_D} \zeta_x \ ,  \qquad 
\frac{\dd \varepsilon_D}{\dd \ell} = \varepsilon_D\ . 
\end{equation}

To derive the higher-order corrections to the flow equations, we need to integrate out ``high-energy'' configurations where distances between adjacent particles are smaller than $x_c$. 
Following the discussions around \cref{eq:dZ2n_prime} in \cref{sec:BKT}, we need to calculate $\delta Z_{2,1}$ and $\delta Z_{4,1}$. 
$\delta Z_{2,1}$ is $\delta Z_2$ where $x_c b^{-1} <\tau_2 - \tau_1<x_c$. 
Since the integral over $\tau_2$ is proportional to $\dd\ell$ and only $\mathcal{O}(\dd\ell)$ terms are of interest, we can neglect the $b$ factors elsewhere. 
It is straightforward to obtain 
\begin{equation} \label{eq:dZ21-singlet}
    \delta Z_{2,1} =  4\frac{\zeta_x^2}{Tx_c}  \cdot \dd\ell + \mathcal{O}(\dd\ell^2)\ . 
\end{equation}
$\delta Z_{4,1}$ consists of three terms, $\delta Z_{4, 1} = \sum_{i=1}^{3} \delta Z_{4, 1}^{(i+1,i)}$, where $\delta Z_{2n+2, 1}^{(i+1,i)}$ has a molecule formed by $\tau_{i+1}$ and $\tau_i$.
The first term is 
\begin{widetext}
{\small
\begin{align}
\delta Z_{4,1}^{(2,1)} = \frac{\zeta_x^4}{x_c^4} e^{2\varepsilon_D}
  \int_{-\frac{1}{2T}}^{\frac{1}{2T}} \dd\tau_4 
    \int_{-\frac{1}{2T}}^{\tau_4-x_c} \dd\tau_3 
    \int_{-\frac{1}{2T}}^{\tau_3 - x_c} \dd\tau_2 
    \int_{\tau_2 - x_c}^{\tau_2 - x_c/b} \dd\tau_1 \  
    e^{-\frac{\varepsilon_D}{x_c}(\tau_4 - \tau_3 + \tau_2 - \tau_1)}
    \sum_{\{\nu\}}' \sum_{\{s\}}'
    \prod_{j>i} \pare{ \frac{x_c}{\tau_j-\tau_i} }^{-\nu_j \nu_i \,\vxi_{s_j}\cdot\vxi_{s_i}}\ .
\end{align}}
\end{widetext}
A complication comes from the summation $\sum_{\{\nu\}}'\sum_{\{s\}}'$.
Unlike the case in doublet regime (\cref{sec:BKT}), the molecule at $(\tau_2,\tau_1)$ is not necessarily charge neutral, {\it i.e.}, $\nu_2\vec\xi_2 + \nu_1 \vec\xi_1 = 0$. 

We now argue that, for generic $\delta Z_{2n+2,1}^{(i+1,i)}$, only the contribution from {\it neutral} molecule (or dipole) at $(\tau_{i+1},\tau_i)$ is relevant. 
We introduce the vector charge $\vec\zeta_i = \nu_i \vec \xi_{s_i}$ for the particle at $\tau_i$, and the total vector charge $\vec\zeta_0 = \vec\zeta_{i+1} + \vec\zeta_i$ for the molecule to be integrated out. 
Since the total charge of the $2n+2$ particles vanishes, there must be $\sum_{j\neq i,i+1} \vec\zeta_j = - \vec\zeta_0$. 
Then a typical value of the integrand, where distances between remaining particles are typically $\sim T$, is $(x_cT)^{n (2-4\rho_z + 4\rho_z^2) + \frac12 \vec\zeta_0^2}$. 
Thus, terms with $\vec\zeta_0\neq 0$ are typically smaller by a factor of $\mathcal{O}((x_cT)^{\frac12\vec\zeta_0^2})$. 
We will only keep neutral molecule in the following calculations.

Given $\nu_2 \vxi_{s_2} + \nu_1 \vxi_{s_1} =0$ and $\nu_2=-\nu_1$, the remaining two particles in $\delta Z_{4,1}^{(2,1)}$ at $\tau_{4}$ and $\tau_{3}$ must carry $\nu_4=-\nu_3$ and $s_4 = s_3$, as if they are variables for a two-particle partition function. 
With $\tau_2 - \tau_1 = x_c + \mathcal{O}(\dd\ell)$, the factor $\prod_{j>i} \pare{ \frac{x_c}{\tau_{j}-\tau_{i}} }^{-\nu_j \nu_i \,\vxi_{s_j}\cdot\vxi_{s_i}}$ becomes 
\begin{widetext}
\begin{equation}
\exp\pare{ 
    \nu_4\nu_3\vec\xi_4\cdot\vec\xi_3 \ln \pare{\frac{\tau_4 - \tau_3}{x_c}}
    - \nu_4 \nu_1 \vec\xi_4\cdot\vec\xi_1  \ln \pare{\frac{\tau_4 - \tau_2}{\tau_4 - \tau_2 + x_c}}
    + \nu_3 \nu_1 \vec\xi_3\cdot\vec\xi_1  \ln \pare{\frac{\tau_3 - \tau_2 + x_c}{\tau_3 - \tau_2}}
}\ ,
\end{equation}
and the factor $e^{-\frac{\varepsilon_D}{x_c}(\tau_4 - \tau_3 + \tau_2 - \tau_1)}$ becomes 
$e^{-\frac{\varepsilon_D}{x_c}(\tau_4 - \tau_3)} \cdot e^{-\varepsilon_D}$. 
We relabel $i=3,4$ as $i=1,2$, respectively, and relabel the original $\nu_1=-\nu_2$ as $\nu'$, and the original $s_1=s_2$ as $s'$. Also, relabel $\tau_2 = \tau' + \frac12 s$, where $x_c b^{-1} <s<x_c$. These primed variables will be integrated out as virtual processes. 
Following the calculations around \cref{eq:dZ41-21-tmp1}, we obtain 
\begin{align}
\delta Z_{4,1}^{(2,1)} =& \frac{\zeta_x^2}{x_c^2} e^{2\varepsilon_D}
    \int_{-\frac1{2T}}^{\frac1{2T}} \dd\tau_2 \int_{-\frac1{2T}}^{\tau_2 - x_c} \dd\tau_1 
    \sum_{\{\nu_2,\nu_1\}}' \sum_{ \{s_2,s_1\} }'
    e^{-\frac{\varepsilon_D}{x_c}(\tau_2 -\tau_1) } \pare{ \frac{x_c}{\tau_2 -\tau_1} }^{-\nu_2 \nu_1 \vxi_{s_2} \cdot \vxi_{s_1}} \nonumber\\
&\times e^{-\varepsilon_D} \zeta_x^2 \dd\ell   \sum_{\nu'=\pm} \sum_{s'=\uparrow\downarrow}  \int_{-\frac1{2T}}^{\tau_1 - \frac{3}{2} x_c} \dd \tau'
    \pare{  \frac1{x_c} - \frac{\nu_2\nu'\vec\xi_2\cdot\vec\xi'}{\tau_2 -\tau'} 
    + \frac{\nu_1\nu'\vec\xi_1\cdot\vec\xi'}{\tau_1-\tau'} + \mathcal{O}(x_c) }\ ,
\end{align}
In terms of the Coulomb gas analog, the second and third terms in the last row describe the interaction between a (virtual) 2D Coulomb dipole and the remaining particles. 
Importantly, there is $\sum_{\nu'=\pm} \sum_{s'=\uparrow\downarrow} \nu' \vxi_{s'} = 0$, namely, this dipole does not have a definite orientation, as opposed to the dipole in the pair-Kondo model (see \cref{sec:BKT}). 
Therefore, summing over all the possible dipole configurations will average out, and the second and third terms will vanish. 
Consequently, the 2D Coulomb interaction between the remaining particles will not be screened, and $\rho_z$ will remain invariant. 
Integrating the non-vanishing terms over $\dd \tau'$ then produces 
{\small
\begin{align}
\delta Z_{4,1}^{(2,1)} =& \frac{\zeta_x^2}{x_c^2} e^{2\varepsilon_D}
    \int_{-\frac1{2T}}^{\frac1{2T}} \dd\tau_2 \int_{-\frac1{2T}}^{\tau_2 - x_c} \dd\tau_1 
    \sum_{\{\nu\}}'\sum_{ \{s\} }'
    e^{-\frac{\varepsilon_D}{x_c}(\tau_2 -\tau_1) } \pare{ \frac{x_c}{\tau_2 -\tau_1} }^{-\nu_2 \nu_1 \vxi_{s_2} \cdot \vxi_{s_1}} 
    \cdot e^{-\varepsilon_D} \zeta_x^2 \dd\ell   \cdot  
   \frac{4}{x_c} \pare{ \frac{1}{2T} + \tau_1} \ .
\end{align}}

Following the calculations around \cref{eq:dZ41-32-tmp1}, we also obtain 
{\small
\begin{align}
\delta Z_{4,1}^{(3,2)} =& \frac{\zeta_x^2}{x_c^2} e^{2\varepsilon_D}
    \int_{-\frac1{2T}}^{\frac1{2T}} \dd\tau_2 \int_{-\frac1{2T}}^{\tau_2 - x_c} \dd\tau_1 
    \sum_{\{\nu\}}' \sum_{ \{ s \} }'
    e^{-\frac{\varepsilon_D}{x_c}(\tau_2 -\tau_1) } \pare{ \frac{x_c}{\tau_2 -\tau_1} }^{-\nu_2 \nu_1 \vxi_{s_2}\cdot\vxi_{s_1}} 
    \cdot e^{\varepsilon_D} \zeta_x^2 \dd\ell   \cdot  
   \frac4{x_c} \pare{ \tau_2-\tau_1 } \ ,
\end{align}}
The major difference of $\delta Z_{4,1}^{(3,2)}$ from $\delta Z_{4,1}^{(2,1)}$ is the second $e^{\varepsilon_D}$ factor, which comes from $e^{\frac{\varepsilon_D}{x_c}(\tau_3 - \tau_2) }$ with $(\tau_3,\tau_2)$ being the integrated molecule before we relabel the variables. 
Following the calculations around \cref{eq:dZ41-43-tmp1}, we obtain 
{\small
\begin{align}
\delta Z_{4,1}^{(4,3)} =& \frac{\zeta_x^2}{x_c^2} e^{2\varepsilon_D}
    \int_{-\frac1{2T}}^{\frac1{2T}} \dd\tau_2 \int_{-\frac1{2T}}^{\tau_2 - x_c} \dd\tau_1 
    \sum_{\{\nu\}}' \sum_{ \{s\} }'
    e^{-\frac{\varepsilon_D}{x_c}(\tau_2 -\tau_1) } 
    \pare{ \frac{x_c}{\tau_2 -\tau_1} }^{-\nu_2 \nu_1 \vxi_{s_2}\cdot\vxi_{s_1}} 
    \cdot e^{-\varepsilon_D} \zeta_x^2 \dd\ell   \cdot  
   \frac4{x_c} \pare{ \frac{1}{2T} - \tau_2  } \ ,
\end{align}}
where the factor $e^{-\varepsilon_D}$ comes from $e^{-\frac{\varepsilon_D}{x_c}(\tau_4 - \tau_3) }$ with $(\tau_4,\tau_3)$ being the integrated molecule. 
Adding up the three terms, we obtain 
{\small
\begin{equation}
\delta Z_{4,1} = \frac{\zeta_x^2}{x_c^2} e^{\varepsilon_D}
    \int_{-\frac1{2T}}^{\frac1{2T}} \dd\tau_2 \int_{-\frac1{2T}}^{\tau_2 - x_c} \dd\tau_1 
    \sum_{\{\nu\}}'\sum_{ \{s\} }'
    e^{-\frac{\varepsilon_D}{x_c}(\tau_2 -\tau_1) } \pare{ \frac{x_c}{\tau_2 -\tau_1} }^{-\nu_2 \nu_1 \vec\xi_{s_2}\cdot\vec\xi_{s_1}} 
     \dd\ell \  \frac{ 4\zeta_x^2 }{x_c}
    \pare{ \frac{1}{T}  + 
    (e^{2\varepsilon_D}-1) (\tau_2 - \tau_1) 
    }\ .
\end{equation}}

According to \cref{eq:dZ2n_prime}, the renormalized two-particle partition function is $\delta Z_2' = \delta Z_{2,0} + \delta Z_{4,1} - \delta Z_{2,1}\delta Z_{2,0}$, where $\delta Z_{2,0}$ is rescaled as explained after \cref{eq:Z2n-rescale-singlet} and $\delta Z_{2,1} = 4\frac{\zeta_x^2}{x_c T}e^{-\varepsilon_D}$ is given in \cref{eq:dZ21-singlet}. 
The $\frac1{T}$ term in $\delta Z_{4,1}$ is exactly canceled by $\delta Z_{2,1}\delta Z_{2,0}$.
Thus, the higher-order correction to $\delta Z_2'$ (in addition to the tree-level contribution) is 
\begin{align}
&\delta Z_{4,1} - \delta Z_{2,1}\delta Z_{2,0} \nonumber\\
=& \frac{\zeta_x^2}{x_c^2} e^{\varepsilon_D} 
    \int_{-\frac1{2T}}^{\frac1{2T}} \dd\tau_2 \int_{-\frac1{2T}}^{\tau_2 - x_c} \dd\tau_1 
    \sum_{\{\nu\}}'\sum_{ \{s\} }'
    e^{-\frac{\varepsilon_D}{x_c}(\tau_2 -\tau_1) } \pare{ \frac{x_c}{\tau_2 -\tau_1} }^{-\nu_2 \nu_1 \vxi_{s_2}\cdot\vxi_{s_1}} 
      \dd\ell \  \frac{ 4\zeta_x^2 }{x_c}
     (e^{2\varepsilon_D } -1 ) (\tau_2 - \tau_1) \ . 
\end{align}
\end{widetext}
The term $\dd\ell \  \frac{ 4\zeta_x^2 }{x_c}  (e^{2\varepsilon_D}-1) (\tau_2 - \tau_1) $ can be absorbed as a correction $-4(e^{2\varepsilon_D}-1) \zeta_x^2 \dd\ell $ to $\varepsilon_D $. 
It describes how a virtual 1D Coulomb dipole screens the 1D Coulomb interaction. 
Adding them and the tree-level contributions up, we obtain the final flow equations
\begin{equation}
    \frac{\dd \varepsilon_D}{\dd \ell} = \varepsilon_D - 
    4 ( e^{2\varepsilon_D}-1) \zeta_x^2 + \mathcal{O}(\zeta_x^3)\ ,
\end{equation}
\begin{equation}
    \frac{\dd \zeta_x}{\dd\ell} = \pare{ 2\rho_z - 2\rho_z^2 - \frac12 \varepsilon_D } \zeta_x + \mathcal{O}(\zeta_x^3)\ . 
\end{equation}
It is worth emphasizing that $\rho_z$ remains invariant up to the second order of $\zeta_x$.

\subsection{Phase diagram and critical exponent}

\begin{figure}[t]
\centering
\includegraphics[width=1.0\linewidth]{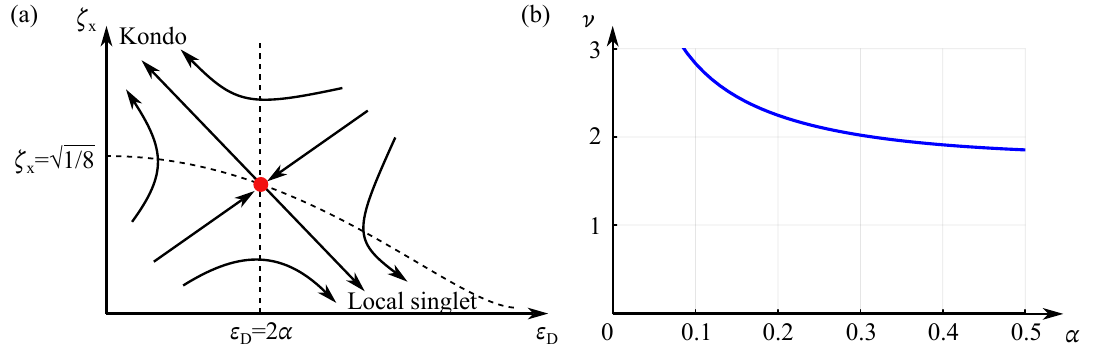}
\caption{RG flow in the singlet regime. 
    (a)  The vertical dashed line indicates $\varepsilon_D = 2\alpha$, and the the dashed curve indicates $\zeta_x = \sqrt{\frac{\varepsilon_D}{4(e^{2\varepsilon_D}-1)}}$. 
    The red dot is the critical point. 
    Here $\alpha=2\rho_z-2\rho_z^2$. 
    (b) The critical exponent $\nu$ as a function of $\alpha$. }
\label{fig:RG-flow2}
\end{figure}

For brevity, we define 
\begin{equation}
    \alpha = 2\rho_z - 2\rho_z^2  \in (0, \frac12) 
\end{equation}
in this subsection. 
$\frac{\dd\zeta_x}{\dd\ell} = 0$ if $\varepsilon_D = 2\alpha$, and 
$\frac{\dd\varepsilon_D}{\dd\ell} = 0$ if $\zeta_x = \sqrt{\frac{\varepsilon_D}{4 (e^{2\varepsilon_D}-1)}}$. 
We then derive the flow diagram shown in \cref{fig:RG-flow2}(a). 
The strong-coupling fixed point at $(\varepsilon_D, \zeta_x)=(0^+,\infty)$ represents the Kondo FL, and the weak-coupling fixed point at $(\varepsilon_D, \zeta_x)=(\infty,0^+)$ represents LS phase. 
They are separated by the critical point at $(\varepsilon_D, \zeta_x)=(2\alpha, \sqrt{\frac{\alpha}{2 (e^{4\alpha}-1)}} )$. 
To extract the critical exponent, we introduce $\delta \varepsilon_D = \varepsilon_D - 2\alpha$ and $\delta \zeta_x = \zeta_x - \sqrt{\frac{\alpha}{2 (e^{4\alpha}-1)}}$, and expand the flow equations to linear order of $\delta\varepsilon_D$ and $\delta \zeta_x$: 
\begin{equation}
\frac{\dd  }{\dd \ell}
\begin{bmatrix}
    \delta \varepsilon_D \\ \delta \zeta_x 
\end{bmatrix}
= \begin{bmatrix}
1-\frac{4 e^{4 \alpha } \alpha }{e^{4 \alpha }-1} & -\frac{4 \sqrt{2} \alpha }{\sqrt{\frac{\alpha }{e^{4 \alpha }-1}}} \\
-\frac{\sqrt{\frac{\alpha }{e^{4 \alpha }-1}}}{2 \sqrt{2}} & 0
\end{bmatrix}
\begin{bmatrix}
    \delta \varepsilon_D \\ \delta \zeta_x 
\end{bmatrix}\ .
\end{equation}
The matrix on the right-hand side has a negative eigenvalue, which corresponds to an irrelevant parameter, and a positive eigenvalue, which corresponds to a relevant parameter $t$. 
The positive eigenvalue is 
{\small
\begin{equation}
    \frac1{\nu} = \frac12 + \!
    \frac{ -4\alpha e^{4 \alpha } \!+ \! \sqrt{e^{8 \alpha } \left(16 \alpha ^2+1\right)+8 \alpha -2 e^{4 \alpha } (4 \alpha +1)+1}}{2 \left(e^{4 \alpha }-1\right)}  .
\end{equation}}
The flow of the relevant parameter is $ \frac{\dd t}{\dd\ell} = \frac1{\nu} t $.
Without loss of generality, we take $t=1$ as the strong-coupling fixed point. 
Then the RG time from a small positive $t$ to the strong-coupling fixed point is $\ell = \nu \cdot \ln \frac{1}{t}$, suggesting a Kondo temperature \begin{equation}
    T_{\rm K} \sim  D_{\rm S} \cdot t^{\nu}\ ,
\end{equation}
with $D_{\rm S}$ being the initial energy scale where the singlet regime is justified. 

$\nu$ does not appear to be a universal constant, as it depends on $\alpha= 2\rho_z - 2\rho_z^2$, as shown in \cref{fig:RG-flow2}(b).
A possible explanation is that $\alpha $ may flow to a fixed point at higher orders of $\zeta_x$, in which case $\nu$ at that fixed point would be a universal constant.
Nevertheless, within a reasonable range of $\alpha$, $0.2 \leq \alpha \leq 0.5$, $\nu$ is approximately 2, consistent with numerical results in Ref. \cite{wang_2025_solution}.

This phase transition is consistent with previous Numerical Renormalization Group (NRG) studies in similar models \cite{fabrizio_nontrivial_2003, leo_spectral_2004, nishikawa_convergence_2012}, where the critical point is found to be described by a non-Fermi liquid with impurity entropy $\ln \! \sqrt{2}$ \cite{fabrizio_nontrivial_2003}. 
When $J_D \!<\! 0$, the low-energy space consists of $S \oplus T$, and the phase transition should be equivalent to that in the two-impurity Kondo problem \cite{Jayaprakash_1981_2IK, Jones_1987_study, Jones_1988_lowT, Jones_1989_critical, Affleck_1992_exactcriticaltheory, Gan_1995_mapping, mitchell_universal_2012, mitchell_2channel_2012}, which was also found to be second-order.

\section{Conclusions and Discussions}
\label{sec:summary}

In this work, we present a comprehensive analytical study on the SVAIM at half-filling using bosonization and refermionization techniques. 
A central result is the identification of the pair-Kondo coupling $\lambda_x$ in the doublet regime, which couples the doublet-flipping of the impurity to a pair-hopping process in the bath electrons. 
Our bosonization RG calculations reveal a BKT phase transition between a weak-coupling fixed line ($\lambda_x=0$) that corresponds to the AD phase and a strong-coupling fixed line ($\rho_z^\star = \frac14$) that corresponds to a FL of pair Kondo resonance. 
Both fixed lines are solvable, allowing for the analytical calculation of the finite-size many-body spectrum and correlation functions. 
Furthermore, in the singlet regime, we find a second order phase transition between the Kondo FL and the LS phases. 

NRG verification of these quantum phase transitions is provided in the companion work \cite{wang_2025_solution}. 
While the analytical RG in this work is carried out under the assumption of PHS, numerical results demonstrate that both the BKT and second order phase transitions are stable against weak PHS breaking terms, e.g., deviation of chemical potential from the PHS point and particle-hole asymmetric hybridization functions. 
Previous works \cite{Affleck_1992_exactcriticaltheory,affleck_conformal-field-theory_1995} found that the critical point in the two-impurity Kondo model, which corresponds to the second-order phase transition in the singlet regime, is unstable against the terms mixing the two effective baths if PHS is absent. 
These terms arise inevitably in the two-impurity problem when integrating out the bath electron, because the two impurities do not couple to independent baths.  
However, the valley-$\Uo$ symmetry of SVAIM forbids such terms and ensures the existence of a critical point even in the absence of PHS \cite{affleck_conformal-field-theory_1995, fabrizio_nontrivial_2003,leo_spectral_2004,zarand_quantum_2006}. Moreover, recent cluster DMFT studies \cite{gleis_emergent_2024,gleis_dynamical_2025} found that the second-order quantum phase transition in the periodic Anderson model, which is similar to our FL-LS transition and the two-impurity Kondo case, can be protected by the DMFT self-consistency even if the particle-hole and orbital (corresponding to valley here) symmetries are lifted.

Our results establish the SVAIM as a fundamental framework for describing quantum impurity physics in moir\'e systems possessing both spin and valley degrees of freedom. 
As an SVAIM is defined only relying on the symmetries of spin-SU(2), valley-U(1), and valley-flipping, our results extend beyond MATBG/MATTG and apply generally to quantum impurities featuring both spin and valley-like degrees of freedom. 
For moir\'e systems, such valley-like degrees of freedom may be realized by different layers, orbital angular momenta, besides the momentum valleys in graphene. 

When applied to MATBG/MATTG with hetero-strain \cite{wang_2025_solution, youn_hundness_2024, youn_hundness_2025}, the SVAIM naturally accounts for the pseudo-gap phenomenon observed in recent experiments \cite{Oh_2021_evidence, park_experimental_2025, kim_2025_resolvingintervalleygapsmanybody}, and, simultaneously, provides a mechanism for pairing potential between quasi-particles. 
While the model is minimal, it is possible to generalize it to further include other degrees of freedom such as sublattice and orbitals for a broader application in moir\'e systems, which we leave for future studies.

\begin{acknowledgments}
We thank Hyunjin Kim and Jeong Min Park for fruitful discussions. 
Z.-D.~S., Y.-J.~W., and G.-D.~Z.~were supported by National Natural Science Foundation of China (General Program No.~12274005), National Key Research and Development Program of China (No.~2021YFA1401900), and Quantum Science and Technology-National Science and Technology Major Project (No.~2021ZD0302403). 
H.~J., S.~Y., and S.-S.~B.~L.~were supported by the National Research Foundation of Korea (NRF) grants funded by the Korean government (MSIT: No.~RS-2023-00214464, No.~RS-2023-00258359, No.~RS-2023-NR119931, No.~RS-2024-00442710; MEST: No.~2019R1A6A1A10073437), the Global-LAMP Program funded by the Ministry of Education (No.~RS-2023-00301976), and Samsung Electronics Co., Ltd.~(No.~IO220817-02066-01).
\end{acknowledgments}

\clearpage
\appendix 

\onecolumngrid

\section{Bosonization-refermionization dictionary}   \label{app:boson}

To apply the bosonization technique to the impurity problem, we linearize the dispersion of bath electrons near the Fermi surface, extend the band width to infinity, and only keep the $s$-wave bath states that interact with the impurity. This effectively reduces the bath to 1-dimensional chiral fermions. 

For the bosonization identities, we follow the constructive approach in Refs.~\cite{von_delft_bosonization_1998, vonDelft_1998_finitesize, zarand_analytical_2000}. 
We will treat the Klein factors that carry the quantum numbers in an exact manner, which helps keep track of the physical states in the enlarged Hilbert space. 
We also keep the $\mathcal{O}(L^{-1})$ terms to analyze the finite-size spectrum, where $L$ denotes the bath system size.
But we will ignore them when calculating physical quantities in the thermodynamic limit, such as the partition functions and correlation functions. 

For all the models studied in this work, the Fermi velocities of all bath flavors will be dictated by symmetries to be degenerate, hence we set them as $v_F = 1$. 
We also set the Planck constant $\hbar=1$, elementary charge $|e|=1$, and the Boltzmann constant $k_B=1$, so all physical quantities can be measured in terms of the energy dimension. 

\subsection{Operator identity}   \label{app:boson-oprt}

Let $\alpha$ label the flavor of bath electrons. 
The chiral fermions can be formally put on a circle of length $L$, hence the finite-size energy spacing between two adjacent single-electron levels is $\DL$. 
The Hamiltonian reads, 
\begin{align}
    H_0 = \sum_{k} \sum_{\alpha} k :d^\dagger_{\alpha}(k) d_{\alpha}(k): \qquad \qquad k \in \DL  \left( \mbb{Z} - \frac{\Pbc}{2} \right) 
\end{align}
with $\left\{ d^\dagger_{\alpha}(k) , d_{\alpha}(k') \right\} = \delta_{kk'} \delta_{\alpha\alpha'}$, $\left\{ d_{\alpha}(k) , d_{\alpha}(k') \right\} = 0$. 
Here, $\Pbc = 0,1$ indicates whether the chemical potential lies exactly within a single-electron level, or between two levels. 
The normal-ordering of chiral fermions $: \cdots :$ is defined with respect to the following background $|0\rangle_0$, 
\begin{align}   \label{eq:def_bath_vacuum}
    d_{\alpha}(k) |0\rangle_0 &= 0   \qquad (\textrm{if}~~ k>0)  \qquad \qquad \qquad 
    d^\dagger_{\alpha}(k) |0\rangle_0 = 0 \qquad (\textrm{if}~~ k \le 0) \ .
\end{align}
Note that $\ket{0}_0$ occupies all {\it non-positive} levels including zero. 
The Fourier transformation to real-space reads
\begin{align} \label{eq:d-psi-fourier-transformation}
    \psi_{\alpha}(x) = \sqrt{\frac{1}{L}} \sum_{k} d_{\alpha}(k) ~ e^{-\ii kx}  , \qquad\qquad
    d_{\alpha}(k) = \sqrt{ \frac{1}{L}}  \int_{-\frac{L}{2}}^{\frac{L}{2}} \mrm{d}x ~ \psi_{\alpha}(x) ~ e^{\ii kx} 
\end{align}
with $\{ \psi^\dagger_{\alpha}(x) , \psi_{\alpha'}(x') \} =  \delta(x-x') \delta_{\alpha\alpha'}$, $\{ \psi_{\alpha}(x) , \psi_{\alpha'}(x') \} = 0$. 
Since $\Pbc = 0,1$ also determines whether the boundary condition at $x = \pm\frac{L}{2}$ is periodic or anti-periodic, we term it as the boundary condition parameter. 
In the real-space, the Hamiltonian reads 
\begin{align} \label{eq:H0-def}
    H_0 &= \int\mrm{d}x \sum_{\alpha} :\psi^\dagger_{\alpha}(x) \left( \ii\partial_x \right) \psi_{\alpha}(x):  \ .
\end{align}
The U(1) charge that counts the total particle number in each flavor $\alpha$ is defined as
\begin{align}
    N_{\alpha} = \sum_{k} : d^\dagger_{\alpha}(k) d_{\alpha}(k) : ~ = \int\mrm{d}x  :\psi^\dagger_\alpha(x) \psi_\alpha(x): ~\in \mbb{Z} \ . 
\end{align}
Note that the chiral fermions are all \textit{left-movers}, and, to keep a consistent notation with Ref.~\cite{von_delft_bosonization_1998}, we have adopted the convention $\psi_\alpha (x)\sim d_\alpha(k) e^{-\ii k x}$ such that $d_\alpha (k)$ is an eigenmode of the energy $k$. 
(If the more common convention $\psi_\alpha(x) \sim d_{\alpha}(k) e^{\ii k x}$ were adopted, $d_\alpha(k)$ would have an eigenenergy of $-k$.)

Bosonization relies on the fact that, any $\vN$-particle Fock state in the physical Hilbert space, where $\vN$ collects all quantum numbers $N_\alpha \in \mbb{Z}$ into a vector, can be constructed from a unique $\vN$-particle ground state $|\vN\rangle_0$ by acting upon it a series of particle-hole excitations that commute with $\vN$. 
All operators within the physical Hilbert space can thus be constructed from two types of elements: 
1) the Klein factors $F_{\alpha}$ that link between $\vN$-particle \textit{ground} states $|\vN\rangle_0$ with different $\vN$, and encode the fermion anti-commutation between different $\alpha$, and 2) the bosonic fields $\phi_{\alpha}(x)$ that generate density fluctuations (i.e. particle-hole excitations) that commute with all $N_{\alpha'}$. 
Also, $[ F_{\alpha}, \phi_{\alpha'}(x)] =0$. 
We refer the detailed derivation of the bosonization procedures to Ref. \cite{von_delft_bosonization_1998}, and only summarize the definitions and key identities below. 

The fermion operator is bosonized to, 
\begin{align}   \label{eq:boson_ferm}
    \psi_{\alpha}(x) &= \frac{F_{\alpha}}{\sqrt{2 \pi x_c}} ~ e^{-\ii \phi_{\alpha}(x)} ~ e^{-\ii\left( N_{\alpha} - \frac{\Pbc}{2} \right) \frac{2\pi x}{L}}  \ .
\end{align}
Here, $x_c \to 0^+$ is an ultraviolet cutoff. 
We remark that the fermion Hilbert space (as well as the boson Hilbert space introduced below) is not truncated, and $x_c\to 0^+$ is only introduced to realize the operator identity. 
Sometimes $x_c^{-1}$ can be interpreted as an ``effective bandwidth'' of the chiral fermion. 
$F_{\alpha}$ are Klein factors that obey
\begin{align}  \label{eq:F_alpha}
    & [N_{\alpha} , F_{\alpha'}] = - F_{\alpha'} \delta_{\alpha\alpha'} \ , \qquad F_{\alpha} F^\dagger_{\alpha} = F^\dagger_{\alpha} F_{\alpha} = 1 \ , \qquad \{ F_{\alpha} , F^\dagger_{\alpha'} \} = 2 \cdot \delta_{\alpha\alpha'} \ , \qquad \{ F_{\alpha} , F_{\alpha'} \} = 2 F^2_{\alpha} \cdot \delta_{\alpha\alpha'} \ .
\end{align}
After specifying a certain ordering of the fermion flavors $\alpha=1,2,\cdots$, we can define the normalized $\vec N$-particle ground states as 
\begin{equation}    \label{eq:boson_vacuum}
\ket{\vec N}_0 = (F_1^\dagger)^{N_1} (F_2^\dagger)^{N_2} \cdots \ket{0}_0\ . 
\end{equation}
where we take the convention that $(F_\alpha^\dagger)^{N_\alpha} = (F_\alpha)^{-N_\alpha}$ if $N_\alpha<0$. 
Correspondingly, the matrix elements of Klein factors under the basis set $\ket{\vec{N}}_0$ read
\begin{align}   \label{eq:F_alpha_def}
    F_{\alpha} |\vN\rangle_0 = (-1)^{\sum_{\alpha'<\alpha} N_{\alpha'}} |\vN - \Delta\vN_{\alpha}\rangle_0 \qquad \qquad \Delta \vN_{\alpha} = (0,\cdots, \underset{\alpha\mrm{-th}}{1}, \cdots, 0) \ ,
\end{align}
where $(-1)^{\sum_{\alpha'<\alpha} N_{\alpha'}}$ is the Jordan-Wigner string due to the anti-commutation between Klein factors. 
It suffices to specify the action of $F_{\alpha}$ on the $\vN$-particle ground states, because all the bosonic operators that generate particle-hole excitations commute with $F_{\alpha}$. 

The bosonic field $\phi_{\alpha}(x)$ is defined as
\begin{align}   \label{eq:def_phi}
    \phi_{\alpha}(x) &= \sum_{q>0} - \sqrt{\frac{2\pi}{qL}} \Big( e^{-\ii qx} b_{\alpha}(q) + e^{\ii qx} b^\dagger_{\alpha}(q) \Big) e^{-\frac{x_c q}{2}} =  \varphi_\alpha(x) + \varphi_\alpha^\dagger(x) \ ,
    \\
    b^\dagger_{\alpha}(q) &= \ii \sqrt{\frac{2\pi}{q L}} \sum_{k} d^\dagger_{\alpha}(k+q) d_{\alpha}(k) \ , \qquad \qquad  q \in \DL \mbb{Z}_+ = \{ \DL, 2\DL, 3\DL, \cdots  \} \ .
\end{align}
By this definition, $\phi_{\alpha}(x)$ is always periodic under $x \to x+L$. 
Boson fields obey $\big[b_{\alpha}(q) ,  b^\dagger_{\alpha'}(q') \big] = \delta_{\alpha\alpha'} \delta_{qq'}$, $\big[b_{\alpha}(q) ,  b_{\alpha'}(q') \big] = 0$, and $[N_{\alpha'} , b_{\alpha}(q) ] = 0$. 
We have also separately defined $\varphi_\alpha(x)$ and $\varphi_\alpha^\dagger(x)$, which are the components of $\phi_\alpha(x)$ that only consist of boson annihilation and creation operators, respectively. 

We first compute the commutator 
\begin{equation}   \label{eq:comm_varphi}
[\varphi_\alpha(x) ,  \varphi^\dagger_{\alpha'}(x')]
= \delta_{\alpha\alpha'} \sum_{n=1}^{\infty}  \frac1{n} e^{(-\ii \frac{2\pi}{L}(x-x')- \frac{2\pi}{L} x_c) n } 
= -\delta_{\alpha\alpha'}  \ln \pare{ 1 - e^{ - \frac{2\pi \ii}{L} (x-x'-\ii x_c) } }\ ,
\end{equation}
where $\sum_{n=1}^\infty \frac1{n} y^n = - \ln(1-y)$ is used for $|y|<1$. 
Hence
\begin{equation}   \label{eq:comm_phi}
    [\phi_\alpha(x) ,  \phi_{\alpha'}(x')] = -\delta_{\alpha\alpha'}  \ln \frac{1 - e^{ - \frac{2\pi \ii}{L} (x-x'-\ii x_c) }}{1 - e^{ - \frac{2\pi \ii}{L} (x'-x-\ii x_c) }} = - 2\ii \cdot \delta_{\alpha\alpha'} \cdot \arg\left(1 - e^{-\ii\DL(x-x')} e^{-\DL x_c} \right) \ .
\end{equation}
Here, the single-valued branch of the above $\ln \frac{\cdots}{\cdots}$ function is always taken such that the commutator equals $0$ if $x=x'$ mod $L$, so that the $\arg(\cdots)$ function takes values in $(-\frac{\pi}{2}, \frac{\pi}{2})$. 
Taylor-expanding \cref{eq:comm_varphi,eq:comm_phi} with regard to $\frac{x-x'}{L}$ yields, respectively, 
\begin{align} \label{eq:comm_varphi_taylor}
[\varphi_\alpha(x),  \varphi^\dagger_{\alpha'}(x')]
&=  \delta_{\alpha\alpha'} \brak{ -\ln \pare{ \frac{2\pi \ii }{L} 
    \big(x-x' - \ii x_c \big) }  + \frac{\pi \ii }{L}(x-x' - \ii x_c ) }
    + \mathcal{O}(L^{-2}) \ . \\\label{eq:phi-phi-comm0}
[\phi_\alpha(x), \phi_{\alpha'}(x')]
&= \delta_{\alpha\alpha'} \bigg[\ln\pare{ \frac{x'-x-\ii x_c}{x-x'-\ii x_c}  }
+ \frac{2\pi \ii}{L} (x-x') \bigg] + \mathcal{O}(L^{-2}) \ . 
\end{align}
Notice that while \cref{eq:comm_varphi,eq:comm_phi} are periodic in $x \to x+L$ and $x' \to x'+L$, the taylor-expanded \cref{eq:comm_varphi_taylor,eq:phi-phi-comm0} are not.
Hereafter we will omit $\mathcal{O}(L^{-2})$ terms in the real-space commutators, unless otherwise specified. 

We write the commutator of the $\phi$ fields in a more commonly used form, 
\begin{align}  \label{eq:boson_comm_phi}
    & [ \phi_{\alpha}(x) , \phi_{\alpha'}(x') ] = \delta_{\alpha\alpha'} \cdot (-\pi \ii) \cdot \left(  \sgn_{x_c}(x-x') - \frac{2}{L} (x-x') \right)   ,  
    \qquad 
    \sgn_{x_c}(x) = \frac{2}{\pi} \arctan \frac{x}{x_c} \\
\label{eq:boson_comm_dphi}
    & [ \phi_{\alpha}(x) , \partial_{x'} \phi_{\alpha'}(x') ] = \delta_{\alpha \alpha'} \cdot (2\pi \ii) \cdot \left( \delta_{x_c}(x-x') - \frac{1}{L} \right) , 
    \qquad 
    \delta_{x_c}(x) = \frac{x_c}{\pi} \frac{1}{x^2 + x_c^2} \ ,
\end{align}
where the single-branch of $\frac{1}{2\ii}\ln \frac{\ii - z}{\ii + z} = \arctan(z)$ is taken in such a way that $\arctan(0) = 0$.

Anti-commutation between fermion operators (\cref{eq:boson_ferm}) with different flavors is guaranteed by the Klein factors. 
We now verify the anti-commutation between fermion operators within the same flavor. 
First, 
\begin{align} \label{eq:anti-commutation-bosonization}
\psi_{\alpha}(x) \psi_{\alpha}(x')
=& \frac1{2\pi x_c} F_\alpha \cdot e^{-\ii(N_\alpha-P_{\rm bc}/2) \frac{2\pi}{L} x} \cdot F_{\alpha} \cdot
 e^{-\ii(N_\alpha-P_{\rm bc}/2)\frac{2\pi}{L}x'} \cdot
 e^{-\ii\phi_{\alpha}(x)} e^{-\ii \phi_{\alpha}(x')} \nonumber\\
=& \frac1{2\pi x_c} F_{\alpha}^2 \cdot 
    e^{-\ii(N_\alpha-P_{\rm bc}/2) \frac{2\pi}{L} (x+x')} \cdot e^{\ii \frac{2\pi}{L} x} \cdot e^{-\ii\phi_{\alpha}(x)} e^{-\ii \phi_{\alpha}(x')} \ ,
\end{align}
where we have made use of $F_\alpha^\dagger N_\alpha F_\alpha = N_\alpha-1$. 
By the Baker-Hausdorff formula
\begin{equation} \label{eq:Baker-Hausdorff}
    e^{A} e^{B} = e^{A+B + \frac12[A,B]} ,\qquad 
    (\text{provided } [A,[A,B]]=[B,[A,B]]=0) \ ,
\end{equation}
and \cref{eq:boson_comm_phi}, we have 
\begin{equation}
\psi_{\alpha}(x) \psi_{\alpha}(x') = 
\frac1{2\pi x_c} F_{\alpha}^2 \cdot 
    e^{-\ii(N_\alpha-P_{\rm bc}/2) \frac{2\pi}{L} (x+x')} \cdot e^{\ii \frac{\pi}{L} (x+x')} \cdot e^{-\ii(\phi_{\alpha}(x)+\phi_{\alpha}(x'))} 
    \cdot e^{\ii \frac{\pi}2 \sgn_{x_c}(x-x')} \ .
\end{equation}
Notice that the second term in the $[\phi_\alpha(x),\phi_\alpha(x')]$ commutator (\cref{eq:boson_comm_phi}) changes the phase factor $e^{\ii \frac{2\pi}{L}x}$ to $e^{\ii \frac{\pi}{L}(x+x')}$. 
Given $x_c\to 0^+$, the above result immediately leads to $\{\psi_{\alpha}(x),\psi_{\alpha}(x')\}=0$. 
One can similarly verify $\{\psi_{\alpha}(x), \psi_{\alpha}^\dagger(x')\}=0$ for $x\neq x'$.

We then consider the operator $\psi_{\alpha}^\dagger(x) \psi_{\alpha}(x')$ in the  $x' \to x$ limit. 
To simplify the calculation, we first rewrite the fermion operator in a normal ordered (with respect to boson vacuum) form 
\begin{align}
\psi_{\alpha}(x)
=& \frac{F_{\alpha}}{\sqrt{L}} e^{-\ii (N_{\alpha} - P_{\rm bc}/2) \frac{2\pi}{L} x} 
    e^{-\ii\varphi_{\alpha}^\dagger(x)} e^{- \ii \varphi_{\alpha}(x)} 
= \frac{F_{\alpha}}{\sqrt{L}} e^{-\ii (N_{\alpha} - P_{\rm bc}/2) \frac{2\pi}{L} x} :e^{-\ii \phi_{\alpha}(x)}: \ ,
\end{align}
where we have made use of 
$e^{-\ii\varphi_{\alpha}^\dagger(x) - \ii \varphi_\alpha(x)}
= e^{-\ii\varphi_{\alpha}^\dagger(x) } e^{- \ii \varphi_\alpha(x)} e^{\frac12 [\varphi_{\alpha}^\dagger(x),\varphi_\alpha(x)]}$
and $[\varphi_{\alpha}^\dagger(x),\varphi_\alpha(x)] = \ln \frac{2\pi x_c}{L}$. 
Here $:\cdots:$ represents normal ordering with respect to the boson vacuum.
Then we have 
\begin{align}
\psi_{\alpha}^\dagger(x) \psi_{\alpha}(x')
=& \frac1{L} e^{\ii(N_\alpha-P_{\rm bc}/2) \frac{2\pi}{L} x} F_{\alpha}^\dagger F_{\alpha} 
 e^{-\ii(N_\alpha-P_{\rm bc}/2)\frac{2\pi}{L}x'} \cdot
 e^{\ii\varphi^\dagger_{\alpha}(x)} e^{\ii\varphi_{\alpha}(x)} e^{-\ii \varphi^\dagger_{\alpha}(x')} e^{-\ii \varphi_{\alpha}(x') } \nonumber\\
=& \frac1{L} e^{\ii(N_\alpha -P_{\rm bc}/2 ) \frac{2\pi}{L} (x-x')} \cdot 
   e^{\ii\varphi^\dagger_{\alpha}(x)}
   \brak{e^{-\ii \varphi^\dagger_{\alpha}(x')} e^{\ii\varphi_{\alpha}(x)}  \cdot e^{[\varphi_{\alpha}(x),\varphi_{\alpha}^\dagger(x')]} }
   e^{-\ii \varphi_{\alpha}(x') } \nonumber\\
=& \frac1{2\pi \ii} \cdot \frac1{x-x'-\ii x_c} e^{\ii(N_\alpha-P_{\rm bc}/2 + 1/2) \frac{2\pi}{L} (x-x')} \cdot 
   e^{\ii(\varphi^\dagger_{\alpha}(x) - \varphi^\dagger_{\alpha}(x'))} 
   e^{\ii (\varphi_{\alpha}(x) - \varphi_{\alpha}(x')) } \ . 
\end{align}
Taking $x_c\to 0^+$ first and then Taylor-expanding $x-x'$, we obtain
\begin{align} \label{eq:point-splitting}
\psi_{\alpha}^\dagger(x) \psi_{\alpha}(x')
=& \frac1{2\pi \ii } \frac{1}{x-x' - \ii x_c} + \frac{N_{\alpha} + 1/2 - P_{\rm bc}/2}{L}+ \frac1{2\pi} \partial_x \phi_{\alpha}(x) + 
 \frac{x-x'}{4\pi}  \pare{ \ii  : (\partial_x \phi_{\alpha}(x))^2 :  +  \partial_x^2 \phi_{\alpha}(x) }  \nonumber\\
& + \ii (x-x') \frac{N_{\alpha}+1/2 - P_{\rm bc}/2}{L} \partial_x \phi_\alpha(x)
  + \mathcal{O}((x-x')^2)  \ . 
\end{align}
Recall that $\OO(L^{-2})$ terms are also omitted. 
The normal ordered density operator, where the constant term $\frac{1}{2\pi\ii } \frac1{ x-x' -\ii x_c} + \frac{1/2 - P_{\rm bc}/2}{L}$ is removed, is then given by 
\begin{equation} 
: \psi_{\alpha}^\dagger(x) \psi_{\alpha}(x) :
= \frac{1}{2\pi} \partial_x \phi_\alpha(x) + \frac{N_{\alpha}}{L}\ ,
\end{equation}
which is consistent with \cref{eq:def_phi}.

We substitute \cref{eq:point-splitting} into the kinetic energy Hamiltonian \cref{eq:H0-def} and obtain
\begin{equation}
H_0 = \sum_{\alpha} \int \frac{\dd x}{4\pi}
    \pare{
    : (\partial_x \phi_{\alpha}(x))^2 : 
    + \mathcal{O}(L^{-2})
    }\ .
\end{equation}
Integral over full derivative terms, {\it e.g.,} $\partial_x\phi_{\alpha}$ and $\partial_x^2\phi_{\alpha}$, vanishes due to the periodic boundary condition of the boson field. 
The omitted $\mathcal{O}(L^{-2})$ term in the integrand will contribute to an $\mathcal{O}(L^{-1})$ term to the total energy, which is of interest. 
To obtain this term, we consider the vacuum $\ket{\vec{N}}_0$ defined in \cref{eq:boson_vacuum}. 
Since the operator $:(\partial_{x} \phi_{\alpha})^2:$ kills $\ket{\vec{N}}_0$, the $\mathcal{O}(L^{-1})$ term determines the energy of  $\ket{\vec{N}}_0$, which can be simply counted as 
$\sum_{\alpha} \sum_{n=1}^{N_{\alpha}}  \frac{2\pi}{L} (n - P_{\rm bc}/2 ) = \frac{2\pi}{L} \frac{N_{\alpha} (N_{\alpha} + 1 - P_{\rm bc})}2 $. 
Therefore, we conclude that the kinetic energy Hamiltonian is 
\begin{align}  \label{eq:boson_H0}
    H_0 &= \sum_{\alpha} \int \frac{\mrm{d}x}{4\pi} : \big(\partial_x \phi_{\alpha}(x) \big)^2:+ \sum_{\alpha} \DL \frac{N_{\alpha} (N_{\alpha} + 1 - \Pbc)}{2} \nonumber\\
    =& \sum_{\alpha} \sum_{q>0}  q : b^\dagger_{\alpha}(q) b_{\alpha}(q) :  + \sum_{\alpha} \DL \frac{N_{\alpha} (N_{\alpha} + 1 - \Pbc)}{2}\ . 
\end{align}
Note that the $\mathcal{O}(L^{-1})$ term relies on the definition of $\ket{0}_0$, which is chosen to occupy all non-positive levels including zero. 

We remark again that \cref{eq:boson_ferm,eq:boson_comm_dphi,eq:boson_comm_phi,eq:boson_H0,eq:boson_dens} contain $\mcl{O}(L^{-1})$ terms.
We will keep these $\mcl{O}(L^{-1})$ terms when discussing the finite-size spectrum and neglect them otherwise.

\subsection{Phase shift due to \texorpdfstring{$\delta$-potential}{delta-potential}}      
\label{app:boson-phase}
\label{app:boson-phase-phase}
\label{app:boson-phase-gauge}

Following Refs.~\cite{Andrei_1983_Solution, vonDelft_1998_finitesize, Krishnan_2024_kondo}, we discuss and compute the phase shift $\pi \rho$ generated by a $\delta$-function potential of strength $\lambda \cdot 2\pi$. 
For this purpose, it suffices to consider a single-flavor problem, and drop the flavor index $\alpha$ in this subsection. 

\paragraph{Phase shift}
The second-quantized Hamiltonian reads, 
\begin{align}
    H_0 + H_1 &= \int\mrm{d}x :\psi^\dagger(x) (\ii\partial_x) \psi(x) : + ~ \lambda \cdot (2\pi)   :\psi^\dagger(0) \psi(0): \ ,
\end{align}
which in the first-quantized language corresponds to an eigenvalue problem
\begin{align}  
    \Big( \ii \partial_x + \lambda \cdot 2\pi \cdot \delta(0) \Big) \psi(x) &= k \cdot \psi(x) \ . 
\end{align}
It can be solved by the following ansatz (with the normalization factor ignored), 
\begin{align}  \label{eq:phase_ansatz}
    \psi(x) \sim e^{-\ii k x} e^{-\ii \pi\rho} \qquad(\mrm{if}~~x<0)  \qquad\qquad \psi(x) \sim e^{-\ii k x} e^{\ii \pi\rho} \qquad( \mrm{if}~~x>0 )
\end{align}
with the phase shift $\pi\rho$ to be determined. 
But the relation between $\rho$ and $\lambda$ depends on how one regularizes the delta potential at the high-energy end. 

Following the discussions in Ref.~\cite{vonDelft_1998_finitesize}, we regularize the delta potential as
\begin{align}  \label{eq:H1}
    H_1 = (\lambda \cdot 2\pi) \int\mrm{d}x ~ \delta_{x_c}(x) \int\mrm{d}x' ~ \delta_{x_c}(x') : \psi^\dagger(x) \psi(x') : . 
\end{align}
$x_c$ restricts us to such processes where the momenta $k'$ and $k$ of both the incoming and outgoing electrons are individually within a cutoff $\mathcal{O}({x_c}^{-1})$, which can be understood as a zero-range potential in a finite-width band. 
If one were to choose another regularization $H_1 = (\lambda \cdot 2\pi) \int\mrm{d}x ~ \delta_{y_c}(x) : \psi^\dagger(x) \psi(x) :$, which only dictates the momentum difference $k-k'$ to be within $\mathcal{O}({y_c}^{-1})$, and can be understood as a finite-range ($y_c$) potential in an infinite-width band ($y_c \gg x_c$), then the $\rho(\lambda)$ relation would be different. 
We will mainly focus on the first scheme.

Corresponding to the regularization scheme \cref{eq:H1}, the first-quantized eigen-value problem is given by
\begin{align}  \label{eq:H1_1}
    \ii \partial_x \psi(x) + (\lambda \cdot 2\pi) \cdot \delta_{x_c}(x) \int\mrm{d}x' ~ \delta_{x_c}(x') ~ \psi(x') &= k \cdot \psi(x) \ . 
\end{align}
For electrons far below the cutoff, $k x_c \to 0^+$, using the ansatz \cref{eq:phase_ansatz}, $\int\mrm{d}x' ~ \delta(x') \psi(x') = \frac{\psi(0^-) + \psi(0^+)}{2}$ is an average of $\psi(0^-)$ and $\psi(0^+)$. 
Therefore, by further integrating \cref{eq:H1_1} over an infinitesimal region containing $x=0$, we can solve the phase shift
\begin{align}  \label{eq:phase-shift-lambda}
    \ii \Big( \psi(0^+) - \psi(0^-) \Big) + (\lambda \cdot 2\pi) \frac{\psi(0^+) + \psi(0^-)}{2} = 0  \qquad 
    \Longrightarrow \qquad \rho=\frac{\arctan(\lambda \pi)}{\pi} \in \left( -\frac{1}{2} , \frac{1}{2} \right) \ . 
\end{align}
In the second regularization scheme, we can similarly derive an eigen-equation, i.e.,  
$ \ii \partial_x \psi(x) + (\lambda\cdot2\pi) \delta_{y_c}(x) \psi(x) = k \cdot \psi(x) $, and obtain the phase shift $\rho = \lambda$. 

In terms of the finite-size spectrum, the phase shift also manifests as a global shift of all single-electron levels. 
Specifically, if one fixes $\psi(-\frac{L}{2}) = e^{-\ii \pi \Pbc} \cdot \psi(\frac{L}{2})$, then the momentum $k$ in \cref{eq:phase_ansatz}, which is also the energy, must be quantized into
\begin{align}  \label{eq:phase_shift}
    k \in \DL \Big( \mbb{Z} - \frac{\Pbc}{2} + \rho \Big)  \ . 
\end{align}
Therefore, if one gradually turns on $\lambda$, all the electron levels, which are equally spaced by $\DL$, will be shifted upward together by an amount of $\rho \cdot \DL$. 
The maximal shift is equal to {\it half} of the level spacing, and is only achieved when $\lambda \to \infty$. 

Due to our regularization scheme \cref{eq:H1}, we should not bosonize $H_1$ by directly applying the point-splitting in \cref{eq:point-splitting,eq:boson_dens}, because the latter relies on the order  $\lim_{x'\to x} \lim_{x_c\to0^+}$ of taking limits, whereas \cref{eq:H1} has $|x-x'|\sim x_c$.
Nevertheless, we can still formally write 
\begin{align} \label{eq:H1-tmp}
    H_1 = \lambda' \int\mrm{d}x ~ \delta(x) \partial_x \phi(x) + \lambda'' \DL N
    = \lambda' ~ \partial_x \phi(x) \Big|_{x=0} + \lambda'' \DL N \ , 
\end{align}
where $N$ counts the fermion number. 
$\lambda',\lambda''$ can be directly determined by the phase shift $\rho$ at large distances.
Importantly, this determination does not depend on the regularization of the $\delta$-potential, which further relates $\rho$ to the potential $\lambda$. 
Suppose we were using the second regularization scheme, where the point-splitting in \cref{eq:point-splitting,eq:boson_dens} applies, then there would be $\lambda'=\lambda''=\rho$ with $\rho=\lambda$ being the phase shift.  
As the relation between $\lambda',\lambda''$ and $\rho$ should not depend on the regularization, $\lambda'=\lambda''=\rho$ must also hold for the first regularization scheme except that now $\rho$ is given by \cref{eq:phase-shift-lambda}.
One can verify this statement by examining the phase shift and finite-size spectrum. 
First, viewing $\phi$ as a classical field, the $\delta$-potential generates a kink $\phi(0^+) - \phi(0^-) = - 2\pi \lambda'$ in its solution, corresponding to a phase shift $e^{\ii 2\pi \lambda'}$ in the fermion field, confirming $\lambda'=\rho$ is the phase shift.
Second, according to the discussion above \cref{eq:boson_H0}, the finite-size ground-state energy with a phase shift $\rho$ is 
$\frac{2\pi}{L} \frac{N (  N +1 + 2\rho - P_{\rm bc} ) }2$, which is changed by $\frac{2\pi}{L} N \rho$ compared to the un-shifted spectrum. 
This confirms $\lambda''=\rho$. 
Therefore, in the first regularization scheme there must be
\begin{equation}
    \lambda'=\lambda''=\rho = \frac1{\pi} \arctan(\lambda\pi)\ . 
\end{equation}
Readers may refer to Ref.~\cite{Kotliar_toulouse_1996, Schiller_phasediagram_2008}) for further discussions.

\paragraph{Gauge transformation canceling the $\delta$-potential}
Due to the above discussions, $H_0 + H_1$ can be readily diagonalized in the original fermion representation, with eigenstates $|G\rangle$ given by the phase-shifted fermion spectrum. 
Now we show that, one can also apply a gauge transformation $U = e^{\ii \rho \phi(0)}$, where $\phi(0)$ is the bosonized field (see \cref{eq:boson_ferm}), so that $\ovl{H} = U H U^\dagger$ reduces to a free Hamiltonian without phase shift. 
The eigenstates of $H$ in the original representation will thus be given by $|G\rangle = U^\dagger |\ovl{G}\rangle$, where $|\ovl{G}\rangle$ is the eigenstate of the free-fermion (free-boson) Hamiltonian $\ovl{H}$. 

By applying the formula
\begin{equation}  \label{eq:BCH}
    e^{A} \cdot B \cdot e^{-A} = B + [A,B] + \frac{1}{2!}[A,[A,B]] + \cdots 
\end{equation}
and \cref{eq:boson_comm_phi}, we obtain
\begin{align}   \label{eq:U_comm_phi}
    U ~ \phi(x) ~ U^\dagger &= \phi(x) - \pi \rho \cdot \sgn_{x_c}(x)  + \frac{2\pi \rho }{L} x  \ ,\\
    \label{eq:U_comm_psi}
    U ~ \psi(x) ~ U^\dagger &= \psi(x) ~ e^{\ii \pi\rho \cdot \sgn_{x_c}(x)} ~ e^{-\ii  \frac{2\pi \rho}{L}x }  \ ,\\
    \label{eq:U_comm_dphi}
    U ~ \partial_x\phi(x) ~ U^\dagger &= \partial_x\phi(x) - 2 \pi \rho \cdot \delta_{x_c}(x)  + \frac{2\pi\rho }{L} \ . 
\end{align}
Applying this gauge transformation to the kinetic energy term $\frac1{4\pi}\int \dd x :(\partial_x \phi)^2:$ generates a $\delta$-potential term  of the form $ - \rho \int \dd x ~\delta_{x_c}(x) \partial_x \phi(x)$, which can be used to cancel the coupling Hamiltonian $H_1$. 
Thus, we expect that $U (H_0 + H_1 ) U^\dagger$ is a free theory. 

However, as the finite-size spectrum is of concern, one cannot naively apply the gauge transformation (\cref{eq:U_comm_dphi}) in real space, which, as well as the commutator \cref{eq:boson_comm_phi},  are derived by taking the $L\to \infty$ limit before $x_c\to 0^+$. 
As explained in Ref.~\cite{von_delft_bosonization_1998} (see discussions around its Eq.~(45)), the two limits do not commute. 
To obtain the {\it exact} form of $U (H_0+H_1) U^\dagger$, we work in momentum space and keep its {\it exact} dependencies on $x_c$ and $L$ until the end. 
Since 
\begin{equation}
    \phi(0) = \sum_{q>0} - \sqrt{\frac{2\pi}{qL}} (   b(q) + b^\dagger(q) ) e^{-x_cq/2}\ ,\qquad 
    [\phi(0), b(q)] = \sqrt{\frac{2\pi}{qL}} e^{-x_cq/2} \ , 
\end{equation}
\cref{eq:BCH} implies the {\it exact} transformation
\begin{align}
    U ~ b(q) ~ U^\dagger = b(q) + \ii \rho \sqrt{\frac{2\pi}{qL}} e^{-\frac{x_cq}{2}}\ .
\end{align}
It follows
\begin{align}
    U \left( \sum_q q ~ b^\dagger(q) b(q) \right) U^\dagger &= \sum_q q \left(  b^\dagger(q) - \ii \rho \sqrt{\frac{2\pi}{qL}} e^{-\frac{x_cq}{2}} \right) \left(  b(q) + \ii \rho \sqrt{\frac{2\pi}{qL}} e^{-\frac{x_cq}{2}} \right) \\\nonumber
    &= \sum_{q>0} q ~ b^\dagger(q) b(q) + \left( \rho \sum_{q>0} \ii q \sqrt{\frac{2\pi}{qL}}  b^\dagger(q) e^{-\frac{x_cq}{2}} + \mrm{H.c.} \right) + \frac{2\pi}{L} \sum_{q>0} \rho^2 e^{-x_c q} \\\nonumber
    &= \sum_{q>0} q ~ b^\dagger(q) b(q) - \rho ~ \partial_x \phi(x) \Big|_{x=0} + \frac{2\pi}{L} \rho^2 \frac{e^{-x_c \DL}}{1 - e^{-x_c \DL}} \\\nonumber
    &= \sum_{q>0} q ~ b^\dagger(q) b(q) - \rho ~ \partial_x \phi(x) \Big|_{x=0} + \frac{\rho^2}{x_c} \left( 1 - \frac{\pi}{L} x_c \right) + \frac{2\pi}{L}\mcl{O}(x_c L^{-1})\ . 
\end{align}
Since $U$ commutes with the electron number operators $N$, for $H_0$ in \cref{eq:boson_H0} (with $\alpha=1$), we have 
\begin{align}
    U H_0 U^\dagger = H_0 - \rho ~ \partial_x \phi(x)\Big|_{x=0} + \frac{\rho^2}{x_c} \left( 1 - \frac{\pi}{L} x_c \right) + \mcl{O}(x_c L^{-2}) \ . 
\end{align}
Let $H_1 = \rho ~ \partial_x\phi(x) \big|_{x=0} + \rho \DL N$ (\cref{eq:H1-tmp}).
Then, since 
{\small
\begin{equation}
[\partial_x\phi(x) \big|_{x=0} , \phi(0)] = \sum_{q>0} \frac{2\pi}{qL} 
    \brak{ -\ii q b(q) + \ii q b^+(q), b(q) + b^\dagger(q) } e^{-x_cq }
= -2\ii \frac{2\pi}{L} \frac{e^{-x_c\frac{2\pi}{L}}}{1 - e^{-x_c\frac{2\pi}{L}}}
= -2\ii \frac{1}{x_c} \pare{ 1 - \frac{\pi}{L}x_c} + \mathcal{O}(x_cL^{-2})\ ,
\end{equation}}
\cref{eq:BCH} leads to 
\begin{align}
    U H_1 U^\dagger =  H_1 - 2 \frac{\rho^2}{x_c}\left(1 - \frac{\pi}{L} x_c \right) + \mcl{O}(x_c L^{-2}) \ . 
\end{align}
To conclude, 
\begin{align} \label{eq:Hbar-single-flavor}
    U (H_0 + H_1) U^\dagger = \sum_{q>0} q ~ b^\dagger(q) b(q) + \DL \frac{N(N+1-\Pbc)}{2} + \rho \DL N - \frac{\rho^2}{x_c} \left( 1 - \frac{\pi}{L} x_c \right)
    + \mathcal{O}(x_c L^{-2}) \ . 
\end{align}

\cref{eq:Hbar-single-flavor} suggests that the ground state energy (for $N=0$)  is changed by $\Delta E= - \frac{\rho^2}{x_c} \pare{1- \frac{\pi}{L}x_c}$ due to the $\delta$-potential. 
The non-divergent energy change is $\frac{\pi}{L} \rho^2$.
We can reproduce this result from the fermion side using a much simpler argument. 
Consider the potential $H_1 = 2\pi \rho \cdot \int \dd x~\delta(x) :\psi^\dagger(x) \psi(x):$ with the second regularization such that it generates the correct phase shift $\rho \in(-\frac12,\frac12)$. 
At the single-particle level, $\rho$ shifts the level $k=\frac{2\pi}{L}(n-\frac{P_{\rm bc}}2)$ to $k=\frac{2\pi}{L}(n + \rho-\frac{P_{\rm bc}}2)$, where $n \in \mbb{Z}$. 
For simplicity, here we assume $\rho$ does not cause a level crossing, i.e., $\rho < \frac{P_{\rm bc}}2$. 
To sum all the energy levels, we introduce an energy truncation factor $e^{-|k|\frac{L}{2\pi}\alpha}$ ($\alpha\to 0^+$) for each level:
\begin{align} \label{eq:level-sum-regularization}
E(\rho)= \frac{2\pi}{L} \sum_{n\le 0} \pare{ n + \rho - \frac{P_{\rm bc}}2 } e^{ ( n + \rho - \frac{P_{\rm bc}}2)\alpha }
\;  \stackrel{\alpha\to 0^+}{=} \; 
\frac{2\pi}{L} \pare{ -\frac1{\alpha^2} + \frac1{12} + \frac{\rho-P_{\rm bc}/2}{2} + \frac{(\rho-P_{\rm bc}/2)^2}{2} } + \mathcal{O}(\alpha) \ . 
\end{align}
The energy change due to level-shift is $E(\rho) - E(0) = \frac{2\pi}{L}[\frac{\rho}2 + \frac{\rho^2 - P_{\rm bc}\rho}2]$. 
Even $E_1(\rho)$ and $E_1(0)$ are individually divergent, the difference is finite. 
After subtracting the constant  
\begin{equation}
    2\pi \rho \cdot \inn{0|\psi^\dagger(0)\psi(0)|0}
= \frac{2\pi}{L} \rho \sum_{n\le 0} e^{(n-\frac{P_{\rm bc}}2)\alpha}
= \frac{2\pi}{L} \pare{ \frac{\rho}{\alpha} + \frac{\rho - \rho \cdot P_{\rm bc}}2 } + \mathcal{O}(\alpha)
\end{equation}
due to the normal ordering in $H_1$, we obtain the total energy change 
\begin{equation}
    \Delta E = - \frac{2\pi}{L} \cdot \frac{\rho}{\alpha} +  \frac{\pi}{L} \cdot \rho^2\ .
\end{equation}
Its non-divergent part is the same as the exact result.

\subsection{Correlation functions}  \label{app:boson-corr}

For a free boson Hamiltonian $H_0$, the Green's function of $\phi(x)$ can be directly computed using the mode expansion \cref{eq:def_phi}. 
Specifically, we define the time-evolved (imaginary or real-time) boson fields by the free $H_0$ as $\phi_\alpha(\tau, x) = e^{\tau H_0} \phi_\alpha(x) e^{-\tau H_0}$ and $\phi_\alpha(t, x) = e^{\ii t H_0} \phi_\alpha(x) e^{- \ii t H_0}$, with expansion
\begin{align} 
    \phi_{\alpha}(\tau,x) &= \sum_{q>0} - \sqrt{\frac{2\pi}{qL}} \Big( e^{-q(\ii x+\tau)} b_{\alpha}(q) + e^{q (\ii x+\tau)} b^\dagger_{\alpha}(q) \Big) e^{-\frac{x_c q}{2}} \ , \\
    \phi_{\alpha}(t,x) &= \sum_{q>0} - \sqrt{\frac{2\pi}{qL}} \Big( e^{-\ii q(x+t)} b_{\alpha}(q) + e^{\ii q(x+t)} b^\dagger_{\alpha}(q) \Big) e^{-\frac{x_c q}{2}} \ . 
\end{align}
Because the bath electrons are left-movers, the time-evolved boson fields only depend on $\tau+\ii x$ and $\ii(t+x)$. So will be the correlation functions. 
It can then be calculated that, at zero temperature, for the free bosonic vacuum $|0\rangle$, 
\begin{align}
    \Big\langle \phi(\tau, x) ~ \phi(0, 0) \Big\rangle_0 &= \sum_{q>0} \frac{2\pi}{qL} e^{-q(\ii x+ \tau + x_c)} =^{1)} - \ln\left[1 - e^{-\DL (\tau + \ii x + x_c) } \right] =^{2)} - \ln \left[\frac{2\pi}{L}\big(\tau+ \ii x + x_c \big) \right] \ ,\\
    \Big\langle \phi(0, 0) ~ \phi(\tau, x) \Big\rangle_0 &= \sum_{q>0} \frac{2\pi}{qL} e^{q(\ii x+ \tau - x_c)} =^{1)} - \ln\left[1 - e^{-\DL (-(\tau + \ii x) + x_c)}  \right] =^{2)} - \ln \left[\frac{2\pi}{L}\big( -(\tau  + \ii x) + x_c \big) \right] \ . 
\end{align}
The same expressions apply to the real-time axis by replacing $\tau \to \ii t$. 
Several remarks are associated with the equal marks. 
1) The series expansion $-\ln(1-x) = \sum_{n=1}^{\infty} \frac{x^n}{n}$ is convergent only for $-1\le |x| <1$. For the correlation functions that we will consider in this work, namely, the $\tau$-ordered, the $t$-ordered, and the $t$-retarded correlation functions, the argument always meets the convergence criterion.  
2) The thermodynamic limit $\DL x, \DL t, \DL \tau \to 0$ is taken and $\mathcal{O}{(L^{-2})}$ terms are omitted, which will be our main focus in this paper for evaluating the correlation functions. 

We tabulate the $\tau$-ordered, $t$-ordered correlation functions in below, with the divergence at $x,\tau,t\to0$ subtracted,  
\begin{align}  
    \Big\langle T_{\tau} ~ \phi(\tau,x) ~ \phi(0,0) \Big\rangle_0  - \Big\langle \phi(0,0)^2 \Big\rangle_0 
    & = \ln \frac{x_c}{(\tau + \ii x) \cdot \sgn(\tau) + x_c}  \qquad (T=0^+) \\\nonumber
    \Big\langle T_{t} ~ \phi(t,x) ~ \phi(0,0) \Big\rangle_0  - \Big\langle \phi(0,0)^2 \Big\rangle_0 
    &= \ln \frac{x_c}{\ii (t + x) \cdot \sgn(t) + x_c}  \qquad (T=0^+)\ .
\end{align}
Following Ref.~\cite{von_delft_bosonization_1998} (see its Eq.~(74) and Appendix H2b), the finite-temperature imaginary time function can also be derived as
{\small
\begin{equation} \label{eq:boson_correlation}
\Big\langle T_{\tau} ~ \phi(\tau,x) ~ \phi(0,0) \Big\rangle_0  - \Big\langle \phi(0,0)^2 \Big\rangle_0 
= \ln  \frac{  \sin (\pi T x_c)  }{  \sin\brak{ \pi T (\tau + \ii x) \cdot \sgn(\tau) + \pi T x_c }  }
\overset{x_c\to 0^+}{=}
\ln  \frac{  \pi T x_c  }{  \sin\brak{ \pi T (\tau + \ii x) \cdot \sgn(\tau) + \pi T x_c }  }\ ,
\end{equation}}
where the limit $L\to\infty$ is taken first. 
It is periodic over the interval $\tau \in [-\frac{1}{2T}, \frac{1}{2T} ]$.
Since it reduces to the zero-temperature result in the $T\to 0^+$ limit, the two orders of limits
\begin{equation} \label{eq:two-limits}
\lim_{T\to 0^+} \lim_{x_c\to 0^+} \lim_{L\to \infty},\qquad 
\lim_{x_c\to 0^+}  \lim_{T\to 0^+} \lim_{L\to \infty}
\end{equation}
give the same correlation functions. 

It is also useful to evaluate various correlation functions of the vertex operators $e^{\ii \kappa \phi}$. A two-point correlation can be calculated by exponentiating the boson correlation function, due to the following identity, 
\begin{align}
    \left\langle e^{\ii \kappa \phi(z_2)} ~ e^{-\ii\kappa \phi(z_1)} \right\rangle_{0} 
    &= e^{ \kappa^2 \big( \langle  \phi(z_2) ~ \phi(z_1) \rangle_0 - \left\langle \phi(0)^2 \right\rangle_0 \big) }    \ ,
\end{align}
where $z_{2}, z_1$ can stand for any type of space-time arguments. 
This identity can be proven in two steps. 
First, using the Baker-Hausdorff formula (\cref{eq:Baker-Hausdorff}), the left hand side equals to $\Inn{ e^{\ii \kappa(\phi(z_2) - \phi(z_1))} }_0 \cdot e^{\frac{\kappa^2}2[\phi(z_2), \phi(z_1) ]}$. 
Second, we use the identity 
\begin{equation}
    \Inn{ e^{B} }_0  = \sum_{n=0,2,4\cdots} \frac{1}{(2n)!} \Inn{B^{2n}}_0
    = e^{\frac12 \Inn{B^2}_0}\ ,
\end{equation}
where $B$ is a linear superposition of boson creation and annihilation operators, and we have made use of Wick's theorem $\langle B^{2n} \rangle_0 = \frac{(2n)!}{2^n} \frac{1}{n!} \left(\langle B^2 \rangle_0\right)^n$. 
Then we have 
$\Inn{ e^{\ii \kappa(\phi(z_2) - \phi(z_1)} }_0 \cdot e^{\frac{\kappa^2}2[\phi(z_2), \phi(z_1) ]} = \exp\pare{\kappa^2 \Inn{\phi(z_2) \phi(z_1) - \phi^2(0) }_0}$. 

We tabulate some useful time-ordered correlation functions:
\begin{align}  \label{eq:vertex_corr_2}
\left\langle T_\tau ~ e^{\ii \kappa \phi(\tau,x)} ~ e^{-\ii\kappa \phi(0,0)} \right\rangle_{0} 
    &= \pare{ \frac{  \pi T x_c  }{  \sin\brak{ \pi T (\tau + \ii x) \cdot \sgn(\tau) + \pi T x_c }  }  }^{\kappa^2}  
    \overset{T\to 0^+}{=} \left( \frac{x_c}{(\tau + \ii x) \cdot \sgn(\tau) + x_c} \right)^{\kappa^2}  \ , \nonumber\\
\left\langle T_t ~ e^{\ii \kappa \phi(t,x)} ~ e^{-\ii\kappa \phi(0,0)} \right\rangle_{0} 
    & \overset{T=0^+}{=} \left( \frac{x_c}{\ii (t + x) \cdot \sgn(t) + x_c} \right)^{\kappa^2}  \ , \nonumber\\
\left\langle \left[ e^{\ii \kappa \phi(t,x)} ,  e^{-\ii\kappa \phi(0,0)} \right] \right\rangle_{0} 
    & {=} \left( \frac{ \pi T x_c}{ \sin\brak{ \ii \pi T (t + x) + \pi T x_c} }\right)^{\kappa^2} 
    - \left( \frac{ \pi T x_c}{ \sin\brak{ - \ii \pi T (t + x) + \pi T x_c} }\right)^{\kappa^2} \nonumber\\
    & \overset{T\to 0^+}{=} \pare{ \frac{x_c}{\ii (t+x) + x_c}  }^{\kappa^2}
        - \pare{ \frac{x_c}{-\ii (t+x) + x_c}  }^{\kappa^2} \ .
\end{align}
The notation $\overset{T\to 0^+}{=}$ means that we take the $L\to\infty$ limit  first and then $T \to 0^+$, and the notation $\overset{T= 0^+}{=}$ means that we take $T\to 0^+$ first and then $L\to \infty$, as specified in \cref{eq:two-limits}.

$\Delta = \frac{\kappa^2}{2}$ is defined as the \textit{scaling dimension} of the vertex operator $Q(t,x)=e^{\ii \kappa \phi}$, because upon the rescaling $t = bt'$, $x= b x'$,  $Q(t,x) = b^{-\Delta} Q'(t',x')$, the correlation function remains unchanged, {\it i.e.,} $\Inn{Q(t,x)Q^\dagger(0,0)}_0 = b^{-2\Delta} \Inn{Q'(t',x'),Q^{\prime\dagger}(0,0)}_0 $. 
We dub $[e^{\ii \kappa \phi}] = \frac{\kappa^2}2$, $[x]=[t]=-1$.

More generically, a $2n$-point correlator is given by 
\begin{align}
    \left\langle e^{\ii \kappa_{2n} \phi(z_{2n})} \cdots e^{\ii \kappa_{i} \phi(z_i)} \cdots e^{\ii \kappa_{1} \phi(z_1)} \right\rangle_{0} 
    &= e^{-\frac{1}{2}(\sum_{i}\kappa_i)^2 \langle\phi(0)^2\rangle_0} \cdot e^{- \sum_{i'>i} \kappa_{i'} \kappa_i \left( \langle\phi(z_{i'}) \phi(z_i)\rangle_0 - \langle\phi(0)^2\rangle_0 \right) } \ .
\end{align}
Since $e^{-\langle\phi(0)^2\rangle_0} = \frac{2\pi x_c}{L} \to 0$, the first factor effectively dictates that the correlation function is non-zero only if $\sum_{j} \kappa_j=0$. 
This is a manifestation of the effective $\rm U(1)$ symmetry $\phi \to \phi + \mrm{const}$ in the free boson theory. 
If we use the imaginary-time and specify the time-ordering as $\tau_{2n} > \cdots > \tau_2 > \tau_1$, 
\begin{align}   \label{eq:vertex_corr_2n}
\left\langle T_\tau e^{\ii \kappa_{2n} \phi(z_{2n})} \cdots 
    e^{\ii \kappa_{j} \phi(z_j)} \cdots e^{\ii \kappa_{1} \phi(z_1)} \right\rangle_{0}  
&= \exp\left[ - \sum_{i'>i} \kappa_{i'} \kappa_{i} \ln \left( 
    \frac{\pi T x_c}{ \sin\brak{ \pi T( z_{i'} - z_{i}) + \pi T x_c }} \right) \right]   \nonumber\\
& \overset{T\to 0^+} 
= \prod_{i'>i} \exp\left[ - \sum_{i'>i} \kappa_{i'} \kappa_{i} \ln \left( 
    \frac{ x_c}{ z_{i'} - z_{i} +  x_c } \right) \right]    \ ,
\end{align}
where $z_i = \tau_i + \ii x_i$, provided $\sum_{i=1}^{2n} \kappa_i = 0$.
For the same reason explained after \cref{eq:boson_correlation}, the $\lim_{T\to 0^+}$  and $\lim_{L\to \infty}$ limits commute with each other for general vertex correlation functions. 
\cref{eq:vertex_corr_2n} will be useful in RG calculations and furnishes the Coulomb gas analog. 

The free-fermion correlation function can also be recovered using \cref{eq:vertex_corr_2}:
{\small
\begin{align} \label{eq:free-fermion-propagator}
G(\tau,x) =& 
    -\frac{\sgn(\tau)}{2\pi x_c} \Inn{T_{\tau} e^{-\ii\phi(\tau,x)} e^{\ii \phi(0,0)} }_0
    {=} - \frac{  \sgn(\tau) T }{  2  \sin\brak{ \pi T (\tau + \ii x) \cdot \sgn(\tau) + \pi T x_c }  } 
    \overset{T\to 0^+}{=} -\frac1{2\pi} \cdot \frac1{ \tau + \ii x + x_c \sgn(\tau) } \ , \nonumber\\
G(t,x) \overset{T= 0^+}{=} & -\ii \Inn{ T_t \psi(t,x) \psi^\dagger(0,0) }_0
    = -\ii\frac{\sgn(t)}{2\pi x_c} \Inn{T_{t} e^{-\ii\phi(t,x)} e^{\ii \phi(0,0)} }_0
    = -\frac{\ii}{2\pi} \cdot \frac1{ \ii(t + x) + x_c \sgn(t) } \ , \nonumber\\
G^R(t,x) =  & -\ii \theta(t) \Inn{ \{ \psi(t,x), \psi^\dagger(0,0)\} }_0
    \overset{T\to 0^+}{=} -\ii \frac{\theta(t)}{2\pi} \left( \frac{1}{\ii (t + x) + x_c} +  \frac{1}{-\ii (t + x) + x_c} \right) 
    = -\ii \theta(t) \cdot \delta_{x_c}(t+x)\ . 
\end{align}}
Note that time-ordering of fermion operators introduces a minus sign when two fermion operators are exchanged, whereas this is not the case for bosonic operators.
The overall signs ($-1$, $-\ii$, $-\ii$) for the imaginary-time, real-time, and retarded Green's functions follow the standard conventions. 

\section{The spin-valley quantum impurity}
\label{app:imp}

\subsection{Symmetry}  \label{app:imp-sym}

We discuss the spin-valley symmetries $\SUt_s \times D_{\infty}$ in more detail. 
Origin of these symmetries in the context of MATBG is reviewed in Ref. \cite{wang_2025_solution}. 

For the spin $\SUt_s$ group, we follow the standard notation of SU(2) groups, and define
\begin{align}
    \hat{S}^\nu = \frac{1}{2} \sum_{ls,l's'} f^\dagger_{ls} [\sigma^0]_{l,l'} [\spin^\nu]_{s,s'} f_{l's'} \qquad \qquad (\nu=x,y,z)   \qquad \qquad \mrm{with~eigenvalues~} S^\nu \in \frac{\mbb{Z}}{2} \ . 
\end{align}
We will also denote $\hat{\mbf{S}} = (\hat{S}^x, \hat{S}^y, \hat{S}^z)$. 
The $\SUt_s$ irreps are uniquely labeled by the spin quantum number $j \in \frac{\mbb{Z}_{\ge 0}}{2}$, which is defined from the eigenvalues of $\hat{\mbf{S}}^2 = \sum_{\nu=x,y,z} (\hat{S}^\nu)^2 = j(j+1)$ in the standard way. 
The degeneracy of an irrep $j$ is $\mrm{DEG}_j = 2j+1$, and the degenerate states are distinguished by $S^z = -j,-j+1,\cdots,j-1,j$. 
The direct product of two irreps is given by $j \otimes j' = |j-j'| \oplus |j-j'|+1 \oplus \cdots \oplus |j+j'|$. 
All the irreps of $\SUt_s$ are also self-conjugate (real if $j$ is integer, pseudo-real if half-integer). 

$D_\infty = \Uo_v \rtimes \Zt$ is the valley symmetry group. 
The notation of a dihedral group follows the algebra relation $\sigma^x \cdot e^{\ii \varphi \sigma^z} \cdot \sigma^x = e^{-\ii \varphi \sigma^z}$, where $\sigma^x$ is the generator of $\Zt$ factor. 
We define the following operator that counts the total $\Uo_v$ charge, 
\begin{align}
    \hat{L}^z = \sum_{ls,l's'} f^\dagger_{ls} [\sigma^z]_{l,l'} [\spin^0]_{s,s'} f_{l's'}  \qquad \qquad \mrm{with~eigenvalues~} L^z \in \mbb{Z} \ .
\end{align}
Since $C_2 = \sigma^x \spin^0$ anti-commutes with $\hat{L}^z$, $+L^z$ and $-L^z$ states must be degenerate if $L^z\not=0$. 
If $L^z=0$, on the other hand, then the irrep will be non-degenerate, and must be eigenstate of $C_2$ action. 
Depending on the $C_2$ eigenvalue $\pm1$, we dub these two irreps as $A_1$ and $A_2$, respectively, following the notation of general $D_n$ groups. 
We introduce the notation $L = A_1, A_2, 1,2,3,\cdots$ to uniquely label the irreps of the $D_{\infty}$ group, and degeneracy
\begin{align}
    \mrm{DEG}_L = \begin{cases}
        1,\qquad & L=A_1,A_2 \\
        2,\qquad & L=1,2,3,\cdots
    \end{cases}  \ . 
\end{align}
For $L=1,2,3,\cdots$, degenerate states are labeled by $L^z =\pm L$. 
Direct product of two irreps of $D_{\infty}$ follows
\begin{align}   \label{eq:Dinfty_product}
    & A_1 \otimes A_1 = A_1 \quad\qquad A_1 \otimes A_2 = A_2 \quad\qquad A_1 \otimes L = L \quad\qquad A_2 \otimes L = L \\\nonumber
    & L \otimes L' = |L-L'| \oplus |L+L'| \quad\qquad\qquad L \otimes L = A_1 \oplus A_2 \oplus 2L
\end{align}
where $L \not=0, L'\not=0$, and $L \not= L'$. 
Also, note that all irreps of $D_{\infty}$ are real, hence self-conjugate. 
This can be testified by the Frobenius-Schur indicator, $\mrm{FSI}[L] = \int\mrm{d}g \cdot \chi^{(L)}(g^2)$, where $\chi^{(L)}(g)$ is the character of group element $g \in D_{\infty}$ in the irrep $L$, and $\int\mrm{d}g \cdot 1 = 1$ is the group measure. 
To be specific, $D_{\infty}$ consists of two connected components, $\Uo_v$ and $C_2 \cdot \Uo_v$. 
For the non-degenerate irreps $L=A_1$ and $A_2$, $\chi^{(L)}(g^2)=1$ for all $g \in D_{\infty}$, hence $\mrm{FSI}[L]=1$. 
For the two-fold degenerate irreps, $L=1,2,3,\cdots$, $\chi^{(L)}\left( \left(e^{\ii \theta \sigma^z} \right)^2 \right)= 2 \cos (L\theta)$, hence the $\Uo_v$ component contributes zero to the FSI, while as $(C_2 \cdot e^{\ii \theta \sigma^z})^2 = 1$, hence $\chi^{(L)} \left(\left(C_{2} \cdot e^{\ii \theta \sigma^z} \right)^2\right) = 2$. After the integral, $\mrm{FSI}[L]=1$ as well.

The irreps of $[D_{\infty} \times \SUt_s]/\mathbb{Z}_2$ are labeled by $[L,j]$, as the valley group commutes with spin group. 
All irreps are self-conjugate. 
The total degeneracy of is given by $\mrm{DEG}_{[L,j]} = \mrm{DEG}_{L} \times \mrm{DEG}_{j}$.

\subsection{The anti-Hund's splitting}

We analyze the general form of $H_{\rm AH}$ [\cref{eq:HAH}] allowed by $\SUt_s \times D_{\infty}$, and obtain the multiplet energies tabulated in \cref{tab:2e}. 

By \cref{app:imp-sym}, the classification of two-electron states into irreps in \cref{tab:2e} is direct. 
There are three independent multiplets $S,D$ and $T$. 
As an overall energy variation of all three multiplets can be absorbed into $U$, it suffices to parametrize $H_{\rm AH}$ with the energy splitting of $S$ and $D$ from $T$, 
\begin{align} \label{eq:Himp_1}
    H_{\rm AH} &= - J_S \frac{f^\dagger_{+\uparrow} f^\dagger_{-\downarrow} - f^\dagger_{+\downarrow} f^\dagger_{-\uparrow}}{\sqrt{2}} \frac{f_{-\downarrow} f_{+\uparrow} - f_{-\uparrow} f_{+\downarrow}}{\sqrt{2}}  - J_D  \sum_{l} f^\dagger_{l\uparrow} f^\dagger_{l\downarrow} f_{l\downarrow} f_{l\uparrow}  \\\nonumber 
    &= - \frac{J_S}{4} \sum_{ll's} f_{ls}^\dagger f_{\bar l \bar s}^\dagger f_{\bar l' \bar s} f_{l's}   - \frac{J_D}{2} \sum_{ls} f_{ls}^\dagger f_{l\bar s}^\dagger f_{l\bar s} f_{ls} \\\nonumber
    &= -\frac{1}{2} \sum_{ss'} \sum_{l_1l_1'l_2l_2'} f^\dagger_{l_1 s} f^\dagger_{l_1' s'}  \begin{bmatrix}
        J_D & 0 & 0 & 0 \\
        0 & \frac{J_S}{2} & \frac{J_S}{2} & 0 \\
        0 & \frac{J_S}{2} & \frac{J_S}{2} & 0 \\
        0 & 0 & 0 & J_D \\
    \end{bmatrix}_{l_1'l_1,l_2'l_2} f_{l_2's'} f_{l_2 s} \ . 
\end{align}
The two terms $J_S$ and $J_D$ in the 1st line are the projectors to $S$ and $D$, respectively, and reorganizing the 1st term leads to \cref{eq:HAH}. 
Symmetrizing the spin indices leads to the 2nd line. 
In the 3rd line, we sort $(l'l) = (++),(+-),(-+),(--)$. 
To see that the 3rd line equals the 2nd line, simply note that the $s=s'$ matrix elements in the 3rd line will be canceled after imposing fermion anti-parity. The $s=\ovl{s'}$ elements recover the 2nd line. 

The energy of the two-electron states is direct to read off, as \cref{eq:HAH} is already in the projector form. 
To evaluate the energy of three and four-electron states, we re-organize $H_{\rm imp}$ as
\begin{align} \label{eq:Himp}
H_{\rm imp} &= \epsilon_f \hat{N} 
+  \pare{ U - \frac14 J_S} \frac{\hat{N}(\hat{N}-1)}2
+  J_S \cdot \hat{\mathbf{S}}_+ \cdot \hat{\mathbf{S}}_-
- \pare{J_D - \frac14 J_S} \sum_l \hat{N}_{l\uparrow} \hat{N}_{l\downarrow}  \ . 
\end{align}
Here, $\hat{S}^{\nu}_l = \frac{1}{2} \sum_{ss'} f^\dagger_{ls} [\spin^\nu]_{s,s'} f_{ls'}$ is the spin operator in valley-$l$, and $\hat{\mbf{S}}_l = (\hat{S}^x_l, \hat{S}^y_l, \hat{S}^z_l)$. 
$\hat{N}_{ls} = f_{ls}^\dagger f_{l s}$. 
Notice that all the three-electron states (similarly, the one-electron states) are dictated to be degenerate, as they form the $[L,j] = [1,\frac{1}{2}]$ irrep. 
It hence suffices to work out the energy for one Fock state that is eigenstate of all $N_{ls}$. 
The four-electron state is also eigenstate of all $N_{ls}$. 
At least one valley is fully filled, hence forming spin-singlet, and the corresponding $\hat{\mathbf{S}}_l$ operator vanishes. 
It then suffices to compute the energies relying on $N_{ls}$. 
The result is given in caption of \cref{tab:2e}.

\subsection{The Kondo couplings}

\begin{table}[tb]
    \centering
    \begin{tabular}{c|c|c}
    \hline
        irrep $[L,j]$ & $\mrm{DEG}_{[L,j]}$ & basis \\
    \hline
        $[A_1, 0]$ & 1 & $\sigma^0 \spin^0$ \\
        $[A_2, 0]$ & 1 & $\sigma^z \spin^0$ \\
        $[2, 0]$ & 2 & $\sigma^{x,y} \spin^0$ \\
        $[A_1, 1]$ & 3 & $\sigma^0 \spin^{x,y,z}$ \\
        $[A_2, 1]$ & 3 & $\sigma^z \spin^{x,y,z}$ \\
        $[2, 1]$ & 6 & $\sigma^{x,y} \spin^{x,y,z}$ \\
    \hline
    \end{tabular}
    \caption{Hermitian bilinear bath operators classified into irreps of $[L,j]$. }
    \label{tab:bath_oprt}
\end{table}

We find the general form for Kondo couplings allowed by the spin-valley $\SUt_s \times D_{\infty}$ group and $C_2T$ symmetry in this subsection. 
Notice that $C_2T$ symmetry is not exploited elsewhere in this work. 

We first sketch the formal procedures of SW transformation $e^{\ii S}$. 

1) Organize the Hilbert space into the low-energy subspace, which contains exactly two $f$-electrons, and the high-energy subspace, which contains $0,1,3$ or $4$ $f$-electrons. 
The projectors to the two subspaces are denoted as $\PP_2$ and $1 - \PP_2 = \PP_0 + \PP_1 + \PP_3 + \PP_4$, respectively. The subscripts indicate the $f$-electron numbers. 
We may further divide $\PP_2 = \PP_S + \PP_D + \PP_T$, where $\PP_S = |S\rangle \langle S|$, $\PP_D = \sum_{L^z=2,\ovl{2}} |D,L^z\rangle \langle D,L^z|$, and $\PP_T = \sum_{S^z=1,0,\ovl{1}} |T,S^z\rangle \langle T,S^z|$. 
$H_0 + H_{\rm imp}$ is already diagonal in the $\PP_2$ and $1 - \PP_2$ subspaces, while $H_{\rm hyb}$ induces off-diagonal elements between $\PP_2$ and $\PP_3$ and between $\PP_2$ and $\PP_1$. 
The energy ``gap'' between the two subspaces is $\mcl{O}(U)$, while the off-diagonal elements are $\mcl{O}(\Delta_0)$. 

2) To eliminate these off-diagonal elements (perturbatively, in powers of $\frac{\Delta_0}{U}$), we devise such a Hermitian operator $S = \sum_{n=1}^{\infty} S^{(n)}$, where $S^{(n)}$ is of order $\mcl{O}\left( (\frac{\Delta_0}{U})^n \right)$. 
If further assuming that $U$ is much larger than the bath electron band width, the leading order $S^{(1)}$ takes the form of
\begin{align}
    S^{(1)} = \left(\sum_{\Gamma=S,D,T}  A_\Gamma \cdot \PP_3 \Big( \sum_{ls} f^\dagger_{ls} \psi_{ls}(0) \Big) \PP_\Gamma + \sum_{\Gamma=S,D,T}  B_\Gamma \cdot \PP_1 \Big( \sum_{ls} \psi^\dagger
    _{ls}(0) f_{ls} \Big) \PP_\Gamma  \right) + \mrm{H.c.} \ ,
\end{align}
where $A_\Gamma, B_{\Gamma}$ are of order $\mcl{O}(\frac{\Delta_0}{U})$ and to be determined. 
$e^{\ii S}$ serves as a slight unitary rotation between the low- and high-energy subspaces. 

3) Compute $\td{H} = e^{\ii S} H e^{-\ii S} = H + [\ii S,H] + \frac{1}{2} [\ii S, [\ii S, H]] + \cdots$ and express each term using the \textit{original} $f$ and $\psi$ operators. 
Unknown parameters in $S$ are determined by requiring that off-diagonal elements vanish, namely, $(1-\PP_2) \td{H} \PP_2 = 0$. 
At the leading order $\mcl{O}(\Delta_0)$, this implies that $(1 - \PP_2) \left( H_{\rm hyb} + [\ii S^{(1)}, H_0 + H_{\rm imp}] \right) \PP_2 = 0$, which fixes $A_\Gamma$ and $B_\Gamma$. 
Here, $\PP_2$ is still defined according to the particle number of $f$-operators; however, after the gauge transformation, this $f$-operator does \textit{not} annihilate a physical electron. 
Instead, the physical electron operator reads $\td{f} = e^{\ii S} f e^{-\ii S} = f + [\ii S^{(1)}, f] + \mcl{O}((\frac{\Delta_0}{U})^2)$. 

4) As $\td{H}$ is now diagonal in the $\PP_2$ and $1 - \PP_2$ subspaces, we simply keep the low-energy one, $\PP_2 \td{H} \PP_2$. 
At the leading $\mcl{O} \left(\frac{\Delta_0^2}{U} \right)$ order, we obtain the Kondo Hamiltonian as $H_{\rm K} = \frac{\ii}{2} \PP_2 [S^{(1)}, H_{\rm hyb}] \PP_2$. 
In general, $H_{\rm K}$ may contain a term that only acts on the impurity; however, this term can be absorbed as a slight shift to the multiplet energies, $E_{\Gamma} \to E_{\Gamma} + \delta E_{\Gamma}$, which are free parameters to begin with. We therefore neglect it. 
Remaining terms in $H_{\rm K}$ will be a coupling between a bilinear operator of bath electrons, and an impurity operator, namely, the Kondo coupling.

The SW transformation carried out for the $\Uf$ symmetric case can be found in previous work \cite{zhou_kondo_2024}. 
We review the result concisely here. $H_{\rm K}$ contains an $\SUf$ moment-moment interaction $\zeta$ (anti-ferromagnetic, $\zeta>0$), and a density-density interaction $\gamma$, 
\begin{align}   \label{eq:HK_U4}
    H_{\rm K} ~ = ~ (2\pi  \zeta) \cdot \sum_{\mu\nu \not= 00} \Theta^{\mu \nu} \cdot \psi^\dagger \sigma^\mu \spin^\nu \psi ~~+~~ (2\pi \gamma) \cdot \PP_2 ~ \cdot~ :\psi^\dagger \sigma^0 \spin^0 \psi: \ .
\end{align}
Here, we have defined the representation of the $\SUf$ generators on the 6 two-electron states as
\begin{align}   \label{eq:Theta}
    \Theta^{\mu \nu}  = \PP_2   \frac{f^\dagger \sigma^{\mu} \spin^{\nu} f }{2} \PP_2 \qquad \qquad \mu\nu \not=00 \ . 
\end{align}
If not specified, bath operators in this section all live at $x=0$. 
$\zeta$ will grow under renormalization, and the system will flow to a Kondo Fermi liquid, where the impurity $\SUf$ moment gets exactly screened by another $\SUf$ moment in the bath. 
Remaining bath electrons sees a $\frac{\pi}{2}$ phase shift at the origin. 

Next we find the general Kondo couplings with $\SUt_{s} \times D_{\infty}$ and $C_2T$, which simply acquires an ``anisotropy'' in the $\SUf$ moment-moment couplings $\zeta$, characterized by 5 independent real-valued parameters, and allows the density-density couplings to $\PP_S$, $\PP_D$, and $\PP_T$ manifolds to be independent. 
The result is summarized in \cref{eq:HK_c}. 

For this sake, a symmetry analysis suffices. 
As $H_{\rm K}$ must be Hermitian, it suffices to separately check the Hermitian impurity operators and the Hermitian bilinear bath operators, and classify them into irreps labeled by $[L,j]$. 
According to the discussions around \cref{eq:Dinfty_product}, if and only if the impurity operators and the bath operators span the same irrep, their tensor product contains an identity irrep ($[L,j] = [A_1, 0]$) that remains invariant under $D_\infty \times \SUt_s$. 
Finally, imposing $C_2 T$ further rules out some choices. 

For the Hermitian bilinear bath operators $\psi^\dagger_{ls} \psi_{l's'}$, which span a $4^2=16$ dimensional Hilbert space, the decomposition is direct. 
As both $\psi^\dagger_{ls}$ and $\psi_{l's'}$ spans the $[1,\frac{1}{2}]$ irrep (all the irreps of $[D_\infty \times \SUt_s]/\mathbb{Z}_2$ are self-conjugate, so we do not need to distinguish the irreps of `bras' from `kets'), the valley part follows $1 \otimes 1 = A_1 \oplus A_2 \oplus 2$ (see \cref{eq:Dinfty_product}), while the spin part follows $\frac{1}{2} \otimes \frac{1}{2} = 0 \oplus 1$. 
The basis operators spanning each irrep are tabulated in \cref{tab:bath_oprt}. 
Crucially, each irrep appears just for once. 

For the impurity operators $|\Xi\rangle \langle \Xi'|$, they span a $6^2=36$ dimensional Hilbert space. 
Both $|\Xi\rangle$ and $\langle \Xi'|$ span a reducible representation $[A_1,0] \oplus [2,0] \oplus [A_2,1]$ (see \cref{tab:2e}). 
To begin with, there are `irrep-diagonal' operators. 
For the `$S$' manifold, $[A_1,0] \otimes [A_1,0] = [A_1, 0]$, and the operator is given by $\PP_S$. 
For the `$D$' manifold, $[2,0] \otimes [2,0] = [A_1,0] \oplus [A_2,0] \oplus [4,0]$, where $[A_1, 0]$ is given by $\PP_D$, and $[A_2, 0]$ is given by 
\begin{align}
    \Theta^{z0} = |D, 2\rangle \langle D, 2| - |D, \ovl{2}\rangle \langle D, \ovl{2}| \ ,
\end{align}
as defined in \cref{eq:Theta}. 
The operators spanning $[4,0]$ are $\Lambda_{\pm}$ defined in \cref{eq:Lambda-def-PK}, which do not belong to \cref{eq:Theta}, and cannot find the corresponding bath bilinear operators to enter the Kondo coupling. 
For the `$T$' manifold, $[A_2, 1] \otimes [A_2, 1] = [A_1, 0] \oplus [A_1, 1] \oplus [A_1, 2]$. 
The $[A_1,0]$ irrep is given by $\PP_T$, while $[A_1, 1]$ is the spin-1 operators $\Theta^{0x,0y,0z}$ in \cref{eq:Theta}. 
The $[A_1,2]$ irrep does not belong to \cref{eq:Theta}, and does not appear in Kondo coupling as well. 

Then, there are `irrep-off-diagonal' operators. Let us take the off-diagonal blocks between $S$ and $D$ manifolds as an example. 
Since there are two blocks that are Hermitian conjugate to each other, $|S\rangle \langle D,L^z|$ and $|D,L^z\rangle \langle S|$, the irrep $[A_1,0] \otimes [2,0] = [2, 0]$ appears twice. 
We can Hermitize the basis for the two irreps as following. One is 
\begin{align}
    \Theta^{x0} =  \frac{|S\rangle \langle D,2| + |S\rangle \langle D,\ovl{2}|}{\sqrt{2}} + \mrm{H.c.} \qquad \qquad 
    \Theta^{y0} = \ii \frac{|S\rangle \langle D,2| - |S\rangle \langle D,\ovl{2}|}{\sqrt{2}} + \mrm{H.c.} \ ,
\end{align}
which follows the definition of \cref{eq:Theta}, and the other is
\begin{align}
    \Phi^{x0} = \ii \frac{|S\rangle \langle D,2| + |S\rangle \langle D,\ovl{2}|}{\sqrt{2}} + \mrm{H.c.} \qquad \qquad 
    \Phi^{y0} = - \frac{|S\rangle \langle D,2| - |S\rangle \langle D,\ovl{2}|}{\sqrt{2}} + \mrm{H.c.}\ ,
\end{align}
which does not belong to \cref{eq:Theta}. 
It can be directly verified that (as $\SUt_s$ actions are all trivial here, they are not listed)
\begin{align}
    e^{\ii \theta \sigma^z \spin^0} \cdot ( \Theta^{x0}, \Theta^{y0} ) \cdot e^{-\ii \theta \sigma^z \spin^0} &= ( \Theta^{x0} , \Theta^{y0} ) \begin{bmatrix}
        \cos(2\theta) &  \sin(2\theta)\\
        -\sin(2\theta) & \cos(2\theta) \\
    \end{bmatrix} \qquad 
    C_2 \cdot ( \Theta^{x0}, \Theta^{y0} ) \cdot C_2 = ( \Theta^{x0} , -\Theta^{y0} ) 
\end{align}
and $(\Phi^{x0}, \Phi^{y0})$ and the bath operators  $(\sigma^x \spin^0, \sigma^y \spin^0)$ transform in the same way. 
Therefore, the coupling of $(\sigma^x \spin^0, \sigma^y \spin^0)$ to $(\Theta^{x0}, \Theta^{y0})$ and $(\Phi^{x0}, \Phi^{y0})$ are both allowed by $D_\infty \times \SUt_s$. 
However, they transform differently under $C_2 T$: $(C_2T) (\Theta^{x0}, \Theta^{y0}) (C_2T)^{-1} = (\Theta^{x0}, -\Theta^{y0})$, while $(C_2T) (\Phi^{x0}, \Phi^{y0}) (C_2T)^{-1} = (-\Phi^{x0}, \Phi^{y0})$, while the bath operator behaves as $(C_2T) (\sigma^x \spin^0, \sigma^y \spin^0) (C_2T)^{-1} = (\sigma^x \spin^0, - \sigma^y \spin^0)$. 
Therefore, the bath operator can only couple to $\Theta^{x0,y0}$. 

The same analysis also applies to the other off-diagonal blocks. 
Between $S$ and $T$, there are two $[A_1,0] \otimes [A_2,1] = [A_2,1]$ irreps, while only the one spanned by $\Theta^{zx,zy,zz}$ is allowed by $C_2 T$ to couple to the bath $(\sigma^z \spin^x, \sigma^z \spin^y, \sigma^z \spin^z)$. 
Between $D$ and $T$, there are two $[2,0] \otimes [A_2,1] = [2,1]$ irreps, while only the one spanned by $\Theta^{xx,xy,xz,yx,yy,yz}$ is allowed by $C_2 T$ to couple to the bath $(\sigma^x \spin^x, \sigma^x \spin^y, \sigma^x \spin^z, \sigma^y \spin^x, \sigma^y \spin^y, \sigma^y \spin^z)$. 

In sum, we obtain \cref{tab:HK}. 
In particular, for the moment-moment couplings, since the $\SUf$ breaking effect is a perturbation ($J_{S,D} \ll U$, hence $\zeta_{\mu\nu} \sim \frac{\Delta_0^2}{U + \mcl{O}(J_{S,D})}$ has the same sign as $\frac{\Delta_0^2}{U}$), we can also expect the signs of the coupling constants to follow the $\Uf$ symmetric case, being anti-ferromagnetic. 

At PHS, $:\psi^\dagger \sigma^0 \spin^0 \psi:$ acquires a minus sign under charge conjugation, yet $\PP_\Gamma$ ($\Gamma=S,D,T$) does not. 
Therefore, the density-density coupling will be forbidden so that all $\gamma_\Gamma=0$. 
Since the density-density coupling is in general not relevant under RG, we will take the advantage of assuming a PHS to ignore it. 

\begin{table}[tb]
    \centering
    \begin{tabular}{c|c|c|c|c|c}
    \hline
        & irrep & $\mrm{DEG}_{[L,j]}$ & impurity operator & coupled to bath bilinear operator & coupling constant \\
    \hline
        Within $S$ & $[A_1, 0]$ & 1 & $\PP_S$ & $\sigma^0 \spin^0$ & $\gamma_S$ \\
        Within $D$ & $[A_1, 0]$ & 1 & $\PP_D$ & $\sigma^0 \spin^0$ & $\gamma_D$ \\
        & $[A_2, 0]$ & 1 & $\Theta^{z0}$ & $\sigma^z \spin^0$ & $\lambda_z$ \\
        & $[4, 0]$ & 2 & $\Lambda_{x,y}$ & $-$ & $-$ \\
        Within $T$ & $[A_1, 0]$ & 1 & $\PP_T$ & $\sigma^{0} \spin^{0}$ & $\gamma_T$ \\
        & $[A_1, 1]$ & 3 & $\Theta^{0x,0y,0z}$ & $\sigma^{0} \spin^{x,y,z}$ & $\zeta_{0z}$ \\
        & $[A_1, 2]$ & 5 & $|T,1\rangle \langle T,\ovl{1}|$, etc & $-$ & $-$ \\
        Between $S,D$ & $[2,0]$ & 2 & $\Theta^{x0,y0}$ & $\sigma^{x,y} \spin^{0}$ & $\zeta_{x}$ \\
        Between $S,T$ & $[A_2,1]$ & 3 & $\Theta^{zx,zy,zz}$ & $\sigma^z \spin^{x,y,z}$ & $\zeta_{zz}$ \\
        Between $D,T$ & $[2,1]$ & 6 & $\Theta^{xx,xy,xz,yx,yy,yz}$ & $\sigma^{x,y} \spin^{x,y,z}$ & $\zeta_{xz}$ \\
    \hline
    \end{tabular}
    \caption{Kondo couplings that are allowed by the $D_{\infty} \times \SUt_s$ group and $C_2T$ symmetries. 
    We also denote $\Theta^{z0}$ as $\Lambda_z$. 
    The couplings $\gamma_S$, $\gamma_D$, $\gamma_T$ break the particle-hole symmetry and are in general not relevant in the low-energy physics. 
    }
    \label{tab:HK}
\end{table}

\section{Susceptibility in the AD phase}   \label{app:exactAD-sus}

The transverse correlation function at the AD fixed point in time domain is derived in \cref{eq:chix_AD}, 
\begin{align}
\chi_x(\tau) 
    \overset{T\to 0^+}{=} - \frac{1}{2} \left( \frac{x_c}{|\tau| + x_c} \right)^{16 \rho_z^2}  \ .
\end{align}
In particular, for $\rho_z<\rho_z^\star=\frac{1}{4}$, where $\rho_z^\star$ is another solvable fixed point, the power $\alpha=16 \rho_z^2<1$.  

\begin{figure}
\centering
\includegraphics[width=0.8\linewidth]{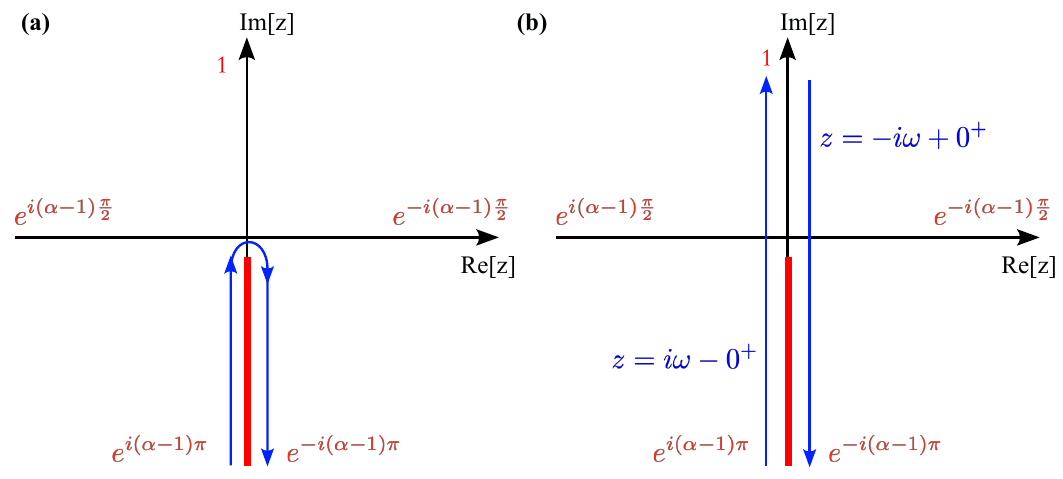}
\caption{Contour integral about $f(z)=(-\ii z + 0^+)^{\alpha-1}$. The bold red line represents the branch-cut of $f(z)$. The red numbers $1$, $e^{\pm (\alpha-1) \frac{\pi}2}$, $e^{\pm (\alpha-1)\pi}$ represent $f(z)/|f(z)|$ in at $z=\ii y$, $\mp x$, $\mp 0^+ -\ii y$, respectively, where $x,y>0$. 
}
\label{fig:contour}
\end{figure}

The transverse susceptibility in the real-frequency domain is defined as $\chi^R_x(\omega) = \int_{-\infty}^{\infty}\mrm{d}t ~ \chi_x^R(t) ~ e^{\ii \omega t} $. 
In numerical calculations such as Numerical Renormalization Group, $\Im [\chi^R_x(\omega)]$ can be computed using the Lehmann spectral representation, where only eigenstates with the energy $\omega$ contribute. 
Thus, $\Im [\chi^R_x(\omega)]$ is important to characterize the low-energy physics. 
Here we first construct a Mastubara $\chi_x(\ii \omega)$ that reproduces $\chi_x(\tau)= -\frac12 \abs{ \frac{x_c}{\tau} }^{\alpha}$ for $|\tau|\gg x_c$, and then derive $\chi^R_x(\omega)$ by analytical continuation. 
We do not concern ourselves with short-time behaviors at $\tau\sim x_c$. 
Consider the integral 
\begin{equation}
    I(\tau) = \lim_{\varepsilon\to 0^+} \int_{-\infty}^{\infty} \frac{\dd \omega}{2\pi}  ~ e^{-\ii \omega \tau} ( -\ii \omega + \varepsilon)^{\alpha-1}\ .
\end{equation}
We introduce the function
\begin{equation}
f(z)=(-\ii z + \varepsilon)^{\alpha-1} 
\end{equation}
and choose it to be analytical in the upper complex $z$-plane ($\Im[z]\ge 0$). 
As shown in \cref{fig:contour}, $f(z)$ has a branch-cut at $z = -\ii y$ ($y\ge \varepsilon$). 
When $\tau<0$, we change the integral in $I(\tau)$ to a contour integral enclosing the upper $z$-plane. Since $f(z)$ is analytical there, $I(\tau)=0$ for $\tau<0$. 
When $\tau>0$, we change integral to a contour integral enclosing the lower $z$-plane, where a branch-cut lies. 
We deform the contour to approach the branch-cut, as illustrated by the blue line in \cref{fig:contour}(a). 
Then we have 
\begin{equation}
I(\tau) = - \frac{\ii}{2\pi} \theta(\tau) \int_\varepsilon^{\infty} d y ~ e^{-y\tau} \pare{ f(\varepsilon-\ii y) - f(-\varepsilon-\ii y) } + \mathcal{O}(\varepsilon f(\varepsilon)) \ ,
\end{equation}
where $\mathcal{O}(\varepsilon f(\varepsilon)) = \mathcal{O}(\varepsilon^\alpha)$ vanishes in the $\varepsilon\to 0^+$ limit as long as $\alpha>0$. 
According to the definition of the branch of $f(\omega)$, we have 
$f(\pm \varepsilon-\ii y) / |f(\pm \varepsilon-\ii y)| = e^{\mp\ii(\alpha-1)\pi}$ for $y\gg \varepsilon$.
Thus, 
\begin{equation}
I(\tau) = - \frac{\ii}{2\pi} \theta(\tau) \lim_{\varepsilon\to 0^+} \int_{\varepsilon}^\infty\dd y ~ e^{-y\tau}  ( (y-\varepsilon)^2 + \varepsilon^2) ^{\frac{\alpha-1}2} \pare{  e^{-\ii \pi (\alpha-1)} - e^{\ii \pi (\alpha-1)} } = \frac{ \sin((1-\alpha)\pi)}{\pi} \frac{\theta(\tau)}{|\tau|^{\alpha}} \Gamma(\alpha)  \ ,
\end{equation}
with $\Gamma(\alpha)$ being the $\Gamma$-function. 
Therefore, the imaginary-time correlation function is reproduced as 
\begin{equation}
    \chi_x(\tau) = -\frac12 \cdot \frac{\pi x_c^\alpha}{\sin((1-\alpha)\pi) \cdot \Gamma(\alpha)} 
    \cdot  \pare{ I(\tau) + I(-\tau) }\ .
\end{equation}
Correspondingly, the Matsubara Green's function is 
\begin{equation}
\chi_x(\ii \omega) = -\frac12 \cdot \frac{\pi x_c^\alpha}{\sin((1-\alpha)\pi) \cdot \Gamma(\alpha)} \pare{ (-\ii \omega + 0^+)^{\alpha-1} +  (\ii \omega + 0^+)^{\alpha-1} }\ .
\end{equation}
$\chi_x(\ii \omega)$ is real and even in $\omega$, as required by the Lehmann spectral representation of bosonic Matsubara Green's function. 
The retarded Green's function can be obtained by analytic continuation ($\ii \omega \to \omega + \ii 0^+$) : 
\begin{align}
\chi^{R}_x(\omega) =& -\frac12 \cdot \frac{\pi\cdot x_c^\alpha} {\sin((1-\alpha)\pi) \cdot \Gamma(\alpha)}
     \pare{ (-\omega -\ii 0^+ )^{\alpha-1} +  ( \omega + \ii 0^+)^{\alpha-1} } \nonumber\\
=& -\frac12 \cdot \frac{\pi\cdot x_c^\alpha} {\sin((1-\alpha)\pi) \cdot \Gamma(\alpha)}
    \cdot |\omega|^{\alpha-1} \cdot \pare{ 1 + e^{\ii \pi (1-\alpha) \sgn(\omega) } }\ . 
\end{align}
One should interpret $(-\omega - \ii 0^+)^{\alpha-1}$ and $(\omega + \ii 0^+)^{\alpha-1}$ as $f(z=-\ii \omega + 0^+)$ and $f(z=\ii \omega - 0^+)$, respectively (\cref{fig:contour}(b)). 
Its imaginary part is 
\begin{equation} \label{eq:chiR-xx-omega}
    \Im[\chi^R_x(\omega)] =  - \frac{x_c^{\alpha} |\omega|^{\alpha-1} \sgn(\omega)}{2\Gamma(\alpha)} \ . 
\end{equation}
It satisfies $\Im[\chi^R_x(\omega>0)] <0$ and $\Im[\chi^R_x(\omega)] =- \Im[\chi^R_x(-\omega)] $, as required by the Lehmann spectral representation.

In a practical numerical calculation, high-energy peaks may appear in $\Im[\chi^R_x(\omega)]$, which, through the Kramers-Kronig relation, will lead to a smooth background to $\Re[\chi^R_x(\omega)]$ for low-energy $\omega$. 
As a result, while the low-energy behavior of $\Im[\chi^R_x(\omega)]$ is universal, that of $\Re[\chi^R_x(\omega)]$ is not.

\section{Evaluating \texorpdfstring{$\delta E'[\Omega]$}{dE}}   \label{app:dE}

We now calculate $\delta E'[\Omega] = \sum_{n=-N}^0 \frac{2\pi}{L} \delta_n $, which is needed for the finite-size many-body spectrum at the $\rho_z= \frac14$ fixed line (\cref{sec:finite-size-spectrum}).  
Introducing an energy cutoff $D$ (bandwidth), we take $N \approx \frac{L}{2\pi} D$. 
According to \cref{eq:delta-n-def-explicit},  
\begin{align}
\frac{2\pi}{L} \delta_n 
=& \frac{2}{L} \arctan\frac{\pi \Gamma}{\frac{2\pi}{L} (n-\frac{1}{2}) - \varepsilon_f - 0^+} 
    - \frac{2\pi}{L} \cdot \theta\pare{ \frac{2\pi}{L}\brak{n-\frac12} > \varepsilon_f } \nonumber\\
& - \pare{\frac{2\pi}{L}}^2 \delta_n \cdot 
    \pare{ \frac{\Gamma \cdot }{\big(\pi \Gamma \big)^2 + \big(\frac{2\pi}{L} (n-\frac{1}{2}) - \varepsilon_f \big)^2} 
    + \delta\pare{  \frac{2\pi}{L}\brak{n-\frac12}- \varepsilon_f  } }
+ \mcl{O}\left( L^{-3} \right) \ . 
\end{align}
The first term in the second row is of the order $\mathcal{O}(L^{-2})$. 
After summing over $N\sim L$ terms, they contribute to an $\mathcal{O}(L^{-1})$ term to $\delta E'[\Omega]$.  
However, the contribution to $\delta E'[\Omega]$ from the $\varepsilon_f$-dependent part of the these terms is of the order $\mathcal{O}(L^{-2})$. 
Since our focus is on the $\varepsilon_f$-dependent $\mathcal{O}(L^{-1})$ terms in $\delta E'[\Omega]$, and constant $\mathcal{O}(L^{-1})$ terms are irrelevant, we can neglect the second row of the above equation. 
We can replace the first row in the above equation by the integral $\int_{n-1}^{n} \dd x ~f(x) + \mathcal{O}(f''(n-\frac12))$, where
\begin{equation}
    f(x) = \frac{2}{L} \arctan \frac{\pi\Gamma}{\frac{2\pi}{L}x -\varepsilon_f }
    - \frac{2\pi}{L} \theta \pare{\frac{2\pi}{L}x-\varepsilon_f } \ .
\end{equation}
The $\mathcal{O}(f''(n-\frac12))$ term is of the order $\mathcal{O}(L^{-3})$ and eventually leads to an $\mathcal{O}(L^{-2})$ term in $\delta E'[\Omega]$. 
We hence can safely omit the $\mathcal{O}(f''(n-\frac12))$ term.
It is worth mentioning that the integral expression also applies when $\tfrac{2\pi}{L}(n-\tfrac12) = \varepsilon_f$ if $\varepsilon_f \in \tfrac{2\pi}{L}(\mathbb{Z}+\tfrac12)$. In this case, the integral reproduces $\tfrac{\pi}{L}$. 
Therefore, $\delta E'[\Omega]$ is given by the integral 
\begin{align}
\delta E'[\Omega] = &
    \theta(\varepsilon_f\le 0) \cdot \varepsilon_f 
    + \frac{2}{L} \int_{-D\frac{L}{2\pi}}^0 \dd x  ~ \arctan \pare{\frac{\pi\Gamma}{\frac{2\pi}{L}x -\varepsilon_f} }  
= \theta(\varepsilon_f\le 0) \cdot \varepsilon_f  
    + \Gamma \int_{\frac{-D-\varepsilon_f}{\pi\Gamma}}^{-\frac{\varepsilon_f}{\pi\Gamma}} \dd \epsilon ~ \arctan \frac1{\epsilon} 
    \nonumber\\
=& \theta(\varepsilon_f\le 0) \cdot \varepsilon_f 
    + \Gamma \cdot \pare{  \frac12 \ln[\epsilon^2 +1] + \epsilon\cdot\arctan\frac{1}{\epsilon} } \Bigg|_{\frac{-D-\varepsilon_f}{\pi\Gamma}}^{-\frac{\varepsilon_f}{\pi\Gamma}} 
    \nonumber\\
=&  \theta(\varepsilon_f\le 0) \cdot \varepsilon_f 
-\Gamma - \Gamma \ln  \pare{\frac{D}{\Gamma}} 
 +\varepsilon_f \cdot \frac{\arctan \frac{\pi\Gamma}{\varepsilon_f}}{\pi} 
 - \frac{\Gamma\varepsilon_f}{D} + \mathcal{O}(L^{-2}) + \mathcal{O}(D^{-2})  \nonumber\\
\stackrel{\Gamma\gg |\varepsilon_f|}{=} & \quad  \frac12 \varepsilon_f - \Gamma - \Gamma \ln  \pare{\frac{D}{\Gamma}} 
    - \frac{\Gamma\varepsilon_f}{D} + \mathcal{O}(L^{-2}) + \mathcal{O}(D^{-2}) \ .
\end{align}
It is a smooth function as $\varepsilon_f$ crosses zero. 
In summary, we have 
\begin{equation}
    \delta E[\Omega] \quad \stackrel{D\to \infty}{=}\quad 
    -\theta(\varepsilon_f\le 0) \cdot \varepsilon_f + 
    \frac12 \varepsilon_f - \Gamma - \Gamma \ln \pare{ \frac{D}{\Gamma}} + \cdots\ . 
\end{equation}
The omitted terms include $\mathcal{O}(L^{-1})$ terms that are independent to $\varepsilon_f$, $\mathcal{O}(D^{-1})$ terms, and $\mathcal{O}(L^{-2})$ terms. 

\twocolumngrid

\end{document}